# Optimal Estimation of Co-heritability in High-dimensional Linear Models

Zijian Guo[1], Wanjie Wang[1,2], T. Tony Cai[1], and Hongzhe Li[2]

[1]Department of Statistics, The Wharton School, University of Pennsylvania
[2]Department of Biostatistics and Epidemiology, Perelman School of Medicine, University of Pennsylvania

**Abstract**

Co-heritability is an important concept that characterizes the genetic associations within pairs of quantitative traits. There has been significant recent interest in estimating the co-heritability based on data from the genome-wide association studies (GWAS). This paper introduces two measures of co-heritability in the high-dimensional linear model framework, including the inner product of the two regression vectors and a normalized inner product by their lengths. Functional de-biased estimators (FDEs) are developed to estimate these two co-heritability measures. In addition, estimators of quadratic functionals of the regression vectors are proposed. Both theoretical and numerical properties of the estimators are investigated. In particular, minimax rates of convergence are established and the proposed estimators of the inner product, the quadratic functionals and the normalized inner product are shown to be rate-optimal. Simulation results show that the FDEs significantly outperform the naive plug-in estimates. The FDEs are also applied to analyze a yeast segregant data set with multiple traits to estimate heritability and co-heritability among the traits.

[0]Zijian Guo is a Ph.D. student (Email: zijguo@wharon.upenn.edu). Wanjie Wang is a Postdoctoral Research Fellow (Email: wanjiew@wharton.upenn.edu). T. Tony Cai is Dorothy Silberberg Professor of Statistics (E-mail: tcai@wharton.upenn.edu). The research of T. Tony Cai was supported in part by NSF Grants DMS-1208982 and DMS-1403708, and NIH Grant R01 CA127334. Hongzhe Li is Professor (E-mail: hongzhe@upenn.edu). The research of Wanjie Wang and Hongzhe Li was supported in part by NIH grants CA127334 and GM097505.

# 1 Introduction

## 1.1 Motivation and Background

Genome-wide association studies (GWAS) have not only led to identification of thousands of genetic variants or single nucleotide polymorphisms (SNPs) that are associated various complex phenotypes (Manolio, 2010), they also provide important information to estimate the genetic heritability, a key population parameter that can help understand the genetic architecture of complex traits (Yang et al., 2010). Results from these GWAS have also shown that many complex phenotype shared comment genetic variants, including various autoimmune diseases (Zhernakova et al., 2009) and psychiatric disorders (Lee et al., 2013). These empirical evidence of shared genetic etiology for various phenotypes can inform nosology and encourages the investigation of common pathophysiologies for related disorders that can be explored for drug repositioning.

There has been a recent interest in estimating the genetic correlations between two traits based on the genome-wide aggregate effects of causal variants affecting both traits. The concept of co-heritability has been proposed to describe the genetic associations within pairs of quantitative traits based on GWAS data. This is in contrast to the traditional approaches estimating co-heritability based on twin or family studies. (Baradat, 1976) first generalized the concept of heritability and defined the coefficient of genetic prediction between two traits as the ratio of additive genetic covariance over the product of the phenotypic standard deviation of each trait. This idea was recently extended to define co-heritability using GWAS data by estimating the genetic risk scores (GRSs) for one phenotype and their association with a second phenotype (Purcell et al., 2009; Wray et al., 2007). This definition is ad hoc and does not account for the fact only a very small set of SNPs are expected to be associated with either of the two traits.

Lee et al. (2012) and Yang et al. (2013) extended the mixed-effect model framework for estimating the heritability to estimate genetic covariance and genetic correlation between two traits. In their models, each individual's phenotype is affected by a genetic random effect, which is correlated across individuals by virtue of sharing some of the genetic variants affecting the traits, and an environmental random effect, which is uncorrelated across individuals. They define co-heritability as the square-root of the ratio of the covariance of the genetic random effects to the product of the total variances. The mixed-effect model approach requires knowledge of the identity of the causal SNPs, and hence the covariance matrix. This is however not available, Lee et al. (2012) and Yang et al. (2013) approximate the genetic correlation between every pair of individuals across the set of causal SNPs by the genetic correlation across the set of all genotyped SNPs. The concept of using an estimated correlation matrix instead of the actual correlation matrix in estimating the heritability is a questionable heuristic (Golan and Rosset, 2011). The very large number of SNPs used for estimating the genetic correlations, most of them likely not causative, might lead to biased estimate of correlations due the set of causal SNPs. This has been clearly demonstrated by Golan and Rosset (2011).

The goal of this paper is to define two quantities that can be used for measuring the co-heritability between a pair of traits based on GWAS data in the framework of high-dimensional linear models. Instead of using the mixed-effects model framework, our definition of the co-heritability is similar in spirit to that of Baradat (1976). The proposed estimation procedure involves selecting the relevant genetic variants, which overcomes the problem of most of the SNPs in GWAS are not relevant to either of the two traits.

## 1.2 Problem Formulation

In this paper, a pair of trait values $(\mathbf{y}, \mathbf{w})$ are modeled as a linear combination of $p$ genetic variants and an error term that includes environmental and unmeasured genetic effects,

$$\boldsymbol{y}_{n_1\times 1} = \boldsymbol{X}_{n_1\times p}\boldsymbol{\beta}_{p\times 1} + \boldsymbol{\epsilon}_{n_1\times 1} \quad \text{and} \quad \boldsymbol{w}_{n_2\times 1} = \boldsymbol{Z}_{n_2\times p}\boldsymbol{\gamma}_{p\times 1} + \boldsymbol{\delta}_{n_2\times 1}, \tag{1}$$

where the rows $\boldsymbol{X}_{i\cdot}$ are i.i.d. $p$-dimensional Sub-gaussian random vectors with mean $\boldsymbol{\mu}_x$ and covariance matrix $\boldsymbol{\Sigma}$, the rows $\boldsymbol{Z}_{i\cdot}$ are i.i.d. $p$-dimensional Sub-gaussian random vectors with mean $\boldsymbol{\mu}_z$ and covariance matrix $\boldsymbol{\Gamma}$ and the error $(\boldsymbol{\epsilon}, \boldsymbol{\delta})^{\intercal}$ follows the multivariate normal distribution with mean zero and covariance

$$\begin{pmatrix} \sigma_1^2 \mathbf{I}_{n_1\times n_1} & \mathbf{0}_{n_1\times n_2} \\ \mathbf{0}_{n_2\times n_1} & \sigma_2^2 \mathbf{I}_{n_2\times n_2} \end{pmatrix}$$

and is assumed to be independent of $\boldsymbol{X}$ and $\boldsymbol{Z}$. Throughout the paper there are no particular assumptions on the relation between the random design matrices $\boldsymbol{X}$ and $\boldsymbol{Z}$, where they can be identical, correlated or independent.

In the study of genetic co-heritability, the pairs of traits $\boldsymbol{y}$ and $\boldsymbol{w}$ are assumed to have mean zero, and the $j$th column of $\boldsymbol{X}$, $\boldsymbol{X}_{\cdot j}$, and the $j$th column of $\boldsymbol{Z}$, $\boldsymbol{Z}_{\cdot j}$, are the numerically coded genetic markers at the $j$th genetic variant and are assumed to have variance 1. Under this model, if the columns of $\boldsymbol{X}$ and $\boldsymbol{Z}$ are independent, for the $i$-th observation,

$$\mathrm{Var}(\boldsymbol{y}_i) = \sum_j \boldsymbol{\beta}_j^2 + \sigma_1^2 = \|\boldsymbol{\beta}\|_2^2 + \sigma_1^2, \quad \text{and} \quad \mathrm{Var}(\boldsymbol{w}_i) = \sum_j \boldsymbol{\gamma}_j^2 + \sigma_2^2 = \|\boldsymbol{\gamma}\|_2^2 + \sigma_2^2,$$

therefore $||\boldsymbol{\beta}||_2^2$ and $||\boldsymbol{\gamma}||_2^2$ can then be interpreted as the narrow sense heritability.

Based on this model, one measure of genetic co-heritability is the inner product of the regression coefficients

$$\mathrm{I}\left(\boldsymbol{\beta}, \boldsymbol{\gamma}\right) = \langle \boldsymbol{\beta}, \boldsymbol{\gamma} \rangle \tag{2}$$

which measures the shared genetic effects between these two traits. Alternatively, a normalized inner product, that is the ratio

$$\mathrm{R}\left(\boldsymbol{\beta}, \boldsymbol{\gamma}\right) = \frac{\langle \boldsymbol{\beta}, \boldsymbol{\gamma} \rangle}{\|\boldsymbol{\beta}\|_2 \|\boldsymbol{\gamma}\|_2} \mathbf{1}(\|\boldsymbol{\beta}\|_2 \|\boldsymbol{\gamma}\|_2 > 0), \tag{3}$$

can also be used. The denominator $\|\boldsymbol{\beta}\|_2$ and $\|\boldsymbol{\gamma}\|_2$ represent the total genetic variants for two traits where the columns of the design matrix are independent. In the case where one of $\|\boldsymbol{\beta}\|_2$ and $\|\boldsymbol{\gamma}\|_2$ is vanishing, the ratio is defined to be zero, which means that there is no correlation between two traits when one of the regression vector is zero. With this normalization, $\mathrm{R}\left(\boldsymbol{\beta}, \boldsymbol{\gamma}\right)$ is always between -1 and 1 and can be used to compare co-heritability among multiple pairs.

The focus of this paper is to develop estimators for $\mathrm{I}\left(\boldsymbol{\beta}, \boldsymbol{\gamma}\right)$ and $\mathrm{R}\left(\boldsymbol{\beta}, \boldsymbol{\gamma}\right)$ based on two independent GWAS data with genotype data measured on the same set of genetic markers, denoted by $(\boldsymbol{y}_i, \boldsymbol{X}_{i\cdot}, i = 1, \cdots, n_1)$ and $(\boldsymbol{w}_i, \boldsymbol{Z}_{i\cdot}, i = 1, \cdots, n_2)$. A naive estimator is to estimate $\boldsymbol{\beta}$ and $\boldsymbol{\gamma}$ first and then plug in the estimators of $\boldsymbol{\beta}$ and $\boldsymbol{\gamma}$ into the expressions (2) and (3). For the problem of interest, usually there are many more genetic markers than the sample size, that is $p \gg \max\{n_1, n_2\}$. However, for any trait, one expect that only a few of these markers have nonzero effects. One can apply any high-dimensional regression methods such as Lasso (Tibshirani, 1996) and scaled Lasso (Sun and Zhang, 2012) and marginal regression with screening (Fan et al., 2012; McCarthy et al., 2008) to estimate these sparse regression coefficients. However, such plug-in estimators have several drawbacks in terms of estimating heritability and co-heritability. The Lasso approach shrinks the estimation towards 0, in particular, some weak effects might be shrunken to 0, yet the accumulation of these weak effects may contribute significant to the trait variability. It is possible that some genetic variants may have strong effects on one trait and weak effects on the other trait. Due to shrinkage, the plug-in of Lasso type estimators fails to capture this part of contribution to co-heritability from such genetic variants. Marginal regression calculates the regression score between the trait and each single marker (i.e., $\boldsymbol{y}$ and $\boldsymbol{X}_{\cdot j}$, $1 \leq j \leq p$), and screen for the large scores. This approach also suffers from the existence of weak effects, as the marginal scores must be large enough to survive in the screening step.

### 1.3 Methods and Main Results

In this paper, a new estimator, functional de-biased estimator of inner product and normalized inner product (FDE) is proposed to estimate the co-heritability measures $\mathrm{I}(\boldsymbol{\beta}, \boldsymbol{\gamma})$ defined in (2) and $\mathrm{R}(\boldsymbol{\beta}, \boldsymbol{\gamma})$ defined in (3). The proposed estimator for $\mathrm{I}(\boldsymbol{\beta}, \boldsymbol{\gamma})$ involves two steps, including estimating the inner product by the plug-in scaled Lasso estimator, and then correcting the plug-in scaled Lasso estimator. Similar two-step procedures are proposed to estimate the quadratic functionals $\|\boldsymbol{\beta}\|_2^2$ and $\|\boldsymbol{\gamma}\|_2^2$. To estimate the normalized inner product, we plug-in the estimators of the inner product and quadratic functionals into the definition (3).

FDE is shown to achieve the minimax optimal convergence rates of estimating $\mathrm{I}(\boldsymbol{\beta}, \boldsymbol{\gamma})$ and $\mathrm{R}(\boldsymbol{\beta}, \boldsymbol{\gamma})$. The correction step in FDE improves the estimation accuracy through correcting the shrinkage from Lasso and is crucial to achieve the optimal convergence rate. As demonstrated in the simulation studies, FDE consistently outperforms the plug-in estimator of the scaled Lasso estimator. In particular, FDE has significant improvement of estimating the co-heritability measures in the existence of some genetic markers having weak effects on one trait and strong effects on the other trait. In addition, FDE does not suffer from dependency among genetic markers. FDE works for a broad class of dependency structure of genetic markers, and it also allows dependency between two design matrices $\boldsymbol{X}$ and $\boldsymbol{Z}$.

The theoretical analysis given in Section 3 shows that the optimal rate of convergence for estimating $\mathrm{I}(\boldsymbol{\beta}, \boldsymbol{\gamma})$ is

$$(\|\boldsymbol{\beta}\|_2 + \|\boldsymbol{\gamma}\|_2)\left(\frac{1}{\sqrt{n}} + \frac{k \log p}{n}\right) + \frac{k \log p}{n},$$

where $p$ is the dimension, $n$ is the sample size, and $k$ is the maximum sparsity of $\boldsymbol{\beta}$ and $\boldsymbol{\gamma}$. The optimal rate depends not only on $p$, $n$ and $k$, but also the signal strength $\|\boldsymbol{\beta}\|_2$ and $\|\boldsymbol{\gamma}\|_2$. Restricting to the relatively strong signals with $\min\{\|\boldsymbol{\beta}\|_2, \|\boldsymbol{\gamma}\|_2\} \geq \eta_0 \geq Ck \log p/n$, where $C$ is some constant, the optimal convergence rate of estimating $\mathrm{R}(\boldsymbol{\beta}, \boldsymbol{\gamma})$ is

$$\frac{1}{\eta_0}\left(\frac{1}{\sqrt{n}} + \frac{k \log p}{n}\right) + \frac{1}{\eta_0^2}\frac{k \log p}{n}.$$

In contrast to estimating $\mathrm{I}(\boldsymbol{\beta}, \boldsymbol{\gamma})$, the optimal rate scales to the inverse of the signal strength, represented by $1/\eta_0$. The estimators $\widehat{\mathrm{I}}(\boldsymbol{\beta}, \boldsymbol{\gamma})$ and $\widehat{\mathrm{R}}(\boldsymbol{\beta}, \boldsymbol{\gamma})$ proposed in Section 3 adaptively achieve the optimal rates for estimating $\mathrm{I}(\boldsymbol{\beta}, \boldsymbol{\gamma})$ and $\mathrm{R}(\boldsymbol{\beta}, \boldsymbol{\gamma})$, respectively. In addition, the optimal rate of convergence for estimating the quadratic functionals $\mathrm{Q}(\boldsymbol{\beta}) = \|\boldsymbol{\beta}\|_2^2$ and $\mathrm{Q}(\boldsymbol{\gamma}) = \|\boldsymbol{\gamma}\|_2^2$ is also established.

### 1.4 Organization of the Paper

The rest of the paper is organized as follows. Section 2 presents the procedures for estimating $\mathrm{I}(\boldsymbol{\beta}, \boldsymbol{\gamma})$, $\mathrm{Q}(\boldsymbol{\beta})$, $\mathrm{Q}(\boldsymbol{\gamma})$, and $\mathrm{R}(\boldsymbol{\beta}, \boldsymbol{\gamma})$ in detail. In Section 3, minimax rates of convergence for the estimation problems are established and the proposed estimators are shown to attain the optimal rates. In Section 4, simulation studies are conducted to evaluate the empirical performance of the FDEs. A yeast cross data is used to illustrate the estimators in Section 5. The proofs of main theorems are present in Section 6. The remaining proofs and the extended simulation studies are given in the supplementary materials.

### 1.5 Notation and Definitions

Basic notation and definitions that will be used in the rest of the paper are defined here. For a matrix $\boldsymbol{X} \in \mathbf{R}^{n\times p}$, $\boldsymbol{X}_{i\cdot}$, $\boldsymbol{X}_{\cdot j}$, and $\boldsymbol{X}_{i,j}$ denote respectively the $i$-th row, $j$-th column, and $(i,j)$-th entry of the matrix $\boldsymbol{X}$, $\boldsymbol{X}_{i,-j}$ denotes the $i$-th row of $\boldsymbol{X}$ excluding the $j$-th coordinate, and $\boldsymbol{X}_{-j}$ denotes the sub-matrix of $\boldsymbol{X}$ excluding the $j$-th column. Let $[p] = \{1, 2, \cdots, p\}$. For a subset $J \subset [p]$, $\boldsymbol{X}_J$ denotes the sub-matrix of $\boldsymbol{X}$ consisting of columns $\boldsymbol{X}_{\cdot j}$ with $j \in J$ and for a vector $\boldsymbol{x} \in \mathbf{R}^p$, $\boldsymbol{x}_J$ is the sub-vector of $\boldsymbol{x}$ with indices in $J$ and $\boldsymbol{x}_{-J}$ is the sub-vector with indices in $J^c$. For a vector $\boldsymbol{x} \in \mathbf{R}^p$, the $\ell_q$ norm of $\boldsymbol{x}$ is defined as $\|\boldsymbol{x}\|_q = (\sum_{i=1}^{q} |\boldsymbol{x}_i|^q)^{\frac{1}{q}}$ for $q \geq 0$ with $\|\boldsymbol{x}\|_0$ denoting the cardinality of non-zero elements of $\boldsymbol{x}$ and $\|\boldsymbol{x}\|_\infty = \max_{1\leq j\leq p} |\boldsymbol{x}_j|$. For a matrix $A$ and $1 \leq q \leq \infty$, $\|\boldsymbol{A}\|_q = \sup_{\|\boldsymbol{x}\|_q=1} \|\boldsymbol{A}\boldsymbol{x}\|_q$ is the matrix $\ell_q$ operator norm. In particular, $\|\boldsymbol{A}\|_2$ is the spectral norm. For a symmetric matrix $\boldsymbol{A}$, $\lambda_{\min}(\boldsymbol{A})$ and $\lambda_{\max}(\boldsymbol{A})$ denote respectively the smallest and largest eigenvalue of $\boldsymbol{A}$. For a set $S$, $|S|$ denotes the cardinality of $S$. For $a \in \mathbf{R}$, $a_+ = \max\{a, 0\}$ and $\text{sign}(a)$ is the sign of $a$, i.e., $\text{sign}(a) = 1$ if $a > 0$, $\text{sign}(a) = -1$ if $a < 0$ and $\text{sign}(0) = 0$. Define the Sub-gaussian norm $\|\boldsymbol{x}\|_{\psi_2}^2$ of $\boldsymbol{x} \in \mathbf{R}^p$ as $\|\boldsymbol{x}\|_{\psi_2}^2 = \sup_{\boldsymbol{v}\in S^{p-1}} \sup_{q\geq 1} (\mathbf{E}|\boldsymbol{v}^\intercal \boldsymbol{x}|^q/q)^{\frac{1}{q}}$ where $S^{p-1}$ is the unit sphere in $\mathbf{R}^p$. The random vector $\boldsymbol{x} \in \mathbf{R}^p$ is defined to be Sub-gaussian if its corresponding Sub-gaussian norm is bounded; see Vershynin (2012) for more on Sub-gaussian random variables. For the design matriices $\boldsymbol{X} \in \mathbf{R}^{n_1\times p}$ and $\boldsymbol{Z} \in \mathbf{R}^{n_2\times p}$, we define the corresponding sample covariance matrices as $\widehat{\boldsymbol{\Sigma}} = \boldsymbol{X}^\intercal \boldsymbol{X}/n_1$ and $\widehat{\boldsymbol{\Gamma}} = \boldsymbol{Z}^\intercal \boldsymbol{Z}/n_2$. Let $z_{\alpha/2}$ denote the upper $\alpha/2$ quantile of the standard normal distribution. For two positive sequences $a_n$ and $b_n$, $a_n \lesssim b_n$ means $a_n \leq C b_n$ for all $n$ and $a_n \asymp b_n$ if $a_n \lesssim b_n$ and $b_n \lesssim a_n$. $c$ and $C$ are used to denote generic positive constants that may vary from place to place.

## 2 Estimation Methods

Details of the FDE estimation procedures are given in this Section. Estimation of $\mathrm{I}(\boldsymbol{\beta}, \boldsymbol{\gamma})$ in Section 2.1 is considered first and then estimation of the quadratic functionals $\mathrm{Q}(\boldsymbol{\beta}) = \|\boldsymbol{\beta}\|_2^2$ and $\mathrm{Q}(\boldsymbol{\gamma}) = \|\boldsymbol{\gamma}\|_2^2$ is given in Section 2.2. An estimator of the normalized inner product $\mathrm{R}(\boldsymbol{\beta}, \boldsymbol{\gamma})$ is given in Section 2.3 by combining the estimators of $\mathrm{I}(\boldsymbol{\beta}, \boldsymbol{\gamma})$, $\mathrm{Q}(\boldsymbol{\beta})$ and $\mathrm{Q}(\boldsymbol{\gamma})$ developed in Sections 2.1 and 2.2.

## 2.1 Estimation of $\mathrm{I}(\boldsymbol{\beta}, \boldsymbol{\gamma})$

Estimation of the inner product $\mathrm{I}(\boldsymbol{\beta}, \boldsymbol{\gamma}) = \langle \boldsymbol{\beta}, \boldsymbol{\gamma} \rangle$ is considered first since $\mathrm{I}(\boldsymbol{\beta}, \boldsymbol{\gamma})$ is of significant interest in its own right. In addition, constructing a good estimator of $\mathrm{I}(\boldsymbol{\beta}, \boldsymbol{\gamma})$ is also a key step in developing an optimal estimator for the normalized inner product $\mathrm{R}(\boldsymbol{\beta}, \boldsymbol{\gamma})$.

The scaled Lasso estimators for high-dimensional linear model (1) are defined through the following optimization algorithm (Sun and Zhang, 2012),

$$\{\widehat{\boldsymbol{\beta}}, \hat{\sigma}_1\} = \arg \min_{\boldsymbol{\beta} \in \mathbf{R}^p, \sigma_1 \in \mathbf{R}^+} \frac{\|\boldsymbol{y} - \boldsymbol{X}\boldsymbol{\beta}\|_2^2}{2n_1\sigma_1} + \frac{\sigma_1}{2} + \frac{\lambda_0}{\sqrt{n_1}} \sum_{j=1}^{p} \frac{\|\boldsymbol{X}_{\cdot j}\|_2}{\sqrt{n_1}} |\boldsymbol{\beta}_j|, \tag{4}$$

and

$$\{\widehat{\boldsymbol{\gamma}}, \hat{\sigma}_2\} = \arg \min_{\boldsymbol{\gamma} \in \mathbf{R}^p, \sigma_2 \in \mathbf{R}^+} \frac{\|\boldsymbol{w} - \boldsymbol{Z}\boldsymbol{\gamma}\|_2^2}{2n_2\sigma_2} + \frac{\sigma_2}{2} + \frac{\lambda_0}{\sqrt{n_2}} \sum_{j=1}^{p} \frac{\|\boldsymbol{Z}_{\cdot j}\|_2}{\sqrt{n_2}} |\boldsymbol{\gamma}_j|, \tag{5}$$

where $\lambda_0 = \sqrt{2.01 \log p}$.

A natural estimator of the inner product $\mathrm{I}(\boldsymbol{\beta}, \boldsymbol{\gamma})$ can then be obtained by simply plugging in the scaled Lasso estimators $\widehat{\boldsymbol{\beta}}$ and $\widehat{\boldsymbol{\gamma}}$ given in (4) and (5) respectively. However, as we will show, $\langle \widehat{\boldsymbol{\beta}}, \widehat{\boldsymbol{\gamma}} \rangle$ is not a good estimate of $\mathrm{I}(\boldsymbol{\beta}, \boldsymbol{\gamma})$ in general due to the bias in the Lasso or scaled Lasso estimators and other reasons. An alternative is to use the de-biased estimator, which was originally introduced for the construction of confidence intervals (Cai and Guo, 2016*a*; Javanmard and Montanari, 2014; van de Geer et al., 2014; Zhang and Zhang, 2014). Plugging in the de-biased estimator does not lead to a good estimator either; it has large variance. See Reamrk 2 for further discussions.

To construct an optimal estimator of $\mathrm{I}(\boldsymbol{\beta}, \boldsymbol{\gamma})$, it is helpful to decompose the difference between $\langle \widehat{\boldsymbol{\beta}}, \widehat{\boldsymbol{\gamma}} \rangle$ and $\langle \boldsymbol{\beta}, \boldsymbol{\gamma} \rangle$ into three terms,

$$\langle \widehat{\boldsymbol{\beta}}, \widehat{\boldsymbol{\gamma}} \rangle - \langle \boldsymbol{\beta}, \boldsymbol{\gamma} \rangle = \langle \widehat{\boldsymbol{\gamma}}, \widehat{\boldsymbol{\beta}} - \boldsymbol{\beta} \rangle + \langle \widehat{\boldsymbol{\beta}}, \widehat{\boldsymbol{\gamma}} - \boldsymbol{\gamma} \rangle - \langle \widehat{\boldsymbol{\beta}} - \boldsymbol{\beta}, \widehat{\boldsymbol{\gamma}} - \boldsymbol{\gamma} \rangle. \tag{6}$$

The last term on the right hand side, $\langle \widehat{\boldsymbol{\beta}} - \boldsymbol{\beta}, \widehat{\boldsymbol{\gamma}} - \boldsymbol{\gamma} \rangle$ is "small", but the first two terms $\langle \widehat{\boldsymbol{\gamma}}, \widehat{\boldsymbol{\beta}} - \boldsymbol{\beta} \rangle$ and $\langle \widehat{\boldsymbol{\beta}}, \widehat{\boldsymbol{\gamma}} - \boldsymbol{\gamma} \rangle$ can be large. Our approach is to first estimate these two terms and then subtract them from $\langle \widehat{\boldsymbol{\beta}}, \widehat{\boldsymbol{\gamma}} \rangle$ to obtain the final estimate of $\mathrm{I}(\boldsymbol{\beta}, \boldsymbol{\gamma})$.

Some intuition for estimating $\langle \widehat{\boldsymbol{\gamma}}, \boldsymbol{\beta} - \widehat{\boldsymbol{\beta}} \rangle$ is given here. Since

$$\frac{1}{n_1} \boldsymbol{X}^{\intercal} \left( \boldsymbol{y} - \boldsymbol{X}\widehat{\boldsymbol{\beta}} \right) = \widehat{\boldsymbol{\Sigma}} \left( \boldsymbol{\beta} - \widehat{\boldsymbol{\beta}} \right) + \frac{1}{n_1} \boldsymbol{X}^{\intercal} \epsilon, \tag{7}$$

multiplying both sides of (7) by a vector $\boldsymbol{u} \in \mathbf{R}^p$ yields

$$\frac{1}{n_1}\boldsymbol{u}^\intercal \boldsymbol{X}^\intercal \left(\boldsymbol{y} - \boldsymbol{X}\widehat{\boldsymbol{\beta}}\right) = \boldsymbol{u}^\intercal \widehat{\boldsymbol{\Sigma}} \left(\boldsymbol{\beta} - \widehat{\boldsymbol{\beta}}\right) + \frac{1}{n_1}\boldsymbol{u}^\intercal \boldsymbol{X}^\intercal \boldsymbol{\epsilon}, \tag{8}$$

which can be written as

$$\frac{1}{n_1}\boldsymbol{u}^\intercal \boldsymbol{X}^\intercal \left(\boldsymbol{y} - \boldsymbol{X}\widehat{\boldsymbol{\beta}}\right) - \langle \widehat{\boldsymbol{\gamma}}, \boldsymbol{\beta} - \widehat{\boldsymbol{\beta}}\rangle = \left(\widehat{\boldsymbol{\Sigma}}\boldsymbol{u} - \widehat{\boldsymbol{\gamma}}\right)^\intercal \left(\boldsymbol{\beta} - \widehat{\boldsymbol{\beta}}\right) + \frac{1}{n_1}\boldsymbol{u}^\intercal \boldsymbol{X}^\intercal \boldsymbol{\epsilon}. \tag{9}$$

If the vector $\boldsymbol{u}$ can be chosen such that the right hand side of (9) is "small", then $\boldsymbol{u}^\intercal \boldsymbol{X}^\intercal \left(\boldsymbol{y} - \boldsymbol{X}\widehat{\boldsymbol{\beta}}\right)/n_1$ is a good estimate of $\langle \widehat{\boldsymbol{\gamma}}, \boldsymbol{\beta} - \widehat{\boldsymbol{\beta}}\rangle$.

To construct $\boldsymbol{u}$, note that the first term on the right hand side of (9) can be further upper bounded as $\left|\left(\widehat{\boldsymbol{\Sigma}}\boldsymbol{u} - \widehat{\boldsymbol{\gamma}}\right)^\intercal \left(\boldsymbol{\beta} - \widehat{\boldsymbol{\beta}}\right)\right| \leq \|\widehat{\boldsymbol{\Sigma}}\boldsymbol{u} - \widehat{\boldsymbol{\gamma}}\|_\infty \|\widehat{\boldsymbol{\beta}} - \boldsymbol{\beta}\|_1$. To estimate $\langle \widehat{\boldsymbol{\gamma}}, \boldsymbol{\beta} - \widehat{\boldsymbol{\beta}}\rangle$, a projection vector $\boldsymbol{u}$ is constructed such that $\left(\widehat{\boldsymbol{\Sigma}}\boldsymbol{u} - \widehat{\boldsymbol{\gamma}}\right)^\intercal \left(\boldsymbol{\beta} - \widehat{\boldsymbol{\beta}}\right)$ is controlled through constraining the term $\|\widehat{\boldsymbol{\Sigma}}\boldsymbol{u} - \widehat{\boldsymbol{\gamma}}\|_\infty$ and the second term of (9) $\boldsymbol{u}^\intercal \boldsymbol{X}^\intercal \boldsymbol{\epsilon}/n_1$, which has mean 0, is controlled through minimizing its variance $\sigma_1^2 \boldsymbol{u}^\intercal \widehat{\boldsymbol{\Sigma}}\boldsymbol{u}/n_1$. This leads to the following convex optimization algorithm for identifying the projection vector $\boldsymbol{u}$ for estimating $\langle \widehat{\boldsymbol{\gamma}}, \boldsymbol{\beta} - \widehat{\boldsymbol{\beta}}\rangle$,

$$\widehat{\boldsymbol{u}}_1 = \underset{\boldsymbol{u}\in\mathbf{R}^p}{\arg\min} \left\{\boldsymbol{u}^\intercal \widehat{\boldsymbol{\Sigma}}\boldsymbol{u} : \|\widehat{\boldsymbol{\Sigma}}\boldsymbol{u} - \widehat{\boldsymbol{\gamma}}\|_\infty \leq \|\widehat{\boldsymbol{\gamma}}\|_2 \frac{\lambda_1}{\sqrt{n_1}}\right\}, \tag{10}$$

where $\lambda_1 = 12\lambda_{\max}^2(\boldsymbol{\Sigma})\sqrt{\log p}$.

Remark 1. The solution of the above optimization problem might not be unique and $\widehat{\boldsymbol{u}}_1$ is defined as any minimizer of the optimization problem. The theory established in Section 3 still holds for any minimizer of (10). The optimization problem (10) is solved through its equivalent Lagrange dual problem, which is computationally efficient and scales well to the high-dimensional problem. See Step 2 in Table 1 for more details.

Once the projection vector $\widehat{\boldsymbol{u}}_1$ is obtained, $\langle \widehat{\boldsymbol{\gamma}}, \boldsymbol{\beta} - \widehat{\boldsymbol{\beta}}\rangle$ is then estimated by $\widehat{\boldsymbol{u}}_1^\intercal \boldsymbol{X}^\intercal \left(\boldsymbol{y} - \boldsymbol{X}\widehat{\boldsymbol{\beta}}\right)/n_1$. Similarly, the projection vector for estimating $\langle \widehat{\boldsymbol{\beta}}, \boldsymbol{\gamma} - \widehat{\boldsymbol{\gamma}}\rangle$ can be obtained via the convex algorithm

$$\widehat{\boldsymbol{u}}_2 = \underset{\boldsymbol{u}\in\mathbf{R}^p}{\arg\min} \left\{\boldsymbol{u}^\intercal \widehat{\boldsymbol{\Gamma}}\boldsymbol{u} : \|\widehat{\boldsymbol{\Gamma}}\boldsymbol{u} - \widehat{\boldsymbol{\beta}}\|_\infty \leq \|\widehat{\boldsymbol{\beta}}\|_2 \frac{\lambda_2}{\sqrt{n_2}}\right\}, \tag{11}$$

where $\lambda_2 = 12\lambda_{\max}^2(\boldsymbol{\Gamma})\sqrt{\log p}$. Then $\langle \widehat{\boldsymbol{\beta}}, \boldsymbol{\gamma} - \widehat{\boldsymbol{\gamma}}\rangle$ is estimated by $\widehat{\boldsymbol{u}}_2^\intercal \boldsymbol{Z}^\intercal (\boldsymbol{w} - \boldsymbol{Z}\widehat{\boldsymbol{\gamma}})/n_2$.

The final estimator $\widehat{\mathrm{I}}(\boldsymbol{\beta}, \boldsymbol{\gamma})$ of $\mathrm{I}(\boldsymbol{\beta}, \boldsymbol{\gamma})$ is given by

$$\widehat{\mathrm{I}}(\boldsymbol{\beta}, \boldsymbol{\gamma}) = \langle \widehat{\boldsymbol{\beta}}, \widehat{\boldsymbol{\gamma}}\rangle + \widehat{\boldsymbol{u}}_1^\intercal \frac{1}{n_1}\boldsymbol{X}^\intercal \left(\boldsymbol{y} - \boldsymbol{X}\widehat{\boldsymbol{\beta}}\right) + \widehat{\boldsymbol{u}}_2^\intercal \frac{1}{n_2}\boldsymbol{Z}^\intercal (\boldsymbol{w} - \boldsymbol{Z}\widehat{\boldsymbol{\gamma}}). \tag{12}$$

It is clear from the above discussion that the key idea for the construction of the final estimator $\widehat{\mathrm{I}}(\boldsymbol{\beta},\boldsymbol{\gamma})$ is to identify the projection vectors $\widehat{\boldsymbol{u}}_1$ and $\widehat{\boldsymbol{u}}_2$ such that $\langle\widehat{\boldsymbol{\gamma}},\boldsymbol{\beta}-\widehat{\boldsymbol{\beta}}\rangle$ and $\langle\widehat{\boldsymbol{\beta}},\boldsymbol{\gamma}-\widehat{\boldsymbol{\gamma}}\rangle$ are well estimated. It will be shown in Section 3 that the estimator $\widehat{\mathrm{I}}(\boldsymbol{\beta},\boldsymbol{\gamma})$ is adaptively minimax rate-optimal.

Remark 2. As mentioned before, simply plugging in the Lasso, scaled Lasso, or de-biased estimator does not lead to a good estimator of $\mathrm{I}(\boldsymbol{\beta},\boldsymbol{\gamma})$. Another natural approach is to first threshold the de-biased estimator to obtain a sparse estimator of the coefficient vectors (see details in Zhang and Zhang (2014, Section 3.3), Guo et al. (2016, equation (10))) and then plug in. This estimator is referred as the thresholded estimator. Simulations in Section 4 demonstrate that the proposed estimator defined in (12) outperforms the three plug-in estimators using the scaled Lasso, de-biased, and thresholded estimators.

## 2.2 Estimation of $\mathrm{Q}(\boldsymbol{\beta})$ and $\mathrm{Q}(\boldsymbol{\gamma})$

In order to estimate the normalized inner product $\mathrm{R}(\boldsymbol{\beta},\boldsymbol{\gamma})$, it is necessary to estimate the quadratic functionals $\mathrm{Q}(\boldsymbol{\beta})=\|\boldsymbol{\beta}\|_2^2$ and $\mathrm{Q}(\boldsymbol{\gamma})=\|\boldsymbol{\gamma}\|_2^2$. To this end, we randomly split the data $(\boldsymbol{y},\boldsymbol{X})$ into two subsamples $\left(\boldsymbol{y}^{(1)},\boldsymbol{X}^{(1)}\right)$ with sample size $n_1/2$ and $\left(\boldsymbol{y}^{(2)},\boldsymbol{X}^{(2)}\right)$ with sample size $n_1/2$ and the data $(\boldsymbol{w},\boldsymbol{Z})$ into two subsamples $\left(\boldsymbol{w}^{(1)},\boldsymbol{Z}^{(1)}\right)$ with sample size $n_2/2$ and $\left(\boldsymbol{w}^{(2)},\boldsymbol{Z}^{(2)}\right)$ with sample size $n_2/2$.

With a slight abuse of notation, let $\widehat{\boldsymbol{\beta}}$ and $\widehat{\boldsymbol{\gamma}}$ denote the optimizers of the scaled Lasso algorithm (4) applied to $\left(\boldsymbol{y}^{(1)},\boldsymbol{X}^{(1)}\right)$ and (5) applied to $\left(\boldsymbol{w}^{(1)},\boldsymbol{Z}^{(1)}\right)$, respectively. For the scaled Lasso algorithms, the sample sizes $n_1$ and $n_2$ are replaced by $n_1/2$ and $n_2/2$, respectively. Again, the simple plug-in estimator $\|\widehat{\boldsymbol{\beta}}\|_2^2$ of $\mathrm{Q}(\boldsymbol{\beta})$ and $\|\widehat{\boldsymbol{\gamma}}\|_2^2$ of $\mathrm{Q}(\boldsymbol{\gamma})$ do not perform well. Note that

$$\|\widehat{\boldsymbol{\beta}}\|_2^2-\|\boldsymbol{\beta}\|_2^2=2\langle\widehat{\boldsymbol{\beta}},\widehat{\boldsymbol{\beta}}-\boldsymbol{\beta}\rangle-\|\widehat{\boldsymbol{\beta}}-\boldsymbol{\beta}\|_2^2. \tag{13}$$

The second term on the right hand side of (13) is "small", but the first can be large. The plug-in estimator $\|\widehat{\boldsymbol{\beta}}\|_2^2$ does not estimate $\|\boldsymbol{\beta}\|_2^2$ well for this reason. Similar ideas as in estimating $\langle\boldsymbol{\beta},\boldsymbol{\gamma}\rangle$ are applied here to estimate $\|\boldsymbol{\beta}\|_2^2$. Specifically, the term $2\langle\widehat{\boldsymbol{\beta}},\boldsymbol{\beta}-\widehat{\boldsymbol{\beta}}\rangle$ is estimated first and then is added it to $\|\widehat{\boldsymbol{\beta}}\|_2^2$ to obtain the final estimate of $\|\boldsymbol{\beta}\|_2^2$. To estimate

$\langle\widehat{\boldsymbol{\beta}}, \boldsymbol{\beta}-\widehat{\boldsymbol{\beta}}\rangle$, a projection vector $\boldsymbol{u}$is identified such that the following difference is controlled,

$$\frac{1}{n_1/2}\boldsymbol{u}^\intercal\left(\boldsymbol{X}^{(2)}\right)^\intercal\left(\boldsymbol{y}^{(2)}-\boldsymbol{X}^{(2)}\widehat{\boldsymbol{\beta}}\right)-\langle\widehat{\boldsymbol{\beta}}, \boldsymbol{\beta}-\widehat{\boldsymbol{\beta}}\rangle=\left(\boldsymbol{u}^\intercal\widehat{\boldsymbol{\Sigma}}^{(2)}-\widehat{\boldsymbol{\beta}}\right)\left(\boldsymbol{\beta}-\widehat{\boldsymbol{\beta}}\right)+\frac{1}{n_1/2}\boldsymbol{u}^\intercal\left(\boldsymbol{X}^{(2)}\right)^\intercal\boldsymbol{\epsilon}, \tag{14}$$

with $\widehat{\boldsymbol{\Sigma}}^{(2)}=\left(\boldsymbol{X}^{(2)}\right)^\intercal\boldsymbol{X}^{(2)}/(n_1/2)$. Let the projection vector $\widehat{\boldsymbol{u}}_3$ be the solution to the following optimization algorithm

$$\widehat{\boldsymbol{u}}_3=\underset{\boldsymbol{u}\in\mathbf{R}^p}{\arg\min}\left\{\boldsymbol{u}^\intercal\widehat{\boldsymbol{\Sigma}}^{(2)}\boldsymbol{u}:\|\widehat{\boldsymbol{\Sigma}}^{(2)}\boldsymbol{u}-\widehat{\boldsymbol{\beta}}\|_\infty\leq\|\widehat{\boldsymbol{\beta}}\|_2\frac{\lambda_1}{\sqrt{n_1/2}}\right\}, \tag{15}$$

where $\lambda_1=12\lambda_{\max}^2\left(\boldsymbol{\Sigma}\right)\sqrt{\log p}$. We then estimate $\langle\widehat{\boldsymbol{\beta}}, \boldsymbol{\beta}-\widehat{\boldsymbol{\beta}}\rangle$ by $\widehat{\boldsymbol{u}}_3^\intercal\left(\boldsymbol{X}^{(2)}\right)^\intercal\left(\boldsymbol{y}^{(2)}-\boldsymbol{X}^{(2)}\widehat{\boldsymbol{\beta}}\right)/(n_1/2)$ and propose the final estimator of $\|\boldsymbol{\beta}\|_2^2$ as

$$\widehat{\mathrm{Q}}\left(\boldsymbol{\beta}\right)=\left(\|\widehat{\boldsymbol{\beta}}\|_2^2+2\widehat{\boldsymbol{u}}_3^\intercal\frac{1}{n_1/2}\left(\boldsymbol{X}^{(2)}\right)^\intercal\left(\boldsymbol{y}^{(2)}-\boldsymbol{X}^{(2)}\widehat{\boldsymbol{\beta}}\right)\right)_+. \tag{16}$$

Similarly, the estimator of $\|\boldsymbol{\gamma}\|_2^2$ is given by

$$\widehat{\mathrm{Q}}\left(\boldsymbol{\gamma}\right)=\left(\|\widehat{\boldsymbol{\gamma}}\|_2^2+2\widehat{\boldsymbol{u}}_4^\intercal\frac{1}{n_2/2}\left(\boldsymbol{Z}^{(2)}\right)^\intercal\left(\boldsymbol{w}^{(2)}-\boldsymbol{Z}^{(2)}\widehat{\boldsymbol{\gamma}}\right)\right)_+, \tag{17}$$

where

$$\widehat{\boldsymbol{u}}_4=\underset{\boldsymbol{u}}{\arg\min}\left\{\boldsymbol{u}^\intercal\widehat{\boldsymbol{\Gamma}}^{(2)}\boldsymbol{u}:\|\widehat{\boldsymbol{\Gamma}}^{(2)}\boldsymbol{u}-\widehat{\boldsymbol{\gamma}}\|_\infty\leq\|\widehat{\boldsymbol{\gamma}}\|_2\frac{\lambda_2}{\sqrt{n_2/2}}\right\}, \tag{18}$$

with $\widehat{\boldsymbol{\Gamma}}^{(2)}=\left(\boldsymbol{Z}^{(2)}\right)^\intercal\left(\boldsymbol{Z}^{(2)}\right)/(n_2/2)$ and $\lambda_2=12\lambda_{\max}^2\left(\boldsymbol{\Gamma}\right)\sqrt{\log p}$.

Remark 3. Sample splitting is used here for the purpose of the theoretical analysis. In the Simulation Section, the performance of the proposed estimator without sample splitting is investigated; see Steps 5 - 7 in Table 1. The proposed estimator without sample splitting performs even better numerically than with sample splitting since more observations are used in constructing the initial estimators $\|\widehat{\boldsymbol{\beta}}\|_2^2$ and $\|\widehat{\boldsymbol{\gamma}}\|_2^2$ and the projection vectors $\widehat{\boldsymbol{u}}_3$ and $\widehat{\boldsymbol{u}}_4$.

## 2.3 Estimation of $\mathrm{R}\left(\boldsymbol{\beta},\boldsymbol{\gamma}\right)$

Given the estimators $\widehat{\mathrm{I}}\left(\boldsymbol{\beta},\boldsymbol{\gamma}\right)$, $\widehat{\mathrm{Q}}\left(\boldsymbol{\beta}\right)$ and $\widehat{\mathrm{Q}}\left(\boldsymbol{\gamma}\right)$ constructed in Sections 2.1 and 2.2, a natural estimator for the normalized inner product $\mathrm{R}\left(\boldsymbol{\beta},\boldsymbol{\gamma}\right)$ is given by

$$\widehat{\mathrm{R}}\left(\boldsymbol{\beta},\boldsymbol{\gamma}\right)=\mathrm{sign}\left(\widehat{\mathrm{I}}\left(\boldsymbol{\beta},\boldsymbol{\gamma}\right)\right)\cdot\min\left\{\frac{|\widehat{\mathrm{I}}\left(\boldsymbol{\beta},\boldsymbol{\gamma}\right)|}{\sqrt{\widehat{\mathrm{Q}}\left(\boldsymbol{\beta}\right)\widehat{\mathrm{Q}}\left(\boldsymbol{\gamma}\right)}}\mathbf{1}\left\{\widehat{\mathrm{Q}}\left(\boldsymbol{\beta}\right)\widehat{\mathrm{Q}}\left(\boldsymbol{\gamma}\right)>0\right\},1\right\}, \tag{19}$$

where $\widehat{\mathrm{I}}(\boldsymbol{\beta},\boldsymbol{\gamma})$, $\widehat{\mathrm{Q}}(\boldsymbol{\beta})$ and $\widehat{\mathrm{Q}}(\boldsymbol{\gamma})$ are estimators of $\langle\boldsymbol{\beta},\boldsymbol{\gamma}\rangle$, $\|\boldsymbol{\beta}\|_2^2$ and $\|\boldsymbol{\gamma}\|_2^2$ defined in (12), (16) and (17), respectively. It is possible that one of $\widehat{\mathrm{Q}}(\boldsymbol{\beta})$ and $\widehat{\mathrm{Q}}(\boldsymbol{\gamma})$ is 0 if $\|\boldsymbol{\beta}\|_2^2$ and $\|\boldsymbol{\gamma}\|_2^2$ are close to zero. In this case, the normalized inner product $\mathrm{R}(\boldsymbol{\beta},\boldsymbol{\gamma})$ is estimated as 0. Since $\mathrm{R}(\boldsymbol{\beta},\boldsymbol{\gamma})$ is always between $-1$ and 1, the estimator $\widehat{\mathrm{R}}(\boldsymbol{\beta},\boldsymbol{\gamma})$ is truncated to ensure that it is within the range.

The FDE algorithm without sample splitting for calculating the estimators $\widehat{\mathrm{I}}(\boldsymbol{\beta},\boldsymbol{\gamma})$, $\widehat{\mathrm{Q}}(\boldsymbol{\beta})$, $\widehat{\mathrm{Q}}(\boldsymbol{\gamma})$, and $\widehat{\mathrm{R}}(\boldsymbol{\beta},\boldsymbol{\gamma})$ is detailed in Table 1.

# 3 Theoretical Analysis

The theoretical properties of the estimators introduced in the last section are investigated in this secrtion. The upper bounds and lower bounds together show that the proposed estimators $\widehat{\mathrm{I}}(\boldsymbol{\beta},\boldsymbol{\gamma})$, $\widehat{\mathrm{Q}}(\boldsymbol{\beta})$, $\widehat{\mathrm{Q}}(\boldsymbol{\gamma})$, and $\widehat{\mathrm{R}}(\boldsymbol{\beta},\boldsymbol{\gamma})$ achieve the optimal rates of convergence.

## 3.1 Upper Bound Analysis

The samples sizes $n_1$ and $n_2$ are assumed to be of the same order, that is $n_1 \asymp n_2$. Let $n = \min\{n_1, n_2\}$ be the smallest of two sample sizes. The following assumptions are needed to facilitate the theoretical analysis.

(A1) The covariance matrices $\boldsymbol{\Sigma}$ and $\boldsymbol{\Gamma}$ satisfy $1/M_1 \le \lambda_{\min}(\boldsymbol{\Sigma}) \le \lambda_{\max}(\boldsymbol{\Sigma}) \le M_1$ and $1/M_1 \le \lambda_{\min}(\boldsymbol{\Gamma}) \le \lambda_{\max}(\boldsymbol{\Gamma}) \le M_1$, and the noise levels $\sigma_1$ and $\sigma_2$ satisfy $\max\{\sigma_1,\sigma_2\} \le M_2$, where $M_1 \ge 1$ and $M_2 > 0$ are positive constants.

(A2) The $\ell_2$ norms of the coefficient vectors $\boldsymbol{\beta}$ and $\boldsymbol{\gamma}$ are bounded away from zero in the sense that

$$\min\{\|\boldsymbol{\beta}\|_2, \|\boldsymbol{\gamma}\|_2\} \ge \eta_0 \ge C\sqrt{\frac{k\log p}{n}}, \quad \text{where } k = \max\{\|\boldsymbol{\beta}\|_0, \|\boldsymbol{\gamma}\|_0\}. \tag{20}$$

Assumption (A1) places a condition on the spectrum of the covariance matrices $\boldsymbol{\Sigma}$ and $\boldsymbol{\Gamma}$ and an upper bound on the noise levels $\sigma_1$ and $\sigma_2$. Assumption (A2) requires that the signal has to be bounded away from zero by $\eta_0$, which is only used in the upper bound analysis of the normalized inner product $\mathrm{R}(\boldsymbol{\beta},\boldsymbol{\gamma})$.

Table 1: FDE algorithm without sample splitting for estimating the inner product, lengths and normalized inner product.

---

Input: design matrices: $\boldsymbol{X}$, $\boldsymbol{Z}$; response vectors: $\boldsymbol{y}$, $\boldsymbol{w}$; tuning parameters $\lambda_0$,$\lambda$.

Output: $\widehat{\mathrm{I}}\,(\boldsymbol{\beta},\boldsymbol{\gamma})$, $\widehat{\mathrm{Q}}\,(\boldsymbol{\beta})$, $\widehat{\mathrm{Q}}\,(\boldsymbol{\gamma})$ and $\widehat{\mathrm{R}}\,(\boldsymbol{\beta},\boldsymbol{\gamma})$.

---

**Initial Lasso estimators:**

1. Scaled Lasso: Calculate $\widehat{\boldsymbol{\beta}}$ and $\widehat{\boldsymbol{\gamma}}$ from (4) and (5) with the tuning parameter $\lambda_0$.

**Inner product calculation:**

2. Projection vector $\widehat{\boldsymbol{u}}_1$: Calculate $\widehat{\boldsymbol{u}}_1 = \arg\min_{\boldsymbol{u}} \boldsymbol{u}^\intercal \boldsymbol{X}^\intercal \boldsymbol{X}\boldsymbol{u}/4n_1 + \boldsymbol{u}^\intercal\widehat{\boldsymbol{\gamma}} + \lambda^t\|\boldsymbol{u}\|_1$, where $\lambda^t = \lambda^{t-1}/1.5$, and $\lambda^0 = \lambda/\sqrt{n_1}$. Repeat until $\widehat{\boldsymbol{u}}_1$ can not be solved with $\lambda^\intercal$ replaced by $\lambda^{t+1}$, or $t \geq 10$.
3. Projection vector $\widehat{\boldsymbol{u}}_2$: Calculate $\widehat{\boldsymbol{u}}_2 = \arg\min_{\boldsymbol{u}} \boldsymbol{u}^\intercal \boldsymbol{Z}^\intercal \boldsymbol{Z}\boldsymbol{u}/4n_2 + \boldsymbol{u}^\intercal\widehat{\boldsymbol{\beta}} + \lambda^\intercal\|\boldsymbol{u}\|_1$, where $\lambda^\intercal = \lambda^{t-1}/1.5$, and $\lambda^0 = \lambda/\sqrt{n_1}$. Repeat until $\widehat{\boldsymbol{u}}_2$ can not be solved with $\lambda^\intercal$ replaced by $\lambda^{t+1}$, or $t \geq 10$.
4. Correction: $\widehat{\mathrm{I}}\,(\boldsymbol{\beta},\boldsymbol{\gamma}) = \langle\widehat{\boldsymbol{\beta}},\widehat{\boldsymbol{\gamma}}\rangle + \widehat{\boldsymbol{u}}_1^\intercal \boldsymbol{X}^\intercal(\boldsymbol{y} - \boldsymbol{X}\widehat{\boldsymbol{\beta}})/n_1 + \widehat{\boldsymbol{u}}_2^\intercal \boldsymbol{Z}^\intercal(\boldsymbol{w} - \boldsymbol{Z}\widehat{\boldsymbol{\gamma}})/n_2$.

**Quadratic functional calculation:**

5. Projection vector $\widehat{\boldsymbol{u}}_3$: Calculate $\widehat{\boldsymbol{u}}_3 = \arg\min_{\boldsymbol{u}} \boldsymbol{u}^\intercal \boldsymbol{X}^\intercal \boldsymbol{X}\boldsymbol{u}/4n_1 + \boldsymbol{u}^\intercal\widehat{\boldsymbol{\beta}} + \lambda^\intercal\|\boldsymbol{u}\|_1$, where $\lambda^\intercal = \lambda^{t-1}/1.5$, and $\lambda^0 = \lambda/\sqrt{n_1}$. Repeat until $\widehat{\boldsymbol{u}}_3$ can not be solved with $\lambda^\intercal$ replaced by $\lambda^{t+1}$, or $t \geq 10$.
6. Projection vector $\widehat{\boldsymbol{u}}_4$: Calculate $\widehat{\boldsymbol{u}}_4 = \arg\min_{\boldsymbol{u}} \boldsymbol{u}^\intercal \boldsymbol{Z}^\intercal \boldsymbol{Z}\boldsymbol{u}/4n_2 + \boldsymbol{u}^\intercal\widehat{\boldsymbol{\gamma}} + \lambda^\intercal\|\boldsymbol{u}\|_1$, where $\lambda^\intercal = \lambda^{t-1}/1.5$, and $\lambda^0 = \lambda/\sqrt{n_2}$. Repeat until $\widehat{\boldsymbol{u}}_1$ can not be solved with $\lambda^\intercal$ replaced by $\lambda^{t+1}$, or $t \geq 10$.
7. Correction: $\widehat{\mathrm{Q}}\,(\boldsymbol{\beta}) = \left(\|\widehat{\boldsymbol{\beta}}\|^2 + 2\widehat{\boldsymbol{u}}_3^\intercal \boldsymbol{X}^\intercal(\boldsymbol{y} - \boldsymbol{X}\widehat{\boldsymbol{\beta}}/n_1)\right)_+$,
$\widehat{\mathrm{Q}}\,(\boldsymbol{\gamma}) = \left(\|\widehat{\boldsymbol{\gamma}}\|^2 + 2\widehat{\boldsymbol{u}}_4^\intercal \boldsymbol{Z}^\intercal(\boldsymbol{w} - \boldsymbol{Z}\widehat{\boldsymbol{\gamma}})/n_2\right)_+$.

**Ratio calculation:**

8. $\widehat{\mathrm{R}}\,(\boldsymbol{\beta},\boldsymbol{\gamma}) = \mathrm{sign}(\widehat{\mathrm{I}}\,(\boldsymbol{\beta},\boldsymbol{\gamma})) \cdot \min\left\{\left(|\widehat{\mathrm{I}}\,(\boldsymbol{\beta},\boldsymbol{\gamma})\,|/\sqrt{\widehat{\mathrm{Q}}\,(\boldsymbol{\beta})\,\widehat{\mathrm{Q}}\,(\boldsymbol{\gamma})}\right)\mathbf{1}\left\{\widehat{\mathrm{Q}}\,(\boldsymbol{\beta})\,\widehat{\mathrm{Q}}\,(\boldsymbol{\gamma}) > 0\right\}, 1\right\}$.

---

The following theorem establishes the convergence rates of the estimators $\widehat{\mathrm{I}}(\boldsymbol{\beta}, \boldsymbol{\gamma})$, $\widehat{\mathrm{Q}}(\boldsymbol{\beta})$ and $\widehat{\mathrm{Q}}(\boldsymbol{\gamma})$, proposed in (12), (16) and (17), respectively.

THEOREM 1. *Suppose the assumption* (A1) *holds and* $k \log p/n \to 0$ *with* $k = \max\{\|\boldsymbol{\beta}\|_0, \|\boldsymbol{\gamma}\|_0\}$. *Then for any fixed constant* $0 < \alpha < 0.5$, *with probability at least* $1 - 2\alpha - p^{-c}$, *we have*

$$\left|\widehat{\mathrm{I}}(\boldsymbol{\beta}, \boldsymbol{\gamma}) - \mathrm{I}(\boldsymbol{\beta}, \boldsymbol{\gamma})\right| \leq C\left(\|\boldsymbol{\beta}\|_2 + \|\boldsymbol{\gamma}\|_2\right)\left(\frac{z_{\alpha/2}}{\sqrt{n}} + \frac{k \log p}{n}\right) + C\frac{k \log p}{n}, \tag{21}$$

$$\left|\widehat{\mathrm{Q}}(\boldsymbol{\beta}) - \mathrm{Q}(\boldsymbol{\beta})\right| \leq C\|\boldsymbol{\beta}\|_2\left(\frac{z_{\alpha/2}}{\sqrt{n}} + \frac{k \log p}{n}\right) + C\frac{k \log p}{n}, \tag{22}$$

*and*

$$\left|\widehat{\mathrm{Q}}(\boldsymbol{\gamma}) - \mathrm{Q}(\boldsymbol{\gamma})\right| \leq C\|\boldsymbol{\gamma}\|_2\left(\frac{z_{\alpha/2}}{\sqrt{n}} + \frac{k \log p}{n}\right) + C\frac{k \log p}{n}, \tag{23}$$

*where* $c$ *and* $C$ *are positive constants.*

The upper bound of estimating $\langle \boldsymbol{\beta}, \boldsymbol{\gamma}\rangle$ not only depends on $k, n$ and $p$, but also scales to the signal strengths $\|\boldsymbol{\beta}\|_2$ and $\|\boldsymbol{\gamma}\|_2$. For the estimation of the quadratic functional $\mathrm{Q}(\boldsymbol{\beta})$, the convergence rate depends on $\|\boldsymbol{\beta}\|_2$. The similar phenomenon holds for estimation of $\mathrm{Q}(\boldsymbol{\gamma})$. The following theorem establishes the convergence rate of the estimator $\widehat{\mathrm{R}}(\boldsymbol{\beta}, \boldsymbol{\gamma})$ proposed in (19).

THEOREM 2. *Suppose the assumptions* (A1) *and* (A2) *hold and* $k \log p/n \to 0$ *with* $k = \max\{\|\boldsymbol{\beta}\|_0, \|\boldsymbol{\gamma}\|_0\}$. *Then for any fixed constant* $0 < \alpha < 0.5$, *with probability at least* $1 - \alpha - p^{-c}$, *we have*

$$\left|\widehat{\mathrm{R}}(\boldsymbol{\beta}, \boldsymbol{\gamma}) - \mathrm{R}(\boldsymbol{\beta}, \boldsymbol{\gamma})\right| \leq C\frac{1}{\eta_0}\left(\frac{z_{\alpha/2}}{\sqrt{n}} + \frac{k \log p}{n}\right) + C\frac{1}{\eta_0^2}\frac{k \log p}{n}, \tag{24}$$

*where* $c$ *and* $C$ *are positive constants.*

In contrast to Theorem 1, Theorem 2 requires the extra assumption (A2) on the signal strengths $\|\boldsymbol{\beta}\|_2$ and $\|\boldsymbol{\gamma}\|_2$. The convergence rate of estimating $\mathrm{R}(\boldsymbol{\beta}, \boldsymbol{\gamma})$ is scaled to the inverse of signal strengths, $1/\eta_0$. This is different from the error bound in Theorem 1, where the estimation accuracy is scaled to the $\ell_2$ norm. The lower bound results established in Theorem 3 will demonstrate the necessity of Assumption (A2) for estimation of $\mathrm{R}(\boldsymbol{\beta}, \boldsymbol{\gamma})$.

## 3.2 Minimax Lower Bound

This section establishes the minimax lower bounds that show the optimality of the proposed estimators in Section 2. We first introduce the parameter spaces for $\boldsymbol{\theta} = (\boldsymbol{\beta}, \boldsymbol{\Sigma}, \sigma_1, \boldsymbol{\gamma}, \boldsymbol{\Gamma}, \sigma_2)$, which is the product of parameter spaces for $\boldsymbol{\theta}_{\boldsymbol{\beta}} = (\boldsymbol{\beta}, \boldsymbol{\Sigma}, \sigma_1)$ and $\boldsymbol{\theta}_{\boldsymbol{\gamma}} = (\boldsymbol{\gamma}, \boldsymbol{\Gamma}, \sigma_2)$. To start with, we introduce the following parameter space for both $\boldsymbol{\theta}_{\boldsymbol{\beta}} = (\boldsymbol{\beta}, \boldsymbol{\Sigma}, \sigma_1)$ and $\boldsymbol{\theta}_{\boldsymbol{\gamma}} = (\boldsymbol{\gamma}, \boldsymbol{\Gamma}, \sigma_2)$,

$$\mathcal{G}(k) = \left\{ (\boldsymbol{\beta}, \boldsymbol{\Sigma}, \sigma_1) : \|\boldsymbol{\beta}\|_0 \le k, \ \frac{1}{M_1} \le \lambda_{\min}(\boldsymbol{\Sigma}) \le \lambda_{\max}(\boldsymbol{\Sigma}) \le M_1, \ \sigma_1 \le M_2 \right\}, \tag{25}$$

where $M_1 \ge 1$ and $M_2 > 0$ are positive constants. The parameter space defined in (25) requires the sparsity $\|\boldsymbol{\beta}\|_0 \le k$. The other conditions $1/M_1 \le \lambda_{\min}(\boldsymbol{\Sigma}) \le \lambda_{\max}(\boldsymbol{\Sigma}) \le M_1$ and $\sigma_1 \le M_2$ are regularity conditions. Based on the definition (25), the parameter space for $(\boldsymbol{\beta}, \boldsymbol{\Sigma}, \sigma_1, \boldsymbol{\gamma}, \boldsymbol{\Gamma}, \sigma_2)$ is defined as a product of two parameter spaces,

$$\Theta(k) = \mathcal{G}(k) \times \mathcal{G}(k) = \{\boldsymbol{\theta} = (\boldsymbol{\beta}, \boldsymbol{\Sigma}, \sigma_1, \boldsymbol{\gamma}, \boldsymbol{\Gamma}, \sigma_2) \ : \ (\boldsymbol{\beta}, \boldsymbol{\Sigma}, \sigma_1) \in \mathcal{G}(k), \ (\boldsymbol{\gamma}, \boldsymbol{\Gamma}, \sigma_2) \in \mathcal{G}(k)\}. \tag{26}$$

Define a proper subspace $\mathcal{G}(k, \eta_0)$ of $\mathcal{G}(k)$ as

$$\mathcal{G}(k, \eta_0) = \mathcal{G}(k) \cap \{\|\boldsymbol{\beta}\|_2 \ge \eta_0\}, \tag{27}$$

where $\eta_0 \ge 0$. Compared with $\mathcal{G}(k)$, the parameter space $\mathcal{G}(k, \eta_0)$ is restricted to signals whose $\ell_2$ norm is bounded away from 0 by $\eta_0$. A proper subspace of $\Theta(k)$ for $\boldsymbol{\theta} = (\boldsymbol{\beta}, \boldsymbol{\Sigma}, \sigma_1, \boldsymbol{\gamma}, \boldsymbol{\Gamma}, \sigma_2)$ can be defined as

$$\Theta(k, \eta_0) = \{\boldsymbol{\theta} = (\boldsymbol{\beta}, \boldsymbol{\Sigma}, \sigma_1, \boldsymbol{\gamma}, \boldsymbol{\Gamma}, \sigma_2) \ : \ (\boldsymbol{\beta}, \boldsymbol{\Sigma}, \sigma_1) \in \mathcal{G}(k, \eta_0), \ (\boldsymbol{\gamma}, \boldsymbol{\Gamma}, \sigma_2) \in \mathcal{G}(k, \eta_0)\}. \tag{28}$$

The following theorem establishes the minimax lower bounds for the convergence rates of estimating the inner product $\langle \boldsymbol{\beta}, \boldsymbol{\gamma} \rangle$, the quadratic functionals $\mathrm{Q}(\boldsymbol{\beta})$ and $\mathrm{Q}(\boldsymbol{\gamma})$ and the normalized inner product $\mathrm{R}(\boldsymbol{\beta}, \boldsymbol{\gamma})$.

Theorem 3. *Suppose $k \le c \min\{n/\log p, p^{\nu}\}$ for some constants $c > 0$ and $0 < \nu < 1/2$, then*

$$\inf_{\widehat{\mathrm{I}}} \sup_{\boldsymbol{\theta} \in \Theta(k)} \mathbb{P}_{\boldsymbol{\theta}} \left( \left| \widehat{\mathrm{I}} - \mathrm{I}(\boldsymbol{\beta}, \boldsymbol{\gamma}) \right| \ge c' \left( (\|\boldsymbol{\beta}\|_2 + \|\boldsymbol{\gamma}\|_2) \left( \frac{1}{\sqrt{n}} + \frac{k \log p}{n} \right) + \frac{k \log p}{n} \right) \right) \ge \frac{1}{4}, \tag{29}$$

$$\inf_{\widehat{\mathrm{Q}}} \sup_{\boldsymbol{\theta}_{\boldsymbol{\beta}} \in \mathcal{G}(k)} \mathbb{P}_{\boldsymbol{\theta}_{\boldsymbol{\beta}}} \left( \left| \widehat{\mathrm{Q}} - \mathrm{Q}(\boldsymbol{\beta}) \right| \geq c' \left( \|\boldsymbol{\beta}\|_2 \left( \frac{1}{\sqrt{n}} + \frac{k \log p}{n} \right) + \frac{k \log p}{n} \right) \right) \geq \frac{1}{4}, \tag{30}$$

$$\inf_{\widehat{\mathrm{Q}}} \sup_{\boldsymbol{\theta}_{\boldsymbol{\gamma}} \in \mathcal{G}(k)} \mathbb{P}_{\boldsymbol{\theta}_{\boldsymbol{\gamma}}} \left( \left| \widehat{\mathrm{Q}} - \mathrm{Q}(\boldsymbol{\gamma}) \right| \geq c' \left( \|\boldsymbol{\gamma}\|_2 \left( \frac{1}{\sqrt{n}} + \frac{k \log p}{n} \right) + \frac{k \log p}{n} \right) \right) \geq \frac{1}{4}, \tag{31}$$

$$\inf_{\widehat{\mathrm{R}}} \sup_{\boldsymbol{\theta} \in \Theta(k, \eta_0)} \mathbb{P}_{\boldsymbol{\theta}} \left( \left| \widehat{\mathrm{R}} - \mathrm{R}(\boldsymbol{\beta}, \boldsymbol{\gamma}) \right| \geq c' \min \left\{ \frac{1}{\eta_0} \left( \frac{1}{\sqrt{n}} + \frac{k \log p}{n} \right) + \frac{1}{\eta_0^2} \frac{k \log p}{n}, 1 \right\} \right) \geq \frac{1}{4}, \tag{32}$$

*where $c' > 0$ is a positive constant.*

Combined with Theorem 1, the above theorem implies that the estimators $\widehat{\mathrm{I}}(\boldsymbol{\beta}, \boldsymbol{\gamma})$, $\widehat{\mathrm{Q}}(\boldsymbol{\beta})$, $\widehat{\mathrm{Q}}(\boldsymbol{\gamma})$ proposed in (12),(16) and (17) achieve the minimax lower bounds (29), (30) and (31), so they are minimax rate-optimal estimators of $\mathrm{I}(\boldsymbol{\beta}, \boldsymbol{\gamma})$, $\mathrm{Q}(\boldsymbol{\beta})$ and $\mathrm{Q}(\boldsymbol{\gamma})$, respectively. For estimation of $\mathrm{R}(\boldsymbol{\beta}, \boldsymbol{\gamma})$, under the assumption that $\eta_0 \geq C\sqrt{k \log p / n}$, Theorem 2 shows that the estimator $\widehat{\mathrm{R}}(\boldsymbol{\beta}, \boldsymbol{\gamma})$ proposed in (19) achieves the minimax lower bound $1/\eta_0 \times (1/\sqrt{n} + k \log p / n) + 1/\eta_0^2 \times k \log p / n$ in (32). Hence $\widehat{\mathrm{R}}(\boldsymbol{\beta}, \boldsymbol{\gamma})$ is the rate-optimal estimator of $\mathrm{R}(\boldsymbol{\beta}, \boldsymbol{\gamma})$ under the assumption that the signal strengths $\|\boldsymbol{\beta}\|_2$ and $\|\boldsymbol{\gamma}\|_2$ are bounded away from zero. However, if $\eta_0 \leq c\sqrt{k \log p / n}$, the estimation of $\mathrm{R}(\boldsymbol{\beta}, \boldsymbol{\gamma})$ is not interesting since the trivial estimator 0 will achieve the constant minimax lower bound in this case, which demonstrates the necessity of Assumption (A2).

# 4 Simulation Evalulations and Comparisons

In this section, the performance of several estimators for $\mathrm{I}(\boldsymbol{\beta}, \boldsymbol{\gamma})$ and $\mathrm{R}(\boldsymbol{\beta}, \boldsymbol{\gamma})$ were compared in two different settings. These estimators included plug-in scaled Lasso estimator (Sun and Zhang, 2012), plug-in de-biased estimator (Cai and Guo, 2016*a*; Javanmard and Montanari, 2014; van de Geer et al., 2014; Zhang and Zhang, 2014), plug-in thresholded estimator (Zhang and Zhang, 2014, Section 3.3) and the proposed estimator FDE. Details about these estimators are listed as follows.

- Plug-in scaled Lasso estimator (Lasso): The inner product $\mathrm{I}(\boldsymbol{\beta}, \boldsymbol{\gamma})$ is estimated by $\langle \widehat{\boldsymbol{\beta}}, \widehat{\boldsymbol{\gamma}} \rangle$ and the normalized inner product $\mathrm{R}(\boldsymbol{\beta}, \boldsymbol{\gamma})$ is estimated by

$$\left[ \langle \widehat{\boldsymbol{\beta}}, \widehat{\boldsymbol{\gamma}} \rangle / \left( \|\widehat{\boldsymbol{\beta}}\|_2 \|\widehat{\boldsymbol{\gamma}}\|_2 \right) \right] \mathbf{1} \left\{ \|\widehat{\boldsymbol{\beta}}\|_2 \|\widehat{\boldsymbol{\gamma}}\|_2 > 0 \right\}.$$

- Plug-in de-biased estimator (De-biased): Denote the de-biased Lasso estimators as $\widetilde{\boldsymbol{\beta}}$ and $\widetilde{\boldsymbol{\gamma}}$. The inner product $\mathrm{I}(\boldsymbol{\beta}, \boldsymbol{\gamma})$ is estimated by $\langle \widetilde{\boldsymbol{\beta}}, \widetilde{\boldsymbol{\gamma}} \rangle$ and the normalized inner product $\mathrm{R}(\boldsymbol{\beta}, \boldsymbol{\gamma})$ is estimated by $\left[\langle \widetilde{\boldsymbol{\beta}}, \widetilde{\boldsymbol{\gamma}} \rangle / \left(\|\widetilde{\boldsymbol{\beta}}\|_2 \|\widetilde{\boldsymbol{\gamma}}\|_2\right)\right] \mathbf{1}\left\{\|\widetilde{\boldsymbol{\beta}}\|_2 \|\widetilde{\boldsymbol{\gamma}}\|_2 > 0\right\}$.

- Plug-in thresholded estimator (Thresholded): Denote the thresholded estimators as $\bar{\boldsymbol{\beta}}$ and $\bar{\boldsymbol{\gamma}}$. The inner product $\mathrm{I}(\boldsymbol{\beta}, \boldsymbol{\gamma})$ is estimated by $\langle \bar{\boldsymbol{\beta}}, \bar{\boldsymbol{\gamma}} \rangle$ and the normalized inner product $\mathrm{R}(\boldsymbol{\beta}, \boldsymbol{\gamma})$ is estimated by $\left[\langle \bar{\boldsymbol{\beta}}, \bar{\boldsymbol{\gamma}} \rangle / \left(\|\bar{\boldsymbol{\beta}}\|_2 \|\bar{\boldsymbol{\gamma}}\|_2\right)\right] \mathbf{1}\left\{\|\bar{\boldsymbol{\beta}}\|_2 \|\bar{\boldsymbol{\gamma}}\|_2 > 0\right\}$.

- FDE: The inner product $\mathrm{I}(\boldsymbol{\beta}, \boldsymbol{\gamma})$ is estimated by $\widehat{\mathrm{I}}(\boldsymbol{\beta}, \boldsymbol{\gamma})$ in (12) and the ratio $\mathrm{R}(\boldsymbol{\beta}, \boldsymbol{\gamma})$ is estimated by $\widehat{\mathrm{R}}(\boldsymbol{\beta}, \boldsymbol{\gamma})$ in (19). We consider FDE with sample splitting (FDE-S) and without sample splitting (FDE-NS) for $\widehat{\mathrm{R}}(\boldsymbol{\beta}, \boldsymbol{\gamma})$.

Implementation of the de-biased, thresholded and FDE estimators requires the scaled Lasso estimators $\widehat{\boldsymbol{\beta}}$ and $\widehat{\boldsymbol{\gamma}}$ in the initial step. The scaled Lasso estimator was implemented by the equivalent square-root Lasso algorithm (Belloni et al., 2011). The theoretical tuning parameter is $\lambda_0 = \sqrt{2.01 \log p/n}$, which may be conservative in the numerical studies. Instead, the tuning parameter was chosen as $\lambda_0 = b\sqrt{2.01 \log p}$. However, the performances of all estimators were evaluated across a grid of tuning parameter values $b \in \{.25, .5, .75, 1\}$ (see Supplementary Material, Section A.1). The results showed that $b = .5$ was a good choice for all the estimators. Hence, $\lambda_0 = .5\sqrt{2.01 \log p/n}$ was used for the numerical studies in this section and Section 5. To implement the FDE algorithm, the other tuning parameter $\lambda$ as $\sqrt{2.01 \log p/n}$ for the correction Steps 2,3,5 and 6 are given in Table 1 .

Comparisons of estimates of $\mathrm{I}(\boldsymbol{\beta}, \boldsymbol{\gamma})$ and $\mathrm{R}(\boldsymbol{\beta}, \boldsymbol{\gamma})$ are presented below. Results on estimating the quadratic functionals are presented in the Supplementary Material, Section A.2. For each setting, with the parameters $(p, n_1, n_2, s, s_1, s_2)$, $\boldsymbol{\Sigma}$, $\boldsymbol{\Gamma}$, $F_{\boldsymbol{\beta}}$, $F_{\boldsymbol{\gamma}}$ specified, the following steps were implemented:

1. Generate sets $S_1 \subset [p]$ and $S_2 \subset [p]$, with $|S_1| = s_1$, $|S_2| = s_2$ and $|S_1 \cap S_2| = s$. For $\boldsymbol{\beta} \in \mathbf{R}^p$ and $\boldsymbol{\gamma} \in \mathbf{R}^p$, generate $\boldsymbol{\beta}_j \sim F_{\boldsymbol{\beta}}$ and $\boldsymbol{\gamma}_k \sim F_{\boldsymbol{\gamma}}$, for $j \in S_1$, $k \in S_2$, and set $\boldsymbol{\beta}_j = 0$ and $\boldsymbol{\gamma}_k = 0$, for $j \notin S_1$, $k \notin S_2$.

2. Generate $\boldsymbol{X}_{i\cdot} \overset{\text{i.i.d}}{\sim} N(\mathbf{0}, \boldsymbol{\Sigma})$, $1 \leq i \leq n_1$, and $\boldsymbol{Z}_{i\cdot} \overset{\text{i.i.d}}{\sim} N(\mathbf{0}, \boldsymbol{\Gamma})$, $1 \leq i \leq n_2$.

3. Generate the noise $\boldsymbol{\epsilon}_i \overset{\text{i.i.d}}{\sim} N(0,1)$, $1 \le i \le n_1$, and $\boldsymbol{\delta}_i \overset{\text{i.i.d}}{\sim} N(0,1)$, $1 \le i \le n_2$. Generate the outcome as $\boldsymbol{y} = \boldsymbol{X}\boldsymbol{\beta} + \boldsymbol{\epsilon}$ and $\boldsymbol{w} = \boldsymbol{Z}\boldsymbol{\gamma} + \boldsymbol{\delta}$.

4. With $\boldsymbol{X}$, $\boldsymbol{y}$, $\boldsymbol{Z}$, and $\boldsymbol{w}$, estimate $\mathrm{I}(\boldsymbol{\beta},\boldsymbol{\gamma})$ and $\mathrm{R}(\boldsymbol{\beta},\boldsymbol{\gamma})$ through different estimators.

5. Repeat 2-4 for *rep* replications.

For a given quantity T, denote the estimator from $l$-th replication as $\widehat{\mathrm{T}}(\boldsymbol{X},\boldsymbol{y},\boldsymbol{Z},\boldsymbol{w};l)$. The mean squared error (MSE) is used to measure the performance of this estimator,

$$\mathrm{MSE}(\mathrm{T}) = \frac{1}{rep}\sum_{l=1}^{rep}(\widehat{\mathrm{T}}(\boldsymbol{X},\boldsymbol{y},\boldsymbol{Z},\boldsymbol{w};l) - \mathrm{T})^2. \tag{33}$$

**Experiment 1**. In this experiment, the parameters were set as follows: $(p, n_1, n_2, rep) = (600, 400, 400, 300)$, the sparsity parameters $(s, s_1, s_2) = (15, 30, 25)$, and the covariance matrices $\boldsymbol{\Sigma}$ and $\boldsymbol{\Gamma}$ satisfy $\boldsymbol{\Sigma}_{ij} = \boldsymbol{\Gamma}_{ij} = (0.8)^{|i-j|}$. For given positive values $\tau_1$ and $\tau_2$, the signals of $\boldsymbol{\beta}$ satisfy that $\boldsymbol{\beta}_{j_i} = (1 + i/s_1)\,\tau_1/2$, for $j_i \in S_1$, $i = 1, 2, \cdots, s_1$, and the signals in $\boldsymbol{\gamma}$ satisfy that $\boldsymbol{\gamma}_j = \tau_2$ for $j \in S_2$. This simulation setting aimed to investigate the case where the coefficients for one regression are much larger than the other by varying the signal strength parameters as $(\tau_1, \tau_2) \in \{(3.0, .1), (2.6, .2), (2.2, .3), (1.8, .4), (.1, 1.6), (.2, 1.4), (.3, 1.2), (.4, 1.0)\}$.

The results are summarized in Table 2. For all combinations of the signal strength parameters, in terms of estimating the inner product $\mathrm{I}(\boldsymbol{\beta},\boldsymbol{\gamma})$, FDE consistently outperformed the plugin estimates with Lasso and Thresholded Lasso. Moreover, with increasing difference between $\tau_1$ and $\tau_2$, the advantage of FDE over the plugin estimate using Lasso or thresholded Lasso became larger. The same results were observed for estimation of the normalized inner product $\mathrm{R}(\boldsymbol{\beta},\boldsymbol{\gamma})$, where FDE-NS had consistent better performance than other methods. Although De-biased performed well in terms of estimating $\mathrm{I}(\boldsymbol{\beta},\boldsymbol{\gamma})$, it performed much worse than FDE-NS for estimating $\mathrm{R}(\boldsymbol{\beta},\boldsymbol{\gamma})$.

As discussed in Section 2, the sample splitting of estimating the normalized inner product is simply proposed to facilitate the theoretical analysis and might not be necessary for the algorithm. Our simulation results indicated that the proposed estimator without sample splitting (FDE-NS) performed quite well in all settings, even better than FDE-S, due to the fact that more samples were used for estimation and correction steps. Such observations led us to use the proposed estimator without sample splitting (FDE-NS) in the real data analysis in Section 5.

Table 2: Estimation of the inner product $\mathrm{I}(\boldsymbol{\beta}, \boldsymbol{\gamma})$ and the normalized inner product $\mathrm{R}(\boldsymbol{\beta}, \boldsymbol{\gamma})$: the truth and mean squared error (MSE) for the plug-in estimator with the scaled Lasso estimator (Lasso), plug-in estimator with the de-biased estimator (De-biased), plug-in estimator with the thresholded estimator (Thresholded), and the proposed estimator $\widehat{\mathrm{I}}(\boldsymbol{\beta}, \boldsymbol{\gamma})$ (FDE), proposed estimators with sample splitting (FDE-S) and without sample splitting (FDE-NS).

| | | Strength parameters, $(\tau_1, \tau_2)$ | | | | | | | |
|---|---|---|---|---|---|---|---|---|---|
| | Method | (1.8, .4) | (2.2, .3) | (2.6, .2) | (3, .1) | (.1, 1.6) | (.2, 1.4) | (.3, 1.2) | (.4, 1) |
| $\mathrm{I}(\boldsymbol{\beta}, \boldsymbol{\gamma})$ | Truth | 8.088 | 7.414 | 5.841 | 3.370 | 1.797 | 3.145 | 4.044 | 4.493 |
| | Lasso | 9.295 | 11.564 | 12.560 | 7.279 | 2.377 | 4.889 | 5.409 | 4.800 |
| | De-biased | 1.733 | 2.191 | 2.324 | 1.386 | .449 | .838 | .985 | .886 |
| | Thresholded | 2.029 | 3.377 | 6.463 | 5.789 | 1.877 | 3.024 | 2.432 | 1.546 |
| | FDE | 1.847 | 2.471 | 2.662 | 2.118 | .734 | .995 | 1.028 | .986 |
| $\mathrm{R}(\boldsymbol{\beta}, \boldsymbol{\gamma})$ | Truth | .5314 | .5314 | .5314 | .5314 | .5314 | .5314 | .5314 | .5314 |
| | Lasso | .0023 | .0075 | .0332 | .1260 | .1574 | .0624 | .0227 | .0087 |
| | De-biased | .0208 | .0415 | .0864 | .1590 | .1736 | .1068 | .0627 | .0373 |
| | Thresholded | .0045 | .0139 | .0585 | .1753 | .0964 | .0981 | .0389 | .0153 |
| | FDE-S | .0337 | .0303 | .0621 | .0678 | .2130 | .1199 | .0694 | .0616 |
| | FDE-NS | .0036 | .0064 | .0163 | .0580 | .0892 | .0237 | .0116 | .0061 |

**Experiment 2**. In this experiment, the parameters were set as follows: $(p, n_1, n_2, rep) = (800, 400, 400, 300)$, signal strength parameters $(\tau_1, \tau_2) = (.2, .1)$ and the covariance matrices $\boldsymbol{\Sigma}$ and $\boldsymbol{\Gamma}$ satisfy $\boldsymbol{\Sigma}_{ij} = \boldsymbol{\Gamma}_{ij} = (0.8)^{|i-j|}$. The signals of $\boldsymbol{\beta}$ follow that $\boldsymbol{\beta}_{j_i} = (1 + i/s_1)\,\tau_1/2$, for $j_i \in S_1$, $i = 1, 2, \cdots, s_1$, and the signals in $\boldsymbol{\gamma}$ satisfy that $\boldsymbol{\gamma}_j = \tau_2$ for $j \in S_2$.

This simulation setting aimed to investigate the relationship between the performance of estimators and the signal sparsity level and vary the number of signals in $\boldsymbol{\beta}$ and $\boldsymbol{\gamma}$ as $(s_1, s_2) \in \{(40, 40), (50, 50), (60, 60), (70, 70), (80, 80), (90, 90), (100, 100), (110, 110)\}$, and fix the number of common signals at $s = 20$. Since the number of the associated variants is very large for both coefficient vectors, large values of $\tau_1$ and $\tau_2$ induce strong signals so that all the methods perform well. Instead, the signal magnitude was fixed at $\tau_1 = .2$ and $\tau_2 = .1$ as weak signals.

The results are summarized in Table 3. Clearly, FDE outperformed the other methods. When the signals became denser, the improvement of FDE over other methods was more pronounced. For estimation of $\mathrm{R}(\boldsymbol{\beta}, \boldsymbol{\gamma})$, the results showed that FDE-NS consistently outperformed other estimators. As the number of signals increased, the MSE corresponding to FDE-NS decreased quickly.

Table 3: Estimation of the inner product $\mathrm{I}(\boldsymbol{\beta}, \boldsymbol{\gamma})$ and the normalized inner product $\mathrm{R}(\boldsymbol{\beta}, \boldsymbol{\gamma})$: the truth and MSE for the plug-in estimator with the scaled Lasso estimator (Lasso), plug-in estimator with the de-biased estimator (De-biased), plug-in estimator with the thresholded estimator (Thresholded), and the proposed estimator $\widehat{\mathrm{I}}(\boldsymbol{\beta}, \boldsymbol{\gamma})$ (FDE), proposed estimators with sample splitting (FDE-S) and without sample splitting (FDE-NS).

| | | Sparsity parameter, $s_1$ ($s_2 = s_1$) | | | | | | | |
|---|---|---|---|---|---|---|---|---|---|
| | Method | 40 | 50 | 60 | 70 | 80 | 90 | 100 | 110 |
| $\mathrm{I}(\boldsymbol{\beta}, \boldsymbol{\gamma})$ | Truth | .190 | .170 | .219 | .212 | .179 | .221 | .183 | .221 |
| | Lasso | .032 | .025 | .039 | .036 | .024 | .035 | .023 | .028 |
| | De-biased | .015 | .015 | .018 | .017 | .024 | .027 | .040 | .066 |
| | Thresholded | .027 | .021 | .031 | .029 | .020 | .025 | .018 | .018 |
| | FDE | .020 | .014 | .021 | .022 | .011 | .013 | .008 | .008 |
| $\mathrm{R}(\boldsymbol{\beta}, \boldsymbol{\gamma})$ | Truth | .4027 | .2908 | .3122 | .2592 | .1914 | .2110 | .1573 | .1725 |
| | Lasso | .1157 | .0517 | .0539 | .0370 | .0166 | .0180 | .0097 | .0079 |
| | De-biased | .1267 | .0601 | .0659 | .0411 | .0160 | .0173 | .0063 | .0059 |
| | Thresholded | .1392 | .0687 | .0732 | .0504 | .0262 | .0277 | .0155 | .0142 |
| | FDE-S | .1154 | .1225 | .0779 | .0574 | .0456 | .0499 | .0450 | .0493 |
| | FDE-NS | .0847 | .0340 | .0368 | .0294 | .0115 | .0091 | .0055 | .0047 |

# 5 Co-heritability Estimation of Yeast Cross Genome Wide Association Data

Bloom et al. (2013) reported a large scale genome-wide association study of 46 quantitative traits based on 1,008 *Saccharomyces cerevisiae* segregants between a laboratory strain and a wine strain. The data set included 11,623 unique genotype markers. Since many of these markers were highly correlated, Bloom et al. (2013) further selected a set of 4,410 markers that are weakly dependent based in the linkage disequilibrium information. These markers were selected by picking one marker closest to each centimorgan position on the genetic map. The maker genotype was coded as 1 or $-1$, according to which strain it came from. For each segregant, the traits of interest were the end-point colony size normalized by the control growth under 46 different growth media, including Hydrogen Peroxide, Diamide, Calcium, Yeast Nitrogen Base (YNB) and Yeast extract Peptone Dextrose (YPD), etc. To demonstrate the co-heritability among these traits, 8 traits were considered, including the normalized colony size under Calcium Chloride (Calcium), Diamide, Hydrogen Peroxide (Hydrogen), Paraquat, Raffinose, 6 Azauracil (Azauracil), YNB, and YPD. Each trait was normalized to have variance 1, so the quadratic norm represents the total genetic effects for each trait.

FDE was applied to every pair of these 8 traits without sample splitting, and the results are summarized in Table 4, including estimates of the inner product for all possible pairs, the quadratic norm for each trait and the normalized inner product $\mathrm{R}(\boldsymbol{\beta}, \boldsymbol{\gamma})$. The genetic heritability of these traits ranged from 0.22 for Raffinose to 0.67 for YPD. About two thirds of these pairs had an estimated co-heritability smaller than 0.1, indicating relatively weak genetic correlations among these traits.

To further demonstrate the genetic relatedness among these pairs, for each trait, a $t$-statistic was calculated based on regressing the trait value $\boldsymbol{y}$ on genetic genetic marker $\boldsymbol{X}_{\cdot j}$, for $i \leq j \leq p$. A larger absolute value of the $t$-statistic implied a stronger effect of the marker on the trait. For any pair of traits, the scatter plot of the $t$-statistics provided a way of revealing the genetic relationship between them. Figure 1 shows the plots of several pairs of the traits, including the pairs with a large positive $\mathrm{R}(\boldsymbol{\beta}, \boldsymbol{\gamma})$, YPD v.s. YNB and Calcium v.s.

Table 4: FDE estimation for the inner product, the quadratic norm (top row of each block) and the normalized inner product (bottom row of each block) among 8 traits of yeast cross.

| Traits | Calcium | Diamide | Hydrogen | Paraquat | Raffinose | Azauracil | YNB | YPD |
|---|---|---|---|---|---|---|---|---|
| Calcium | .3314 | -.0189 | -.1003 | .0084 | .0927 | .0095 | -.0656 | .-0134 |
| | 1 | -.0286 | -.1579 | .0117 | .1972 | .0172 | -.0968 | -.0164 |
| Diamide | | .4390 | .0598 | -.0039 | .0500 | .0446 | -.0159 | .0803 |
| | | 1 | .0942 | -.0053 | .1065 | .0809 | -.0235 | .0983 |
| Hydrogen | | | .4033 | .0576 | -.1040 | .0601 | .0672 | .0637 |
| | | | 1 | .0799 | -.2213 | .1089 | .0991 | .0779 |
| Paraquat | | | | .5199 | .0023 | .0365 | .1148 | .1029 |
| | | | | 1 | .0049 | .0661 | .1693 | .1259 |
| Raffinose | | | | | .2208 | .0137 | .0830 | .0331 |
| | | | | | 1 | .0248 | .1224 | .0405 |
| Azauracil | | | | | | .3045 | -.0259 | .0703 |
| | | | | | | 1 | -.0383 | .0860 |
| YNB | | | | | | | .4594 | .4246 |
| | | | | | | | 1 | .5195 |
| YPD | | | | | | | | .6680 |
| | | | | | | | | 1 |

YNB, pairs with a large negative $\mathrm{R}(\boldsymbol{\beta}, \boldsymbol{\gamma})$, Raffinose v.s. Hydrogen and Calcium v.s. Hydrogen, and pairs with $\mathrm{R}(\boldsymbol{\beta}, \boldsymbol{\gamma})$ near to 0, including Paraquat v.s. Diamide and Paraquat v.s. Raffinose. The plot clearly indicates a strong positive genetic correlation between YPD and YNB. The correlation between Calcium and YNB is slightly smaller. Raffinose/Hydrogen and Calcium/Hydrogen pairs clearly showed negative genetic correlation. There were several genetic variants with very large effects on Hydrogen, but they were not associated with the other traits such as Raffinose and Calcium. The shared genetic variants were relatively weak, leading to a correlation around -0.2. The two plots on the right show the pairs of trats with weak correlations. These plots indicated that the proposed co-heritability measures can indeed capture the genetic sharing among different related traits.

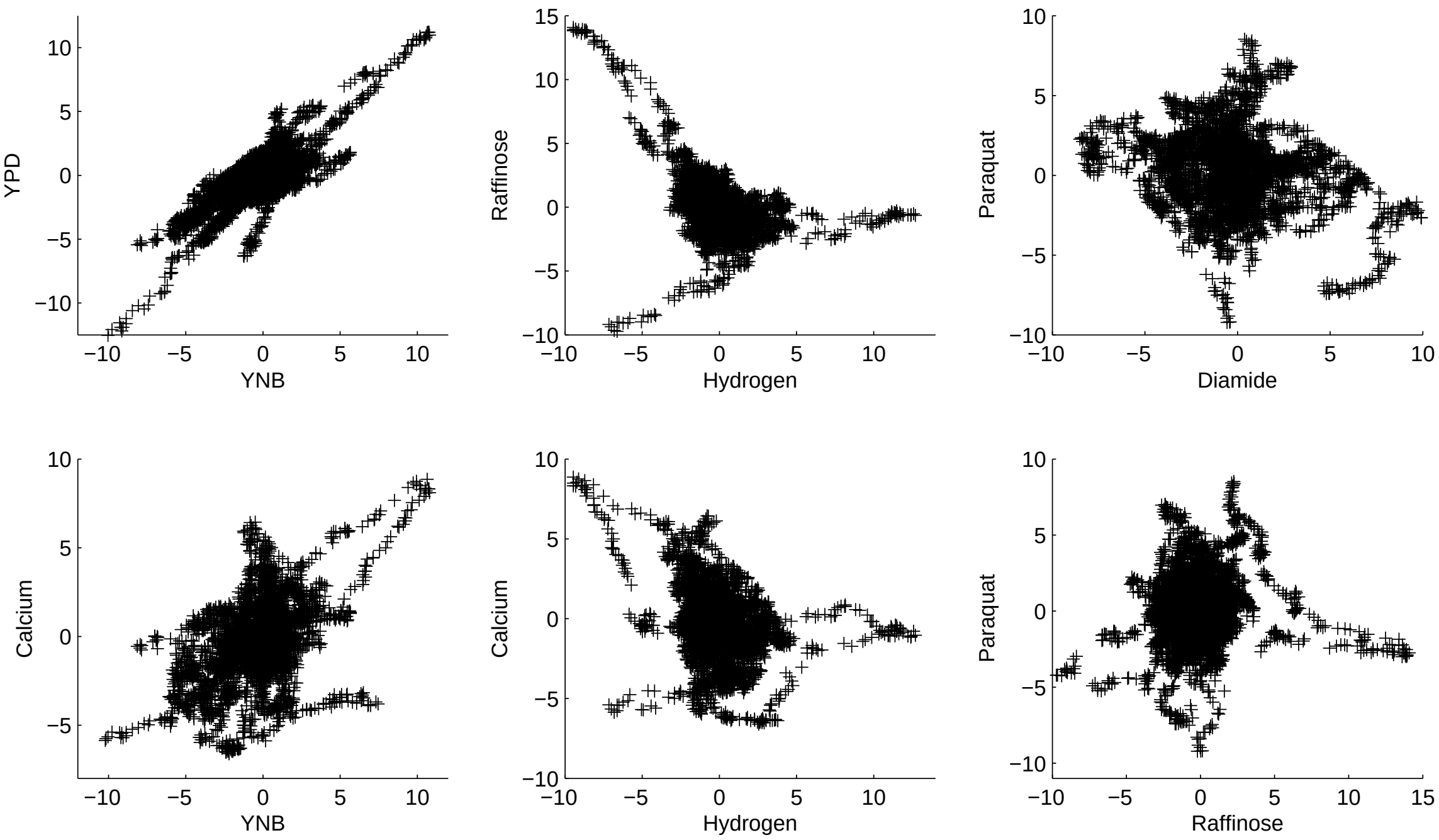

Figure 1: Scatter plots of marginal regression $t$ statistics for six pairs of traits, including the pairs with positive correlation case (left panel), negative correlation (middle panel), and no correlation (right panel).

# 6 Proofs

## 6.1 Proof of Theorem 1

For simplicity of notation, we assume $n_1 = n_2$ and use $n = n_1 = n_2$ to represent the sample size throughout the proof. The proofs can be easily generalized to the case $n_1 \asymp n_2$. Without loss of generality, we assume that the Sub-gaussian norm of random vectors $\boldsymbol{X}_{i\cdot}$ and $\boldsymbol{Z}_{i\cdot}$ are also upper bounded by $M_1$, that is, $\max\left\{\|\boldsymbol{X}_{i\cdot}\|_{\psi_2}^2, \|\boldsymbol{Z}_{i\cdot}\|_{\psi_2}^2\right\} \leq M_1$.

**Proof of** (21)

Note the following important decomposition,

$$\begin{aligned}
&\widehat{\mathrm{I}}\left(\boldsymbol{\beta},\boldsymbol{\gamma}\right)-\mathrm{I}\left(\boldsymbol{\beta},\boldsymbol{\gamma}\right)=\langle\widehat{\boldsymbol{\beta}},\widehat{\boldsymbol{\gamma}}\rangle+\widehat{\boldsymbol{u}}_1^{\intercal}\frac{1}{n}\boldsymbol{X}^{\intercal}\left(\boldsymbol{y}-\boldsymbol{X}\widehat{\boldsymbol{\beta}}\right)+\widehat{\boldsymbol{u}}_2^{\intercal}\frac{1}{n}\boldsymbol{Z}^{\intercal}\left(\boldsymbol{w}-\boldsymbol{Z}\widehat{\boldsymbol{\gamma}}\right)-\langle\boldsymbol{\beta},\boldsymbol{\gamma}\rangle\\
=&\left(\widehat{\boldsymbol{u}}_1^{\intercal}\frac{1}{n}\boldsymbol{X}^{\intercal}\left(\boldsymbol{y}-\boldsymbol{X}\widehat{\boldsymbol{\beta}}\right)-\langle\widehat{\boldsymbol{\gamma}},\boldsymbol{\beta}-\widehat{\boldsymbol{\beta}}\rangle\right)+\left(\widehat{\boldsymbol{u}}_2^{\intercal}\frac{1}{n}\boldsymbol{Z}^{\intercal}\left(\boldsymbol{w}-\boldsymbol{Z}\widehat{\boldsymbol{\gamma}}\right)-\langle\widehat{\boldsymbol{\beta}},\boldsymbol{\gamma}-\widehat{\boldsymbol{\gamma}}\rangle\right)-\langle\widehat{\boldsymbol{\beta}}-\boldsymbol{\beta},\widehat{\boldsymbol{\gamma}}-\boldsymbol{\gamma}\rangle\\
=&\widehat{\boldsymbol{u}}_1^{\intercal}\frac{1}{n}\boldsymbol{X}^{\intercal}\boldsymbol{\epsilon}+\left(\widehat{\boldsymbol{u}}_1^{\intercal}\widehat{\boldsymbol{\Sigma}}-\widehat{\boldsymbol{\gamma}}^{\intercal}\right)\left(\boldsymbol{\beta}-\widehat{\boldsymbol{\beta}}\right)+\widehat{\boldsymbol{u}}_2^{\intercal}\frac{1}{n}\boldsymbol{Z}^{\intercal}\boldsymbol{\delta}+\left(\widehat{\boldsymbol{u}}_2^{\intercal}\widehat{\boldsymbol{\Gamma}}-\widehat{\boldsymbol{\beta}}^{\intercal}\right)\left(\boldsymbol{\gamma}-\widehat{\boldsymbol{\gamma}}\right)-\langle\widehat{\boldsymbol{\beta}}-\boldsymbol{\beta},\widehat{\boldsymbol{\gamma}}-\boldsymbol{\gamma}\rangle.
\end{aligned} \tag{34}$$

The following lemmas are introduced to control the terms in (34) and similar results were established in the analysis of Lasso, scaled Lasso and de-biased Lasso (Cai and Guo, 2016*a*; Ren et al., 2013; Sun and Zhang, 2012; Ye and Zhang, 2010). The proofs of the following lemmas can be found in the supplementary material, Section C.

Lemma 1. *Suppose the assumption* (A1) *holds and* $k\log p/n\to 0$ *with* $k=\max\{\|\boldsymbol{\beta}\|_0,\|\boldsymbol{\gamma}\|_0\}$. *Then with probability at least* $1-p^{-c}$, *we have*

$$\|\widehat{\boldsymbol{\beta}}-\boldsymbol{\beta}\|_1\le Ck\sqrt{\frac{\log p}{n}},\quad \|\widehat{\boldsymbol{\gamma}}-\boldsymbol{\gamma}\|_1\le Ck\sqrt{\frac{\log p}{n}}, \tag{35}$$

*and*

$$\|\widehat{\boldsymbol{\beta}}-\boldsymbol{\beta}\|_2\le C\sqrt{\frac{k\log p}{n}},\quad \|\widehat{\boldsymbol{\gamma}}-\boldsymbol{\gamma}\|_2\le C\sqrt{\frac{k\log p}{n}}, \tag{36}$$

*where* $c$ *and* $C$ *are positive constants.*

Lemma 2. *Suppose the assumption* (A1) *holds and* $k\log p/n\to 0$ *with* $k=\max\{\|\boldsymbol{\beta}\|_0,\|\boldsymbol{\gamma}\|_0\}$. *Then with probability at least* $1-p^{-c}$, *we have*

$$\left|\left(\widehat{\boldsymbol{u}}_1^{\intercal}\widehat{\boldsymbol{\Sigma}}-\widehat{\boldsymbol{\gamma}}^{\intercal}\right)\left(\boldsymbol{\beta}-\widehat{\boldsymbol{\beta}}\right)\right|\le C\|\widehat{\boldsymbol{\gamma}}\|_2\frac{\|\boldsymbol{\beta}\|_0\log p}{n}\ \text{and}\ \left|\left(\widehat{\boldsymbol{u}}_2^{\intercal}\widehat{\boldsymbol{\Gamma}}-\widehat{\boldsymbol{\beta}}^{\intercal}\right)\left(\boldsymbol{\gamma}-\widehat{\boldsymbol{\gamma}}\right)\right|\le C\|\widehat{\boldsymbol{\beta}}\|_2\frac{\|\boldsymbol{\gamma}\|_0\log p}{n}. \tag{37}$$

*With probability at least* $1-2\alpha-p^{-c}$,

$$\left|\widehat{\boldsymbol{u}}_1^{\intercal}\frac{1}{n}\boldsymbol{X}^{\intercal}\boldsymbol{\epsilon}\right|\le C\|\widehat{\boldsymbol{\gamma}}\|_2\frac{z_{\alpha/2}}{\sqrt{n}},\quad \text{and}\quad \left|\widehat{\boldsymbol{u}}_2^{\intercal}\frac{1}{n}\boldsymbol{Z}^{\intercal}\boldsymbol{\delta}\right|\le C\|\widehat{\boldsymbol{\beta}}\|_2\frac{z_{\alpha/2}}{\sqrt{n}}, \tag{38}$$

*and hence*

$$\begin{aligned}
\left|\widehat{\boldsymbol{u}}_1^{\intercal}\frac{1}{n}\boldsymbol{X}^{\intercal}\left(\boldsymbol{y}-\boldsymbol{X}\widehat{\boldsymbol{\beta}}\right)-\langle\widehat{\boldsymbol{\gamma}},\boldsymbol{\beta}-\widehat{\boldsymbol{\beta}}\rangle\right|&\le C\|\widehat{\boldsymbol{\gamma}}\|_2\frac{z_{\alpha/2}}{\sqrt{n}}+C\|\widehat{\boldsymbol{\gamma}}\|_2\frac{k\log p}{n},\\
\left|\widehat{\boldsymbol{u}}_2^{\intercal}\frac{1}{n}\boldsymbol{Z}^{\intercal}\left(\boldsymbol{w}-\boldsymbol{Z}\widehat{\boldsymbol{\gamma}}\right)-\langle\widehat{\boldsymbol{\beta}},\boldsymbol{\gamma}-\widehat{\boldsymbol{\gamma}}\rangle\right|&\le C\|\widehat{\boldsymbol{\beta}}\|_2\frac{z_{\alpha/2}}{\sqrt{n}}+C\|\widehat{\boldsymbol{\beta}}\|_2\frac{k\log p}{n},
\end{aligned} \tag{39}$$

*where* $c$ *and* $C$ *are positive constants.*

By the decomposition (34) and the inequalities (36), (37) and (38), we obtain that

$$\left|\widehat{\mathrm{I}}\left(\boldsymbol{\beta},\boldsymbol{\gamma}\right)-\mathrm{I}\left(\boldsymbol{\beta},\boldsymbol{\gamma}\right)\right| \leq C\left(\|\widehat{\boldsymbol{\beta}}\|_2+\|\widehat{\boldsymbol{\gamma}}\|_2\right)\left(\frac{z_{\alpha/2}}{\sqrt{n}}+\frac{k\log p}{n}\right)+C\frac{k\log p}{n}.$$

By (36), we establish (21).

<u>**Proof of (22) and (23)**</u>

The proof of (23) is similar to that of (22) and only the proof of (22) is present in the following. We introduce the estimator $\bar{\mathrm{Q}}(\boldsymbol{\beta}) = \|\widehat{\boldsymbol{\beta}}\|_2^2 + 2\widehat{\boldsymbol{u}}_3^\intercal\left(\boldsymbol{X}^{(2)}\right)^\intercal\left(\boldsymbol{y}^{(2)}-\boldsymbol{X}^{(2)}\widehat{\boldsymbol{\beta}}\right)/(n/2)$ and due to the fact that $\mathrm{Q}\left(\boldsymbol{\beta}\right)$ is non-negative, we have $\left|\widehat{\mathrm{Q}}\left(\boldsymbol{\beta}\right)-\mathrm{Q}\left(\boldsymbol{\beta}\right)\right| \leq \left|\bar{\mathrm{Q}}(\boldsymbol{\beta})-\mathrm{Q}\left(\boldsymbol{\beta}\right)\right|$. We decompose the difference between $\bar{\mathrm{Q}}(\boldsymbol{\beta})$ and $\mathrm{Q}\left(\boldsymbol{\beta}\right)$,

$$\begin{aligned}\bar{\mathrm{Q}}(\boldsymbol{\beta})-\mathrm{Q}\left(\boldsymbol{\beta}\right) &= \|\widehat{\boldsymbol{\beta}}\|_2^2-\|\boldsymbol{\beta}\|_2^2+2\widehat{\boldsymbol{u}}_3^\intercal\frac{1}{n/2}\left(\boldsymbol{X}^{(2)}\right)^\intercal\left(\boldsymbol{y}^{(2)}-\boldsymbol{X}^{(2)}\widehat{\boldsymbol{\beta}}\right)\\ &= 2\left(\widehat{\boldsymbol{u}}_3^\intercal\widehat{\boldsymbol{\Sigma}}^{(2)}-\widehat{\boldsymbol{\beta}}^\intercal\right)\left(\boldsymbol{\beta}-\widehat{\boldsymbol{\beta}}\right)+2\widehat{\boldsymbol{u}}_3^\intercal\frac{1}{n/2}\left(\boldsymbol{X}^{(2)}\right)^\intercal\boldsymbol{\epsilon}^{(2)}-\|\widehat{\boldsymbol{\beta}}-\boldsymbol{\beta}\|_2^2.\end{aligned}$$

Combined with the above argument, the upper bound (22) follows from (36) and the following lemma. The proof of the following lemma can be found in the supplementary material Section C.

LEMMA 3. *Suppose the assumption* (A1) *holds and* $k\log p/n \to 0$ *with* $k=\max\left\{\|\boldsymbol{\beta}\|_0,\|\boldsymbol{\gamma}\|_0\right\}$. *Then with probability at least* $1-p^{-c}-\alpha$,

$$\left|\widehat{\boldsymbol{u}}_3^\intercal\frac{1}{n/2}\left(\boldsymbol{X}^{(2)}\right)^\intercal\boldsymbol{\epsilon}^{(2)}\right| \leq C\|\widehat{\boldsymbol{\beta}}\|_2\frac{z_{\alpha/2}}{\sqrt{n}}, \tag{40}$$

*and*

$$\left|\left(\widehat{\boldsymbol{u}}_3^\intercal\widehat{\boldsymbol{\Sigma}}^{(2)}-\widehat{\boldsymbol{\beta}}^\intercal\right)\left(\boldsymbol{\beta}-\widehat{\boldsymbol{\beta}}\right)\right| \leq C\|\widehat{\boldsymbol{\beta}}\|_2\frac{k\log p}{n}, \tag{41}$$

*where* $c$ *and* $C$ *are positive constants.*

## 6.2 Proof of Theorem 2

Let $\widetilde{\mu}_1 = \widehat{\boldsymbol{u}}_1^\intercal\boldsymbol{X}^\intercal\left(\boldsymbol{y}-\boldsymbol{X}\widehat{\boldsymbol{\beta}}\right)/n$ and $\widetilde{\mu}_2 = \widehat{\boldsymbol{u}}_2^\intercal\boldsymbol{Z}^\intercal\left(\boldsymbol{w}-\boldsymbol{Z}\widehat{\boldsymbol{\gamma}}\right)/n$. We introduce the new estimator

$$\bar{\mathrm{R}}(\boldsymbol{\beta},\boldsymbol{\gamma}) = \frac{\widehat{\mathrm{I}}\left(\boldsymbol{\beta},\boldsymbol{\gamma}\right)}{\sqrt{\widehat{\mathrm{Q}}\left(\boldsymbol{\beta}\right)\widehat{\mathrm{Q}}\left(\boldsymbol{\beta}\right)}} = \frac{\langle\widehat{\boldsymbol{\beta}},\widehat{\boldsymbol{\gamma}}\rangle+\widetilde{\mu}_1+\widetilde{\mu}_2}{\sqrt{\widehat{\mathrm{Q}}\left(\boldsymbol{\beta}\right)}\sqrt{\widehat{\mathrm{Q}}\left(\boldsymbol{\gamma}\right)}}.$$

Since $-1 \leq \mathrm{R}(\boldsymbol{\beta},\boldsymbol{\gamma}) \leq 1$, we have $\left|\widehat{\mathrm{R}}(\boldsymbol{\beta},\boldsymbol{\gamma}) - \mathrm{R}(\boldsymbol{\beta},\boldsymbol{\gamma})\right| \leq \left|\bar{\mathrm{R}}(\boldsymbol{\beta},\boldsymbol{\gamma}) - \mathrm{R}(\boldsymbol{\beta},\boldsymbol{\gamma})\right|$. We decompose the difference between $\bar{\mathrm{R}}(\boldsymbol{\beta},\boldsymbol{\gamma})$ and $\mathrm{R}(\boldsymbol{\beta},\boldsymbol{\gamma})$,

$$\frac{\langle\widehat{\boldsymbol{\beta}},\widehat{\boldsymbol{\gamma}}\rangle + \widetilde{\mu}_1 + \widetilde{\mu}_2}{\sqrt{\widehat{\mathrm{Q}}(\boldsymbol{\beta})}\sqrt{\widehat{\mathrm{Q}}(\boldsymbol{\gamma})}} - \frac{\langle\boldsymbol{\beta},\boldsymbol{\gamma}\rangle}{\sqrt{\|\boldsymbol{\beta}\|_2^2\|\boldsymbol{\gamma}\|_2^2}} = \left\langle \frac{\widehat{\boldsymbol{\beta}}}{\sqrt{\widehat{\mathrm{Q}}(\boldsymbol{\beta})}}, \frac{\widehat{\boldsymbol{\gamma}}}{\sqrt{\widehat{\mathrm{Q}}(\boldsymbol{\gamma})}} \right\rangle - \left\langle \frac{\boldsymbol{\beta}}{\|\boldsymbol{\beta}\|_2}, \frac{\boldsymbol{\gamma}}{\|\boldsymbol{\gamma}\|_2} \right\rangle + \frac{\widetilde{\mu}_1 + \widetilde{\mu}_2}{\sqrt{\widehat{\mathrm{Q}}(\boldsymbol{\beta})}\sqrt{\widehat{\mathrm{Q}}(\boldsymbol{\gamma})}}$$

$$= \left\langle \frac{\widehat{\boldsymbol{\beta}}}{\sqrt{\widehat{\mathrm{Q}}(\boldsymbol{\beta})}} - \frac{\boldsymbol{\beta}}{\|\boldsymbol{\beta}\|_2}, \frac{\widehat{\boldsymbol{\gamma}}}{\sqrt{\widehat{\mathrm{Q}}(\boldsymbol{\gamma})}} - \frac{\boldsymbol{\gamma}}{\|\boldsymbol{\gamma}\|_2} \right\rangle + \left( \left\langle \frac{\boldsymbol{\beta}}{\|\boldsymbol{\beta}\|_2}, \frac{\widehat{\boldsymbol{\gamma}}}{\sqrt{\widehat{\mathrm{Q}}(\boldsymbol{\gamma})}} - \frac{\boldsymbol{\gamma}}{\|\boldsymbol{\gamma}\|_2} \right\rangle + \frac{\widetilde{\mu}_2}{\sqrt{\widehat{\mathrm{Q}}(\boldsymbol{\beta})}\sqrt{\widehat{\mathrm{Q}}(\boldsymbol{\gamma})}} \right)$$
$$+ \left( \left\langle \frac{\widehat{\boldsymbol{\beta}}}{\sqrt{\widehat{\mathrm{Q}}(\boldsymbol{\beta})}} - \frac{\boldsymbol{\beta}}{\|\boldsymbol{\beta}\|_2}, \frac{\boldsymbol{\gamma}}{\|\boldsymbol{\gamma}\|_2} \right\rangle + \frac{\widetilde{\mu}_1}{\sqrt{\widehat{\mathrm{Q}}(\boldsymbol{\beta})}\sqrt{\widehat{\mathrm{Q}}(\boldsymbol{\gamma})}} \right). \tag{42}$$

The following lemma controls the terms in the above equation and establishes the theorem. The proof of the following lemma can be found in the supplementary material Section C.

Lemma 4. *Suppose the assumptions* (A1) *and* (A2) *hold and* $k\log p/n \to 0$ *with* $k = \max\{\|\boldsymbol{\beta}\|_0, \|\boldsymbol{\gamma}\|_0\}$. *Then with probability at least* $1 - p^{-c} - \alpha$, *we have*

$$\left|\left\langle \frac{\widehat{\boldsymbol{\beta}}}{\sqrt{\widehat{\mathrm{Q}}(\boldsymbol{\beta})}} - \frac{\boldsymbol{\beta}}{\|\boldsymbol{\beta}\|_2}, \frac{\widehat{\boldsymbol{\gamma}}}{\sqrt{\widehat{\mathrm{Q}}(\boldsymbol{\gamma})}} - \frac{\boldsymbol{\gamma}}{\|\boldsymbol{\gamma}\|_2} \right\rangle\right| \leq C\frac{1}{\|\boldsymbol{\beta}\|_2\|\boldsymbol{\gamma}\|_2}\frac{k\log p}{n}, \tag{43}$$

$$\left|\left\langle \frac{\widehat{\boldsymbol{\beta}}}{\sqrt{\widehat{\mathrm{Q}}(\boldsymbol{\beta})}} - \frac{\boldsymbol{\beta}}{\|\boldsymbol{\beta}\|_2}, \frac{\boldsymbol{\gamma}}{\|\boldsymbol{\gamma}\|_2} \right\rangle + \frac{\widetilde{\mu}_1}{\sqrt{\widehat{\mathrm{Q}}(\boldsymbol{\beta})}\sqrt{\widehat{\mathrm{Q}}(\boldsymbol{\gamma})}}\right| \leq C\frac{1}{\|\boldsymbol{\beta}\|_2}\left(\frac{z_{\alpha/2}}{\sqrt{n}} + \frac{k\log p}{n}\right) + C\left(\frac{1}{\|\boldsymbol{\beta}\|_2^2} + \frac{1}{\|\boldsymbol{\beta}\|_2\|\boldsymbol{\gamma}\|_2}\right)\frac{k\log p}{n}, \tag{44}$$

*and*

$$\left|\left\langle \frac{\boldsymbol{\beta}}{\|\boldsymbol{\beta}\|_2}, \frac{\widehat{\boldsymbol{\gamma}}}{\sqrt{\widehat{\mathrm{Q}}(\boldsymbol{\gamma})}} - \frac{\boldsymbol{\gamma}}{\|\boldsymbol{\gamma}\|_2} \right\rangle + \frac{\widetilde{\mu}_3}{\sqrt{\widehat{\mathrm{Q}}(\boldsymbol{\beta})}\sqrt{\widehat{\mathrm{Q}}(\boldsymbol{\gamma})}}\right| \leq C\frac{1}{\|\boldsymbol{\gamma}\|_2}\left(\frac{z_{\alpha/2}}{\sqrt{n}} + \frac{k\log p}{n}\right) + C\left(\frac{1}{\|\boldsymbol{\gamma}\|_2^2} + \frac{1}{\|\boldsymbol{\beta}\|_2\|\boldsymbol{\gamma}\|_2}\right)\frac{k\log p}{n}, \tag{45}$$

*where* $c$ *and* $C$ *are positive constants.*

By the decomposition (42), combining (43), (44) and (45), we have

$$\left|\frac{\widehat{\mathrm{I}}(\boldsymbol{\beta},\boldsymbol{\gamma})}{\sqrt{\widehat{\mathrm{Q}}(\boldsymbol{\beta})\,\widehat{\mathrm{Q}}(\boldsymbol{\beta})}} - \frac{\langle\boldsymbol{\beta},\boldsymbol{\gamma}\rangle}{\sqrt{\|\boldsymbol{\beta}\|_2^2\|\boldsymbol{\gamma}\|_2^2}}\right|$$
$$\leq C\left(\frac{1}{\|\boldsymbol{\beta}\|_2} + \frac{1}{\|\boldsymbol{\gamma}\|_2}\right)\left(\frac{z_{\alpha/2}}{\sqrt{n}} + \frac{k\log p}{n}\right) + C\left(\frac{1}{\|\boldsymbol{\beta}\|_2^2} + \frac{1}{\|\boldsymbol{\gamma}\|_2^2} + \frac{1}{\|\boldsymbol{\beta}\|_2\|\boldsymbol{\gamma}\|_2}\right)\frac{k\log p}{n}.$$

Under the assumption (A2), the upper bound (24) follows from the above inequality.

## 6.3 Proof of (30) and (31) in Theorem 3

We first introduce the notations used in the proof of lower bound results. Let $\pi$ denote the prior distribution supported on the parameter space $\mathcal{H}$. Let $f_{\pi}(z)$ denote the density function of the marginal distribution of the random variable $Z$ with the prior $\pi$ on $\mathcal{H}$. More specifically, $f_{\pi}(z) = \int f_{\theta}(z)\, \pi(\theta)\, d\theta$. We define the $\chi^2$ distance between two density functions $f_1$ and $f_0$ by

$$\chi^2(f_1, f_0) = \int \frac{(f_1(z) - f_0(z))^2}{f_0(z)} dz = \int \frac{f_1^2(z)}{f_0(z)} dz - 1 \tag{46}$$

and the $L_1$ distance by $L_1(f_1, f_0) = \int |f_1(z) - f_0(z)|\, dz$. It is well known that

$$L_1(f_1, f_0) \leq \sqrt{\chi^2(f_1, f_0)}. \tag{47}$$

The proof of the lower bound is based on the following version of Le Cam's Lemma (LeCam (1973); Ren et al. (2013); Yu (1997)).

LEMMA 5. *Let* $\mathrm{T}(\boldsymbol{\theta})$ *denote a functional on* $\boldsymbol{\theta}$. *Suppose that* $\mathcal{H}_0 = \{\boldsymbol{\theta}_0\}$, $\mathcal{H}_0, \mathcal{H}_1 \subset \Theta$ *and* $d = \min_{\boldsymbol{\theta} \in \mathcal{H}_1} |\mathrm{T}(\boldsymbol{\theta}) - \mathrm{T}(\boldsymbol{\theta}_0)|$. *Let* $\pi$ *denote a prior on the parameter space* $\mathcal{H}_1$. *Then we have*

$$\inf_{\widehat{\mathrm{T}}} \sup_{\boldsymbol{\theta} \in \mathcal{H}_0 \cup \mathcal{H}_1} \mathbb{P}_{\boldsymbol{\theta}} \left( \left| \widehat{\mathrm{T}} - \mathrm{T}(\boldsymbol{\theta}) \right| \geq \frac{d}{2} \right) \geq \frac{1 - L_1(f_{\pi}, f_{\boldsymbol{\theta}_0})}{2}. \tag{48}$$

The proofs of (30) and (31) are applications of Lemma 5. The key is to construct the parameter spaces $\mathcal{H}_0 = \{\boldsymbol{\theta}_0\}$, $\mathcal{H}_1$ and the prior on $\mathcal{H}_1$ such that (i) $\mathcal{H}_0, \mathcal{H}_1 \subset \Theta$,(ii)the $L_1$ distance $L_1(f_{\pi}, f_{\boldsymbol{\theta}_0})$ is controlled and (iii)the distance $d = \min_{\boldsymbol{\theta} \in \mathcal{H}_1} |\mathrm{T}(\boldsymbol{\theta}) - \mathrm{T}(\boldsymbol{\theta}_0)|$ is maximized. In the following, we provide the detailed proof of (30). The proof of (31) is very similar to that of (30) and is omitted here. In the discussion of lower bound results, we will assume $\boldsymbol{\mu}_x = \boldsymbol{\mu}_z = 0$ and the design $\boldsymbol{X}_{i\cdot}$ and $\boldsymbol{Z}_{i\cdot}$ follow joint normal distribution.

We define the following two parameter spaces to facilitate the discussion,

$$\mathcal{G}^S(k) = \mathcal{G}(k) \cap \left\{ \|\beta\|_2 \geq \frac{M_2}{4} \right\} \quad \text{and} \quad \mathcal{G}^W(k) = \mathcal{G}(k) \cap \left\{ \|\beta\|_2 \leq C\sqrt{\frac{k \log p}{n}} \right\}. \tag{49}$$

The lower bound (30) can be decomposed into the following three lower bounds,

$$\inf_{\widehat{\mathrm{Q}}} \sup_{\boldsymbol{\theta}_{\boldsymbol{\beta}} \in \mathcal{G}^S(k)} \mathbb{P}_{\boldsymbol{\theta}_{\boldsymbol{\beta}}} \left( \left| \widehat{\mathrm{Q}} - \mathrm{Q}(\boldsymbol{\beta}) \right| \geq c' \|\boldsymbol{\beta}\|_2 \frac{k \log p}{n} \right) \geq \frac{1}{4}, \tag{50}$$

$$\inf_{\widehat{\mathrm{Q}}} \sup_{\boldsymbol{\theta}_{\boldsymbol{\beta}} \in \mathcal{G}^{S}(k)} \mathbb{P}_{\boldsymbol{\theta}_{\boldsymbol{\beta}}} \left( \left| \widehat{\mathrm{Q}} - \mathrm{Q}(\boldsymbol{\beta}) \right| \geq c' \|\boldsymbol{\beta}\|_2 \frac{1}{\sqrt{n}} \right) \geq \frac{1}{4}, \tag{51}$$

and

$$\inf_{\widehat{\mathrm{Q}}} \sup_{\boldsymbol{\theta}_{\boldsymbol{\beta}} \in \mathcal{G}^{W}(k)} \mathbb{P}_{\boldsymbol{\theta}_{\boldsymbol{\beta}}} \left( \left| \widehat{\mathrm{Q}} - \mathrm{Q}(\boldsymbol{\beta}) \right| \geq c' \frac{k \log p}{n} \right) \geq \frac{1}{4}, \tag{52}$$

where $c' > 0$ is a positive constant. Combining (50) and (51), we have

$$\inf_{\widehat{\mathrm{Q}}} \sup_{\boldsymbol{\theta}_{\boldsymbol{\beta}} \in \mathcal{G}^{S}(k)} \mathbb{P}_{\boldsymbol{\theta}_{\boldsymbol{\beta}}} \left( \left| \widehat{\mathrm{Q}} - \mathrm{Q}(\boldsymbol{\beta}) \right| \geq c' \|\boldsymbol{\beta}\|_2 \left( \frac{1}{\sqrt{n}} + \frac{k \log p}{n} \right) \right) \geq \frac{1}{4}. \tag{53}$$

On $\mathcal{G}^{S}(k)$, we have

$$\|\boldsymbol{\beta}\|_2 \left( \frac{1}{\sqrt{n}} + \frac{k \log p}{n} \right) = \max \left\{ \|\boldsymbol{\beta}\|_2 \left( \frac{1}{\sqrt{n}} + \frac{k \log p}{n} \right), \frac{k \log p}{n} \right\}$$

and on $\mathcal{G}^{W}(k)$, we have

$$\frac{k \log p}{n} = \max \left\{ \|\boldsymbol{\beta}\|_2 \left( \frac{1}{\sqrt{n}} + \frac{k \log p}{n} \right), \frac{k \log p}{n} \right\}.$$

Since both $\mathcal{G}^{S}(k)$ and $\mathcal{G}^{W}(k)$ are proper subspace of $\mathcal{G}(k)$, we can establish (30) by combining (53) and (52). In the following, we will establish the lower bounds (50), (51) and (52) separately.

**Proof of** (50) Under the Gaussian random design model, $\mathbf{V}_i = (\boldsymbol{y}_i, \boldsymbol{X}_{i\cdot}) \in \mathbf{R}^{p+1}$ follows a joint Gaussian distribution with mean $\mathbf{0}$. Let $\boldsymbol{\Sigma}^v$ denote the covariance matrix of $\mathbf{V}_i$. For the indices of $\boldsymbol{\Sigma}^v$, we use 0 as the index of $\boldsymbol{y}_i$ and $\{1, \cdots, p\}$ as the indices for $(\boldsymbol{X}_{i1}, \cdots, \boldsymbol{X}_{ip}) \in \mathbf{R}^p$. Decompose $\boldsymbol{\Sigma}^v$ into blocks $\begin{pmatrix} \boldsymbol{\Sigma}^v_{yy} & (\boldsymbol{\Sigma}^v_{xy})^\intercal \\ \boldsymbol{\Sigma}^v_{xy} & \boldsymbol{\Sigma}^v_{xx} \end{pmatrix}$, where $\boldsymbol{\Sigma}^v_{yy}$, $\boldsymbol{\Sigma}^v_{xx}$ and $\boldsymbol{\Sigma}^v_{xy}$ denote the variance of $\boldsymbol{y}_i$, the variance of $\boldsymbol{X}_{i\cdot}$ and the covariance of $\boldsymbol{y}_i$ and $\boldsymbol{X}_{i\cdot}$, respectively. Let $\boldsymbol{\Omega} = \boldsymbol{\Sigma}^{-1}$ denote the precision matrix. There exists a bijective function $h : \boldsymbol{\Sigma}^v \to (\boldsymbol{\beta}, \boldsymbol{\Omega}, \sigma_1)$ and the inverse mapping $h^{-1} : (\boldsymbol{\beta}, \boldsymbol{\Omega}, \sigma_1) \to \boldsymbol{\Sigma}^v$, where $h^{-1}((\boldsymbol{\beta}, \boldsymbol{\Omega}, \sigma_1)) = \begin{pmatrix} \boldsymbol{\beta}^\intercal \boldsymbol{\Omega}^{-1} \boldsymbol{\beta} + \sigma_1^2 & \boldsymbol{\beta}^\intercal \boldsymbol{\Omega}^{-1} \\ \boldsymbol{\Omega}^{-1} \boldsymbol{\beta} & \boldsymbol{\Omega}^{-1} \end{pmatrix}$ and

$$h(\boldsymbol{\Sigma}^v) = \left( (\boldsymbol{\Sigma}^v_{xx})^{-1} \boldsymbol{\Sigma}^v_{xy}, (\boldsymbol{\Sigma}^v_{xx})^{-1}, \boldsymbol{\Sigma}^v_{yy} - \left(\boldsymbol{\Sigma}^v_{xy}\right)^\intercal (\boldsymbol{\Sigma}^v_{xx})^{-1} \boldsymbol{\Sigma}^v_{xy} \right). \tag{54}$$

Based on the bijection, it is sufficient to control the $\chi^2$ distance between two multivariate Gaussian distributions.

We introduce the null parameter space

$$\mathcal{G}_0 = \left\{ \boldsymbol{\theta}_0 = (\boldsymbol{\beta}^*, \mathrm{I}, \sigma_0) : \boldsymbol{\beta}^* = (\boldsymbol{\beta}_1^*, \eta_0, 0, \cdots, 0) \text{ with } \boldsymbol{\beta}_1^* > \frac{M_2}{2} \text{ and } \sigma_0 = \sqrt{\frac{M_2}{8}} \right\} \subset \mathcal{G}^S(k),$$

where $\eta_0 \geq 0$. Define $p_1 = p - 2$. Based on the mapping $h$, we have the corresponding null parameter space for $\boldsymbol{\Sigma}^v$, $\mathcal{F}_0 = \{\boldsymbol{\Sigma}_0^v\}$ where

$$\boldsymbol{\Sigma}_0^v = \left( \begin{array}{c|c|c|c} (\boldsymbol{\beta}_1^*)^2 + \eta_0^2 + \sigma_0^2 & \boldsymbol{\beta}_1^* & \eta_0 & \mathbf{0} \\ \hline \boldsymbol{\beta}_1^* & 1 & 0 & \mathbf{0} \\ \hline \eta_0 & 0 & 1 & \mathbf{0} \\ \hline \mathbf{0} & \mathbf{0} & \mathbf{0} & \mathbf{I}_{p_1 \times p_1} \end{array} \right).$$

We then introduce the alternative parameter space for $\boldsymbol{\Sigma}^v$, which will induce a parameter space for $(\boldsymbol{\beta}, \boldsymbol{\Omega}, \sigma_1)$ through the mapping $h$. Define $\mathcal{F}_1 = \{\boldsymbol{\Sigma}_{\boldsymbol{\alpha}}^v : \boldsymbol{\alpha} \in \ell(p_1, k, \rho)\}$, where

$$\boldsymbol{\Sigma}_{\boldsymbol{\alpha}}^v = \left( \begin{array}{c|c|c|c} (\boldsymbol{\beta}_1^*)^2 + \eta_0^2 + \sigma_0^2 & \boldsymbol{\beta}_1^* & \eta_0 & \rho_0 \boldsymbol{\alpha}^\intercal \\ \hline \boldsymbol{\beta}_1^* & 1 & 0 & \boldsymbol{\alpha}^\intercal \\ \hline \eta_0 & 0 & 1 & \mathbf{0} \\ \hline \rho_0 \boldsymbol{\alpha} & \boldsymbol{\alpha} & \mathbf{0} & \mathbf{I}_{p_1 \times p_1} \end{array} \right), \tag{55}$$

with $\rho_0 = \boldsymbol{\beta}_1^* + \sigma_0$ and

$$\ell(p_1, k, \rho) = \{\boldsymbol{\alpha} : \boldsymbol{\alpha} \in \mathbf{R}^{p_1}, \ \|\boldsymbol{\alpha}\|_0 = k - 2, \ \boldsymbol{\alpha}_i \in \{0, \rho\} \text{ for } 1 \leq i \leq p_1\}. \tag{56}$$

Then we construct the corresponding parameter space $\mathcal{G}_1$ for $(\boldsymbol{\beta}, \boldsymbol{\Omega}, \sigma_1)$, which is induced by the mapping $h$ and the parameter space $\mathcal{F}_1$,

$$\mathcal{G}_1 = \{(\boldsymbol{\beta}, \boldsymbol{\Omega}, \sigma_1) \ : \ (\boldsymbol{\beta}, \boldsymbol{\Omega}, \sigma_1) = h(\boldsymbol{\Sigma}^v) \quad \text{for } \boldsymbol{\Sigma}^v \in \mathcal{F}_1\}. \tag{57}$$

For $\boldsymbol{\Sigma}_{\boldsymbol{\alpha}}^v$, the corresponding $(\boldsymbol{\beta}, \boldsymbol{\Omega}, \sigma_1)$ defined as $h(\boldsymbol{\Sigma}_{\boldsymbol{\alpha}}^v)$ satisfies

$$\boldsymbol{\beta}_1 = \frac{-\|\boldsymbol{\alpha}\|_2^2 \rho_0 + \boldsymbol{\beta}_1^*}{1 - \|\boldsymbol{\alpha}\|_2^2}, \ \boldsymbol{\beta}_2 = \eta_0, \ \boldsymbol{\beta}_{-\{1,2\}} = (\rho_0 - \boldsymbol{\beta}_1) \boldsymbol{\alpha} = \frac{\sigma_0}{1 - \|\boldsymbol{\alpha}\|_2^2} \boldsymbol{\alpha}, \tag{58}$$

$$\sigma_1^2 = \sigma_0^2 - \frac{\|\boldsymbol{\alpha}\|_2^2 (\rho_0 - \boldsymbol{\beta}_1^*)^2}{1 - \|\boldsymbol{\alpha}\|_2^2} \leq \sigma_0^2 \leq M_2, \tag{59}$$

$$\max\{|\lambda_{\min}(\boldsymbol{\Omega}) - 1|, |\lambda_{\max}(\boldsymbol{\Omega}) - 1|\} \leq \|\boldsymbol{\alpha}\|_2. \tag{60}$$

When $\|\boldsymbol{\alpha}\|_2$ is chosen such that $\|\boldsymbol{\alpha}\|_2 \leq \min\{1-1/M_1, M_1-1\}$, then we have

$$\frac{1}{M_1} \leq \lambda_{\min}(\boldsymbol{\Omega}) \leq \lambda_{\max}(\boldsymbol{\Omega}) \leq M_1. \tag{61}$$

The difference between $\boldsymbol{\beta}_1$ and $\boldsymbol{\beta}_1^*$ is

$$\boldsymbol{\beta}_1 - \boldsymbol{\beta}_1^* = \frac{\|\boldsymbol{\alpha}\|_2^2 (\boldsymbol{\beta}_1^* - \rho_0)}{1-\|\boldsymbol{\alpha}\|_2^2} = \frac{-\sigma_0\|\boldsymbol{\alpha}\|_2^2}{1-\|\boldsymbol{\alpha}\|_2^2}.$$

By taking $\rho = \sqrt{\log(4p_1/k^2)/2n}$, we have $|\boldsymbol{\beta}_1 - \boldsymbol{\beta}_1^*| \leq Ck\log p/n$ and hence $\boldsymbol{\beta}_1 \geq \boldsymbol{\beta}_1^* - Ck\log p/n \geq M_2/4$. Combined with (59), (58) and (61), we show that $\mathcal{G}_1 \subset \mathcal{G}^S(k)$. Let $\pi$ denote the uniform prior over the parameter space $\mathcal{G}_1$ induced by the uniform prior of $\boldsymbol{\alpha}$ over $\ell(p_1, k, \rho)$ where $\rho = \sqrt{\log(4p_1/k^2)/2n}$. The control of $L_1(f_\pi, f_{\boldsymbol{\theta}_0})$ is established in the following lemma, which follows from Lemma 2 of Cai and Guo (2016*a*) and is established in (7.21) of Cai and Guo (2016*a*).

LEMMA 6. *Suppose that* $k \leq c\{n/\log p, p^{\gamma}\}$, *where* $0 \leq \boldsymbol{\gamma} < \frac{1}{2}$ *and* $c$ *is a sufficient small positive constant. For* $\rho = \sqrt{\log(4p_1/k^2)/2n}$, *we establish that* $\|\boldsymbol{\alpha}\|_2 \leq \min\{1-1/M_1, M_1-1\}$ *and*

$$L_1(f_\pi, f_{\boldsymbol{\theta}_0}) \leq \frac{1}{4}. \tag{62}$$

To apply Lemma 5, we consider the functional $\mathrm{T}(\boldsymbol{\theta}) = \|\boldsymbol{\beta}\|_2^2$ and calculate the distance

$$d = \left|\boldsymbol{\beta}_1^2 + (\frac{\sigma_0}{1-\|\boldsymbol{\alpha}\|_2^2})^2\|\boldsymbol{\alpha}\|_2^2 - (\boldsymbol{\beta}_1^*)^2\right| = \frac{\sigma_0}{1-\|\boldsymbol{\alpha}\|_2^2}\|\boldsymbol{\alpha}\|_2^2\left|\boldsymbol{\beta}_1 + \boldsymbol{\beta}_1^* - \frac{\sigma_0}{1-\|\boldsymbol{\alpha}\|_2^2}\right|. \tag{63}$$

By taking $\boldsymbol{\beta}_1^* = M_2/2 > 2\sigma_0$ and $\eta_0 = 0$, we have $\boldsymbol{\beta}_1 \geq M_2/2 - Ck\log p/n$ and

$$\frac{\sigma}{1-\|\boldsymbol{\alpha}\|_2^2}\|\boldsymbol{\alpha}\|_2^2\left|\boldsymbol{\beta}_1 + \boldsymbol{\beta}_1^* - \frac{\sigma_0}{1-\|\boldsymbol{\alpha}\|_2^2}\right| \geq ck\frac{\log p}{n}\sigma_0\max\{|\boldsymbol{\beta}_1|, |\boldsymbol{\beta}_1^*|\} \geq ck\frac{\log p}{n}\sigma_0\max\{\|\boldsymbol{\beta}\|_2, \|\boldsymbol{\beta}^*\|_2\},$$

where the last inequality follows from the fact that $\|\boldsymbol{\beta}_{-1}\|_2 \leq Ck\log p\sigma_0/n$ and $\boldsymbol{\beta}_{-1}^* = 0$. Combined with (62), an application of Lemma 5 leads to (50).

**Proof of** (51) We construct the following parameter spaces,

$$\begin{aligned}\mathcal{G}_0 &= \left\{\boldsymbol{\theta}_0 = (\boldsymbol{\beta}^*, \mathrm{I}, \sigma_0) : \boldsymbol{\beta}^* = (\boldsymbol{\beta}_1^*, \eta_0, 0\cdots, 0), \text{ and } \boldsymbol{\beta}_1^* \geq \frac{M_2}{2}\right\} \subset \mathcal{G}^S(k) \\ \mathcal{G}_1 &= \left\{\boldsymbol{\theta}_1 = (\boldsymbol{\beta}, \mathrm{I}, \sigma_0) : \boldsymbol{\beta} = \left(\boldsymbol{\beta}_1^* + \sigma_0\frac{\bar{\epsilon}}{\sqrt{n}}, \eta_0, 0\cdots, 0\right)\right\} \subset \mathcal{G}^S(k),\end{aligned} \tag{64}$$

where $\bar{\epsilon}$ is a small positive constant to be chosen later.

The proof of the following lemma can be found in the supplementary material Section C.

LEMMA 7. *If* $\bar{\epsilon} = \sqrt{\log(17/16)/2}$*, then we have*

$$L_1(f_\pi, f_{\boldsymbol{\theta}_0}) \leq \frac{1}{4}. \tag{65}$$

To apply Lemma 5, we take $\eta_0 = 0$ and calculate the distance

$$d = \left|\boldsymbol{\beta}_1^2 - (\boldsymbol{\beta}_1^*)^2\right| = \left|2\boldsymbol{\beta}_1^*\sigma\frac{\bar{\epsilon}}{\sqrt{n}} + \sigma^2\frac{\bar{\epsilon}^2}{n}\right| \geq c\max\{\|\boldsymbol{\beta}\|_2, \|\boldsymbol{\beta}^*\|_2\}\frac{1}{\sqrt{n}}.$$

Applying Lemma 5, we establish (51).

**Proof of** (52) We introduce the following null and alternative parameter spaces,

$$\begin{aligned} \mathcal{G}_0 &= \{(\boldsymbol{\beta}, \mathbf{I}, \sigma_1) : \boldsymbol{\beta} = (\eta_0, 0, 0, \cdots, 0)\} \\ \mathcal{G}_1 &= \{(\boldsymbol{\beta}, \mathbf{I}, \sigma_1) : \boldsymbol{\beta} = (\eta_0, 0, \boldsymbol{\alpha}) \text{ with } \boldsymbol{\alpha} \in \ell_1(p, k, \rho)\}, \end{aligned} \tag{66}$$

where

$$\ell_1(p, k, \rho) = \left\{\boldsymbol{\alpha} : \boldsymbol{\alpha} \in \mathbf{R}^{p-2}, \|\boldsymbol{\alpha}\|_0 = k, \boldsymbol{\alpha}_i \in \{0, \rho\}\right\}. \tag{67}$$

Let $\pi$ denote the prior over the parameter space $\mathcal{G}_1$ induced by the uniform prior of $\boldsymbol{\alpha}$ over $\ell_1(p, k, \rho)$ where $\rho = \sqrt{\log(4p_1/k^2)/2n}$. The control of $L_1(f_\pi, f_{\boldsymbol{\theta}_0})$ is established in the following lemma, which follows from Lemma 7 of Cai and Guo (2016*b*) and is established in (1.6) of Cai and Guo (2016*b*).

LEMMA 8. *Suppose that* $k \leq c\{n/\log p, p^\gamma\}$*, where* $0 \leq \boldsymbol{\gamma} < \frac{1}{2}$ *and* $c$ *is a sufficient small positive constant. For* $\rho = \sqrt{\log(4p_1/k^2)/2n}$*, we establish*

$$L_1(f_\pi, f_{\boldsymbol{\theta}_0}) \leq \frac{1}{8}. \tag{68}$$

By specifying $\eta_0 = 0$, we have $\mathcal{G}_0$ and $\mathcal{G}_1$ defined in (66) are proper subspaces of the parameter space $\mathcal{G}^W(k)$. To apply Lemma 5, we calculate the distance $d = \|\boldsymbol{\alpha}\|_2^2 \geq ck\log p/n$. Applying Lemma 5, we establish (52).

# SUPPLEMENTARY MATERIAL

**Title:** Supplement to "Optimal Estimation of Co-heritability in High-dimensional Linear Regressions". (.pdf file)

# Supplement to “Optimal Estimation of Co-heritability in High-dimensional Linear Regressions”*

Zijian Guo[1] and Wanjie Wang[1,2]T. Tony Cai[1]Hongzhe Li[2]

[1]Department of Statistics, The Wharton School, University of Pennsylvania
[2]Department of Biostatistics and Epidemiology, Perelman School of Medicine, University of Pennsylvania

**Abstract**

The supplementary materials to the paper “Optimal Estimation of Co-heritability in High-dimensional Linear Regressions” present extended simulations in Section A. In Section B, we prove (29) and (32) in Theorem 3. We also provide detailed proofs of extra lemmas in Section C.

# A Simulation Results

## A.1 Choice of $\lambda_0$

In this section, we explore the choice of the tuning parameter $\lambda_0$ for the scaled Lasso estimators $\widehat{\boldsymbol{\beta}}$ and $\widehat{\boldsymbol{\gamma}}$, which is required in the Lasso approach, and also in the initial step of De-biased, Thresholded and FDE estimators. For these approaches, the theoretical tuning parameter $\lambda_0 = \sqrt{2.01 \log p}$ could be conservative in numerical studies. We would like to choose a tuning parameter such that all the proposed estimators have a reasonable performance.

In this experiment, we implemented the same steps as Section 4, with estimators Lasso, De-biased, Thresholded and FDE (FDE-S and FDE-NS for $\mathrm{R}(\boldsymbol{\beta}, \boldsymbol{\gamma})$) with tuning parameters $\lambda_0$. Parameters were set the same as Experiment 1, that $(p, n_1, n_2, rep) = (600, 400, 400, 300)$, $(s, s_1, s_2) = (15, 30, 25)$, and the covariance matrices $\boldsymbol{\Sigma}$ and $\boldsymbol{\Gamma}$ satisfy $\boldsymbol{\Sigma}_{ij} = \boldsymbol{\Gamma}_{ij} = (0.8)^{|i-j|}$. For given values $\tau_1$ and $\tau_2$, the signals of $\boldsymbol{\beta}$ satisfy that $\boldsymbol{\beta}_{j_i} = (1 + i/s_1)\,\tau_1/2$, for $j_i \in S_1$, $i = 1, 2, \cdots, s_1$, and the signals in $\boldsymbol{\gamma}$ satisfy that $\boldsymbol{\gamma}_j = \tau_2$ for $j \in S_2$. We consider the signal strength parameters as $(\tau_1, \tau_2) \in \{(3.0, .1), (1.8, .4)\}$, including the settings that one coefficient vector has much stronger signals than the other, and that the signals in two coefficient vectors are close. To investigate the effect of tuning paramters, we take $\lambda_0 = b\sqrt{2.01 \log p}$, and vary $b \in \{.25, .5, .75, 1\}$.

The results are summarized in Figure A1. For the estimation of $\mathrm{I}(\boldsymbol{\beta}, \boldsymbol{\gamma})$, the results show that as $b$ increased from .25 to .5, the MSE of most approaches decreased except Lasso. When $b$ increased from .5 to 1, the MSE of Lasso increased significantly, and the MSE of other estimators are relatively stable. For the estimation of $\mathrm{R}(\boldsymbol{\beta}, \boldsymbol{\gamma})$, all the approaches perform worse as $b$ increases from .5 to 1. When $b$ increases from .25 to .5, some approaches work better, and the other approaches do not change much. Combining the results for both $\mathrm{I}(\boldsymbol{\beta}, \boldsymbol{\gamma})$ and $\mathrm{R}(\boldsymbol{\beta}, \boldsymbol{\gamma})$, $b = .5$ is a generally good choice of tuning parameter such that most approaches have almost the best performance. Therefore, we take $\lambda_0 = .5\sqrt{2.01 \log p}$ in our simulation studies.

## A.2 Simulation results for quadratic functionals

In this section, we examine the performance of several estimators of the quadratic functional $Q(\boldsymbol{\beta})$, including the plug-in scaled Lasso estimator (Lasso), the plug-in de-biased estimator (De-biased), the plug-in thresholded Lasso estimator (Thresholded), and the proposed estimator FDE with sample splitting (FDE-S) and without sample splitting (FDE-NS). Details about these estimator are listed as follows.

- Lasso: The quadratic functional $\|\boldsymbol{\beta}\|_2^2$ is estimated by $\|\widehat{\boldsymbol{\beta}}\|_2^2$.

- De-biased: Denote the de-biased Lasso estimators as $\widetilde{\boldsymbol{\beta}}$. The quadratic functional $\|\boldsymbol{\beta}\|_2^2$ is estimated by $\|\widetilde{\boldsymbol{\beta}}\|_2^2$.

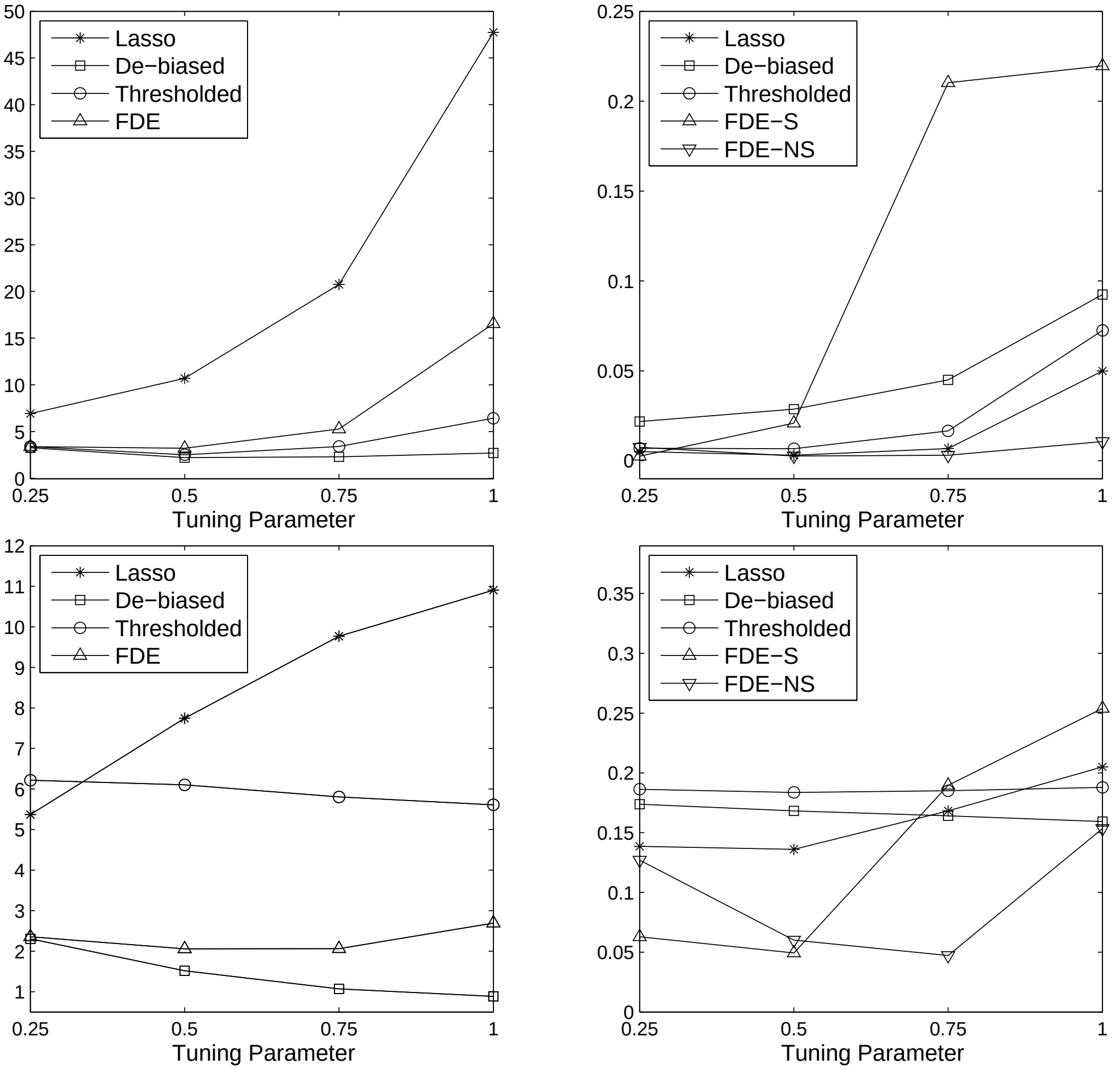

Figure A1: MSE for estimators Lasso (star), De-biased (square), Thresholded (circle), FDE (upward-pointing triangle), FDE-S (upward-pointing triangle) and FDE-NS (downward-pointing triangle) for tuning parameters with $b \in \{0.25, 0.5, 0.75, 1\}$. Top left: Estimation of $\mathrm{I}(\boldsymbol{\beta}, \boldsymbol{\gamma})$ for $(\tau_1, \tau_2) \in (1.8, 0.4)$; Top right: Estimation of $\mathrm{R}(\boldsymbol{\beta}, \boldsymbol{\gamma})$ for $(\tau_1, \tau_2) = (1.8, 0.4)$; Bottom left: Estimation of $\mathrm{I}(\boldsymbol{\beta}, \boldsymbol{\gamma})$ for $(\tau_1, \tau_2) = (3, 0.1)$; Bottom right: Estimation of $\mathrm{R}(\boldsymbol{\beta}, \boldsymbol{\gamma})$ for $(\tau_1, \tau_2) = (3, 0.1)$.

- Thresholded: Denote the thresholded estimator as $\bar{\boldsymbol{\beta}}$. The quadratic functional $\|\boldsymbol{\beta}\|_2^2$ is estimated by $\|\bar{\boldsymbol{\beta}}\|_2^2$.

- FDE: Estimate $\|\boldsymbol{\beta}\|_2^2$ as (16) in the main paper. We consider the method with data splitting (FDE-S) and without data splitting (FDE-NS).

For each setting, with parameters $(p, n, s)$, $\boldsymbol{\Sigma}$ and $F_{\boldsymbol{\beta}}$ specified, the following steps were implemented:

1. Generate sets $S \subset [p]$, with $|S| = s$. For $\boldsymbol{\beta} \in \mathbf{R}^p$, generate $\boldsymbol{\beta}_j \sim F_{\boldsymbol{\beta}}$ for $j \in S_1$, and set $\boldsymbol{\beta}_j = 0$ for $j \notin S_1$.

2. Generate $\boldsymbol{X}_{i\cdot} \overset{\text{i.i.d}}{\sim} N(\mathbf{0}, \boldsymbol{\Sigma})$, $1 \leq i \leq n$. Generate the noise $\boldsymbol{\epsilon}_i \overset{\text{i.i.d}}{\sim} N(0, 1)$, $1 \leq i \leq n$, and generate the outcome as $\boldsymbol{y} = \boldsymbol{X}\boldsymbol{\beta} + \boldsymbol{\epsilon}$.

3. With $\boldsymbol{X}$ and $\boldsymbol{y}$, estimate $\|\boldsymbol{\beta}\|_2^2$ through different estimators.

4. Repeat 2-3 for *rep* repetitions.

We use MSE in (33) in the main paper to measure the performance of each estimator. Similar to Section 4, we implement Experiments 1 and 2.

**Experiment 1**. This experiment aimed to explore the effect of signal strength for two types of $F_{\boldsymbol{\beta}}$. It contains two sub-experiments.

In Experiment 1a, the parameters were set as follows: $(p, n, s, rep) = (600, 400, 30, 300)$ and the covariance matrix $\boldsymbol{\Sigma}$ satisfies $\boldsymbol{\Sigma}(i, j) = (0.8)^{|i-j|}$. The signals of $\boldsymbol{\beta}$ follows that $\boldsymbol{\beta}_{j_i} = (1 + i/s)\tau/2$, $i = 1, 2, \cdots, s$. To investigate how the signal strength affects the estimation, vary the signal strength parameter $\tau \in \{.1, .2, .3, .4, 1.8, 2.2, 2.6, 3\}$.

In Experiment 1b, we use the same setting, except that $s = 25$, $F_{\boldsymbol{\beta}}$ is a point mass at $\tau$, and $\tau \in \{.1, .2, .3, .4, 1.0, 1.2, 1.4, 1.6\}$.

The results are summarized in Tables 5 and 6. FDE-NS has consistently outperformed Lasso, De-biased and Thresholded when the signals are weak. The performances of FDE-NS, Thresholded and De-biased were very similar when the signals are strong.

**Experiment 2**. This experiment aimed to examine the effect of signal sparsity on the estimators. The parameters were set as follows: $(p, n, rep) = (800, 400, 300)$, and the covariance matrix $\boldsymbol{\Sigma}$ has elements $\boldsymbol{\Sigma}_{ij} = (0.8)^{|i-j|}$. Two settings of $F_{\boldsymbol{\beta}}$ were considered: I. the signals of $\boldsymbol{\beta}$ follows that $\boldsymbol{\beta}_{j_i} = (1 + i/s)\tau/2$, $i = 1, 2, \cdots, s$, $\tau = .2$; and II. $\boldsymbol{\beta}_j = \tau$, $j \in S$, $\tau = .1$. To investigate the effect of signal sparsity, vary the number of signals $s \in \{40, 50, 60, 70, 80, 90, 100, 110\}$.

Table 5: Estimation of the quadratic functional $\mathrm{Q}(\boldsymbol{\beta})$ in Experiment 1a: the truth and MSE for the plug-in estimator with the scaled Lasso estimator (Lasso), plug-in estimator with the de-biased estimator (De-biased), plug-in estimator with the thresholded estimator (Thresholded), and the proposed estimators with sample splitting (FDE-S) and without sample splitting (FDE-NS).

| | Strength parameter, $\tau$ | | | | | | | |
|---|---|---|---|---|---|---|---|---|
| Method | .1 | .2 | .3 | .4 | 1.8 | 2.2 | 2.6 | 3.0 |
| Truth | .179 | .715 | 1.609 | 2.860 | 57.920 | 86.522 | 120.844 | 160.888 |
| Lasso | .014 | .210 | .882 | 2.224 | 95.226 | 147.032 | 210.047 | 273.340 |
| De-biased | 4.430 | 4.333 | 3.773 | 3.044 | 4.953 | 10.301 | 17.790 | 26.021 |
| Thresholded | .096 | .062 | .055 | .101 | 14.025 | 22.824 | 33.764 | 44.588 |
| FDE-S | .024 | .287 | 1.079 | 2.702 | 152.593 | 192.361 | 337.549 | 351.409 |
| FDE-NS | .012 | .015 | .100 | .277 | 9.725 | 21.173 | 21.610 | 40.177 |

Table 6: Estimation of the quadratic functional $\mathrm{Q}(\boldsymbol{\beta})$ in Experiment 1b: the truth and MSE for the plug-in estimator with the scaled Lasso estimator (Lasso), plug-in estimator with the de-biased estimator (De-biased), plug-in estimator with the thresholded estimator (Thresholded), and the proposed estimators with sample splitting (FDE-S) and without sample splitting (FDE-NS).

| | Strength parameter, $\tau$ | | | | | | | |
|---|---|---|---|---|---|---|---|---|
| Method | .1 | .2 | .3 | .4 | 1.0 | 1.2 | 1.4 | 1.6 |
| Truth | .250 | 1.000 | 2.000 | 4.000 | 25.000 | 36.000 | 49.000 | 64.000 |
| Lasso | .025 | .370 | 1.423 | 3.290 | 31.125 | 46.710 | 65.449 | 87.341 |
| De-biased | 4.400 | 4.246 | 3.463 | 2.679 | .822 | 1.324 | 2.387 | 4.011 |
| Thresholded | .091 | .057 | .063 | .146 | 3.553 | 5.812 | 8.632 | 12.012 |
| FDE-S | .025 | .370 | 1.423 | 3.290 | 31.125 | 46.710 | 65.449 | 87.341 |
| FDE-NS | .009 | .031 | .132 | .298 | 2.913 | 5.576 | 6.407 | 8.067 |

The results are summarized in Table 7. FDE-NS has consistently outperformed other methods for Settings I and II. In Setting II, where the signals were weaker than Setting I, the improvement of FDE-NS over other methods was larger. Compared to other methods, De-biased works much worse, which results from the large noise of the de-biased estimator.

Table 7: Estimation of the quadratic functional $\mathrm{Q}(\boldsymbol{\beta})$: the truth and MSE for the plug-in estimator with the scaled Lasso estimator (Lasso), plug-in estimator with the thresholded estimator (Thresholded), and the proposed estimators with sample splitting (FDE-S) and without sample splitting (FDE-NS).

| | | Sparsity parameter, $s$ | | | | | | | |
|---|---|---|---|---|---|---|---|---|---|
| | Method | 40 | 50 | 60 | 70 | 80 | 90 | 100 | 110 |
| I | Truth | .5535 | .6868 | .8201 | .9534 | 1.0868 | 1.2201 | 1.3534 | 1.4867 |
| | Lasso | .1347 | .2179 | .2688 | .3609 | .4634 | .5285 | .6236 | .7331 |
| | De-biased | 7.1867 | 7.2886 | 7.4523 | 7.4733 | 7.6875 | 8.0140 | 8.1476 | 8.3544 |
| | Thresholded | .2012 | .1717 | .1997 | .1827 | .1779 | .2052 | .2072 | .2331 |
| | FDE-S | .1788 | .2843 | .3573 | .4949 | .6078 | .7237 | .8509 | 1.0035 |
| | FDE-NS | .0134 | .0264 | .0296 | .0527 | .0790 | .0919 | .0954 | .1074 |
| II | Truth | .4000 | .5000 | .6000 | .7000 | .8000 | .9000 | 1.0000 | 1.1000 |
| | Lasso | .0683 | .1150 | .1401 | .1747 | .2326 | .2622 | .2893 | .2988 |
| | De-biased | 7.2395 | 7.5467 | 7.6953 | 7.9302 | 8.1280 | 8.5039 | 8.6979 | 8.9836 |
| | Thresholded | .2078 | .1973 | .2143 | .2212 | .2370 | .2589 | .2999 | .3806 |
| | FDE-S | .0683 | .1150 | .1401 | .1747 | .2326 | .2622 | .2893 | .2988 |
| | FDE-NS | .0057 | .0105 | .0113 | .0113 | .0148 | .0189 | .0264 | .0234 |

# B Proof of (29) and (32) in Theorem 3

We will apply Lemma 5 and establish the lower bounds (29) and (32) in this section. Let $f(\boldsymbol{X}, \boldsymbol{y}, \boldsymbol{Z}, \boldsymbol{w})$ denote the joint density function of $\boldsymbol{X}, \boldsymbol{y}, \boldsymbol{Z}, \boldsymbol{w}$, $f(\boldsymbol{X}, \boldsymbol{y})$ denote the joint density function of $\boldsymbol{X}, \boldsymbol{y}$ and $f(\boldsymbol{Z}, \boldsymbol{w})$ denote the joint density function of $\boldsymbol{Z}, \boldsymbol{w}$. The following lemma reduces the two sample $L_1$ distance to one sample $L_1$ distance. The proof of the following lemma can be found in the supplementary material Section C.

Lemma 9. *Assuming independence between $(\boldsymbol{X}, \boldsymbol{y})$ and $(\boldsymbol{Z}, \boldsymbol{w})$, then we have*

$$L_1\left(f_\pi\left(\boldsymbol{X}, \boldsymbol{y}, \boldsymbol{Z}, \boldsymbol{w}\right), f_{\boldsymbol{\theta}_0}\left(\boldsymbol{X}, \boldsymbol{y}, \boldsymbol{Z}, \boldsymbol{w}\right)\right) \leq L_1\left(f_\pi\left(\boldsymbol{X}, \boldsymbol{y}\right), f_{\boldsymbol{\theta}_0}\left(\boldsymbol{X}, \boldsymbol{y}\right)\right) + L_1\left(f_\pi\left(\boldsymbol{Z}, \boldsymbol{w}\right), f_{\boldsymbol{\theta}_0}\left(\boldsymbol{Z}, \boldsymbol{w}\right)\right). \tag{69}$$

Hence, by restricting to the cases where $(\boldsymbol{X}, \boldsymbol{y})$ and $(\boldsymbol{Z}, \boldsymbol{w})$ are independent, it is sufficient to control the $L_1$ distances $L_1\left(f_\pi\left(\boldsymbol{X}, \boldsymbol{y}\right), f_{\boldsymbol{\theta}_0}\left(\boldsymbol{X}, \boldsymbol{y}\right)\right)$ and $L_1\left(f_\pi\left(\boldsymbol{Z}, \boldsymbol{w}\right), f_{\boldsymbol{\theta}_0}\left(\boldsymbol{Z}, \boldsymbol{w}\right)\right)$. In the following, we construct the parameter space $\mathcal{H}_0$ and $\mathcal{H}_1$ and apply Lemma 5 and Lemma 9. The lower bound (32) in the main paper can be decomposed into the following three lower bounds,

$$\inf_{\widehat{\mathrm{R}}} \sup_{\boldsymbol{\theta} \in \Theta(k, \eta_0)} \mathbb{P}_{\boldsymbol{\theta}}\left(\left|\widehat{\mathrm{R}} - \mathrm{R}\left(\boldsymbol{\beta}, \boldsymbol{\gamma}\right)\right| \geq c' \frac{1}{\eta_0} \frac{k \log p}{n}\right) \geq \frac{1}{4}, \quad \text{for} \quad \eta_0 \geq C\sqrt{k \log p / n}, \tag{70}$$

$$\inf_{\widehat{\mathrm{R}}} \sup_{\boldsymbol{\theta} \in \Theta(k, \eta_0)} \mathbb{P}_{\boldsymbol{\theta}}\left(\left|\widehat{\mathrm{R}} - \mathrm{R}\left(\boldsymbol{\beta}, \boldsymbol{\gamma}\right)\right| \geq c' \frac{1}{\eta_0} \frac{1}{\sqrt{n}}\right) \geq \frac{1}{4}, \quad \text{for} \quad \eta_0 \geq C\sqrt{k \log p / n}, \tag{71}$$

$$\inf_{\widehat{\mathrm{R}}} \sup_{\boldsymbol{\theta} \in \Theta(k, \eta_0)} \mathbb{P}_{\boldsymbol{\theta}}\left(\left|\widehat{\mathrm{R}} - \mathrm{R}\left(\boldsymbol{\beta}, \boldsymbol{\gamma}\right)\right| \geq c' \min\left\{\frac{1}{\eta_0^2} \frac{k \log p}{n}, 1\right\}\right) \geq \frac{1}{4}, \tag{72}$$

where $c' > 0$ is a positive constant. Assuming $\eta_0 \geq C\sqrt{k \log p / n}$, combining (70), (71) and (72), we can establish the following lower bound

$$\inf_{\widehat{\mathrm{R}}} \sup_{\boldsymbol{\theta} \in \Theta(k, \eta_0)} \mathbb{P}_{\boldsymbol{\theta}}\left(\left|\widehat{\mathrm{R}} - \mathrm{R}\left(\boldsymbol{\beta}, \boldsymbol{\gamma}\right)\right| \geq c' \frac{1}{\eta_0}\left(\frac{1}{\sqrt{n}} + \frac{k \log p}{n}\right) + c' \frac{1}{\eta_0^2} \frac{k \log p}{n}\right) \geq \frac{1}{4}. \tag{73}$$

In the case $\eta_0 \leq C\sqrt{k \log p / n}$, by (72), we establish the lower bound

$$\inf_{\widehat{\mathrm{R}}} \sup_{\boldsymbol{\theta} \in \Theta(k, \eta_0)} \mathbb{P}_{\boldsymbol{\theta}}\left(\left|\widehat{\mathrm{R}} - \mathrm{R}\left(\boldsymbol{\beta}, \boldsymbol{\gamma}\right)\right| \geq c'\right) \geq \frac{1}{4}. \tag{74}$$

Combining (73) and (74), we establish (32) in the main paper.

To facilitate the discussion of the lower bound (29) in the main paper, we introduce the following three disjoint parameter spaces,

$$\begin{aligned}
\Theta_1(k) &= \Theta(k) \cap \left\{ \|\boldsymbol{\beta}\|_2 \leq C\sqrt{\frac{k\log p}{n}} \right\} \cap \left\{ \|\boldsymbol{\gamma}\|_2 \geq \frac{M_2}{4} \right\}, \\
\Theta_2(k) &= \Theta(k) \cap \left\{ \|\boldsymbol{\beta}\|_2 \geq \frac{M_2}{4} \right\} \cap \left\{ \|\boldsymbol{\gamma}\|_2 \leq C\sqrt{\frac{k\log p}{n}} \right\}, \\
\Theta_3(k) &= \Theta(k) \cap \left\{ \|\boldsymbol{\beta}\|_2 \leq C\sqrt{\frac{k\log p}{n}} \right\} \cap \left\{ \|\boldsymbol{\gamma}\|_2 \leq C\sqrt{\frac{k\log p}{n}} \right\}.
\end{aligned} \tag{75}$$

The lower bound (29) in the main paper can be decomposed into the following three lower bounds,

$$\inf_{\widehat{\mathrm{I}}} \sup_{\boldsymbol{\theta}\in\Theta_1(k)\cup\Theta_2(k)} \mathbb{P}_{\boldsymbol{\theta}}\left( \left|\widehat{\mathrm{I}} - \mathrm{I}(\boldsymbol{\beta},\boldsymbol{\gamma})\right| \geq c'\left(\|\boldsymbol{\beta}\|_2 + \|\boldsymbol{\gamma}\|_2\right)\frac{k\log p}{n} \right) \geq \frac{1}{4}, \tag{76}$$

$$\inf_{\widehat{\mathrm{I}}} \sup_{\boldsymbol{\theta}\in\Theta_1(k)\cup\Theta_2(k)} \mathbb{P}_{\boldsymbol{\theta}}\left( \left|\widehat{\mathrm{I}} - \mathrm{I}(\boldsymbol{\beta},\boldsymbol{\gamma})\right| \geq c'\left(\|\boldsymbol{\beta}\|_2 + \|\boldsymbol{\gamma}\|_2\right)\frac{1}{\sqrt{n}} \right) \geq \frac{1}{4}, \tag{77}$$

$$\inf_{\widehat{\mathrm{I}}} \sup_{\boldsymbol{\theta}\in\Theta_3(k)} \mathbb{P}_{\boldsymbol{\theta}}\left( \left|\widehat{\mathrm{I}} - \mathrm{I}(\boldsymbol{\beta},\boldsymbol{\gamma})\right| \geq c'\frac{k\log p}{n} \right) \geq \frac{1}{4}, \tag{78}$$

where $c' > 0$ is a positive constant. Combining (76) and (77), we can establish

$$\inf_{\widehat{\mathrm{I}}} \sup_{\boldsymbol{\theta}\in\Theta_1(k)\cup\Theta_2(k)} \mathbb{P}_{\boldsymbol{\theta}}\left( \left|\widehat{\mathrm{I}} - \mathrm{I}(\boldsymbol{\beta},\boldsymbol{\gamma})\right| \geq c'\left(\|\boldsymbol{\beta}\|_2 + \|\boldsymbol{\gamma}\|_2\right)\left(\frac{1}{\sqrt{n}} + \frac{k\log p}{n}\right) \right) \geq \frac{1}{4}. \tag{79}$$

On $\Theta_1(k) \cup \Theta_2(k)$, we have

$$\left(\|\boldsymbol{\beta}\|_2 + \|\boldsymbol{\gamma}\|_2\right)\left(\frac{1}{\sqrt{n}} + \frac{k\log p}{n}\right) = \max\left\{ \left(\|\boldsymbol{\beta}\|_2 + \|\boldsymbol{\gamma}\|_2\right)\left(\frac{1}{\sqrt{n}} + \frac{k\log p}{n}\right), \frac{k\log p}{n} \right\}$$

and on $\Theta_3(k)$, we have

$$\frac{k\log p}{n} = \max\left\{ \left(\|\boldsymbol{\beta}\|_2 + \|\boldsymbol{\gamma}\|_2\right)\left(\frac{1}{\sqrt{n}} + \frac{k\log p}{n}\right), \frac{k\log p}{n} \right\}.$$

Since both $\Theta_1(k) \cup \Theta_2(k)$ and $\Theta_3(k)$ are proper subspace of $\Theta(k)$, we can establish (29) by combining (79) and (78).

We will first present the proofs of (70) and (76), which are similar in terms of construction of the null and alternative parameter spaces. Then we will present proofs for (71) and (77) and for (72) and (78). Define $\boldsymbol{\gamma}^* = (\boldsymbol{\gamma}_1^*, 0, 0 \cdots, 0)$ for some $\boldsymbol{\gamma}_1^* \geq 0$.

**Proof of (70) and (76)** We define the null and alternative spaces as

$$\begin{aligned} \mathcal{H}_0 &= \{(\boldsymbol{\beta}^*, \mathbf{I}, \sigma_0, \boldsymbol{\gamma}^*, \mathbf{I}, \sigma_2) : \boldsymbol{\beta}^* = (\boldsymbol{\beta}_1^*, \eta_0, 0 \cdots, 0)\}, \\ \mathcal{H}_1 &= \{(\boldsymbol{\beta}, \boldsymbol{\Omega}, \sigma_1, \boldsymbol{\gamma}^*, \mathbf{I}, \sigma_2) : (\boldsymbol{\beta}, \boldsymbol{\Omega}, \sigma_1) \in \mathcal{G}_1\}, \end{aligned} \tag{80}$$

where $\mathcal{G}_1$ is defined in (57) in the main paper. Let $\pi$ denote the uniform prior over the parameter space $\mathcal{H}_1$, which is induced by the uniform prior of $\boldsymbol{\alpha}$ over $\ell_1(p, k, \rho)$ defined in (56) in the main paper. Taking $\rho = \sqrt{\log(4p_1/k^2)/2n}$, by Lemma 9 and Lemma 6, we establish that

$$L_1(f_\pi, f_{\boldsymbol{\theta}_0}) \leq \frac{1}{4}. \tag{81}$$

1. To establish the lower bound for the ratio, we take $\boldsymbol{\beta}_1^* = 0$ and then we have $\boldsymbol{\beta}_1 = -\|\boldsymbol{\alpha}\|_2^2 \sigma_0/(1 - \|\boldsymbol{\alpha}\|_2^2)$ and specify $\boldsymbol{\gamma}_1^* = \eta_0$ and hence $\mathcal{H}_0, \mathcal{H}_1 \subset \Theta(k, \eta_0)$. We calculate the distance as
$$d = \left| \frac{\boldsymbol{\beta}_1 \eta_0}{\|\boldsymbol{\beta}\|_2 |\eta_0|} - 0 \right| \geq c \frac{k \frac{\log p}{n}}{\sqrt{\eta_0^2 + k \frac{\log p}{n}}}.$$
For $\eta_0 \geq C\sqrt{k \log p/n}$, we establish (70).

2. To establish the lower bound for the inner product, we take $\boldsymbol{\beta}_1^* = 0$ and $\eta_0 = 0$ and then we have $\boldsymbol{\beta}_1 = -\|\boldsymbol{\alpha}\|_2^2 \sigma_0/(1 - \|\boldsymbol{\alpha}\|_2^2)$, $\|\boldsymbol{\beta}^*\|_2 = 0$ and $\|\boldsymbol{\beta}\|_2 \leq \sqrt{k \log p/n}$. We take $\boldsymbol{\gamma}_1^* = M_2/2$ and hence $\mathcal{H}_0, \mathcal{H}_1 \subset \Theta_1(k)$. We calculate
$$d = |\boldsymbol{\beta}_1 \boldsymbol{\gamma}_1^* - 0| = \|\boldsymbol{\gamma}^*\|_2 \frac{\|\boldsymbol{\alpha}\|_2^2 \sigma_0}{1 - \|\boldsymbol{\alpha}\|_2^2} \geq c \|\boldsymbol{\gamma}^*\|_2 \frac{k \log p}{n}.$$
Applying Lemma 5, we establish
$$\inf_{\widehat{\mathrm{I}}} \sup_{\boldsymbol{\theta} \in \Theta_1(k)} \mathbb{P}_{\boldsymbol{\theta}} \left( \left| \widehat{\mathrm{I}} - \mathrm{I}(\boldsymbol{\beta}, \boldsymbol{\gamma}) \right| \geq c \|\boldsymbol{\gamma}\|_2 \frac{k \log p}{n} \right) \geq \frac{1}{4}. \tag{82}$$
By symmetry, we can also establish
$$\inf_{\widehat{\mathrm{I}}} \sup_{\boldsymbol{\theta} \in \Theta_2(k)} \mathbb{P}_{\boldsymbol{\theta}} \left( \left| \widehat{\mathrm{I}} - \mathrm{I}(\boldsymbol{\beta}, \boldsymbol{\gamma}) \right| \geq c \|\boldsymbol{\beta}\|_2 \frac{k \log p}{n} \right) \geq \frac{1}{4}. \tag{83}$$
Combining (82) and (83), we establish (76).

**Proof of (71) and (77)** We define the null and alternative spaces as

$$\begin{aligned}\mathcal{H}_0 &= \left\{(\boldsymbol{\beta}^*, \mathbf{I}, \sigma_0, \boldsymbol{\gamma}^*, \mathbf{I}, \sigma_2) : \boldsymbol{\beta}^* = (\boldsymbol{\beta}_1^*, \eta_0, 0 \cdots, 0)\right\}, \\ \mathcal{H}_1 &= \left\{(\boldsymbol{\beta}, \mathbf{I}, \sigma_0, \boldsymbol{\gamma}^*, \mathbf{I}, \sigma_2) : \boldsymbol{\beta} = \left(\boldsymbol{\beta}_1^* + \sigma_0 \frac{\bar{\epsilon}}{\sqrt{n}}, \eta_0, 0 \cdots, 0\right)\right\}.\end{aligned} \tag{84}$$

By Lemma 9 and Lemma 7, we establish that

$$L_1 \left(f_\pi, f_{\boldsymbol{\theta}_0}\right) \leq \frac{1}{4}. \tag{85}$$

1. To establish the lower bound for the ratio, we take $\boldsymbol{\beta}_1^* = 0$ and $\boldsymbol{\gamma}_1^* = \eta_0$ and hence $\mathcal{H}_0, \mathcal{H}_1 \subset \Theta(k, \eta_0)$. We calculate the distance as

$$d = \left| \frac{\boldsymbol{\beta}_1 \eta_0}{\|\boldsymbol{\beta}\|_2 |\eta_0|} - 0 \right| \geq c \frac{\frac{1}{\sqrt{n}}}{\sqrt{\eta_0^2 + c\frac{1}{n}}}.$$

   For $\eta_0 \geq C\sqrt{k \log p / n} \gg 1/\sqrt{n}$, we establish (71).

2. To establish the lower bound for the inner product, we take $\boldsymbol{\beta}_1^* = 0$ and $\eta_0 = 0$ and then we have $\|\boldsymbol{\beta}^*\|_2 = 0$ and $\|\boldsymbol{\beta}\|_2 \leq 1/\sqrt{n}$. We take $\boldsymbol{\gamma}_1^* = M_2/2$ and hence $\mathcal{H}_0, \mathcal{H}_1 \subset \Theta_1(k)$. We calculate the distance as

$$d = |\boldsymbol{\beta}_1 \boldsymbol{\gamma}_1^* - 0| \geq c \|\boldsymbol{\gamma}^*\|_2 \frac{\sigma_0}{\sqrt{n}}.$$

   Applying Lemma 5, we establish

$$\inf_{\widehat{\mathrm{I}}} \sup_{\boldsymbol{\theta} \in \Theta_1(k)} \mathbb{P}_{\boldsymbol{\theta}} \left( \left| \widehat{\mathrm{I}} - \mathrm{I}\left(\boldsymbol{\beta}, \boldsymbol{\gamma}\right) \right| \geq c \|\boldsymbol{\gamma}\|_2 \frac{1}{\sqrt{n}} \right) \geq \frac{1}{4}. \tag{86}$$

   By symmetry, we can also establish

$$\inf_{\widehat{\mathrm{I}}} \sup_{\boldsymbol{\theta} \in \Theta_2(k)} \mathbb{P}_{\boldsymbol{\theta}} \left( \left| \widehat{\mathrm{I}} - \mathrm{I}\left(\boldsymbol{\beta}, \boldsymbol{\gamma}\right) \right| \geq c \|\boldsymbol{\beta}\|_2 \frac{1}{\sqrt{n}} \right) \geq \frac{1}{4}. \tag{87}$$

   Combining (86) and (87), we establish (77).

**Proof of (72) and (78)** We introduce the following null and alternative,

$$\begin{aligned}\mathcal{H}_0 &= \left\{(\boldsymbol{\beta}, \mathbf{I}, \sigma_1, \boldsymbol{\gamma}, \mathbf{I}, \sigma_2) : \boldsymbol{\beta} = (\eta_0, 0, 0, \cdots, 0), \ \boldsymbol{\gamma} = (0, \eta_0, 0, \cdots, 0)\right\} \subset \Theta(k, \eta_0) \subset \Theta(k) \\ \mathcal{H}_1 &= \left\{(\boldsymbol{\beta}, \mathbf{I}, \sigma_1, \boldsymbol{\gamma}, \mathbf{I}, \sigma_2) : \boldsymbol{\beta} = (\eta_0, 0, \boldsymbol{\alpha}), \ \boldsymbol{\gamma} = (0, \eta_0, \boldsymbol{\alpha}), \text{ with } \boldsymbol{\alpha} \in \ell_1\left(p, k, \rho\right)\right\} \subset \Theta(k, \eta_0) \subset \Theta(k),\end{aligned} \tag{88}$$

where $\ell_1(p,k,\rho)$ is defined in (67) in the main paper. Let $\pi$ denote the prior over the parameter space $\mathcal{H}_1$ induced by the uniform prior of $\boldsymbol{\alpha}$ over $\ell_1(p,k,\rho)$ defined in (67) in the main paper. Taking $\rho = \sqrt{\log(4p_1/k^2)/2n}$, by Lemma 8, we have

$$L_1\left(f_\pi(\boldsymbol{y},\boldsymbol{X}), f_{\boldsymbol{\theta}_0}(\boldsymbol{y},\boldsymbol{X})\right) \leq \frac{1}{8} \quad \text{and} \quad L_1\left(f_\pi(\boldsymbol{w},\boldsymbol{Z}), f_{\boldsymbol{\theta}_0}(\boldsymbol{w},\boldsymbol{Z})\right) \leq \frac{1}{8}.$$

By Lemma 9, we have

$$L_1\left(f_\pi(\boldsymbol{y},\boldsymbol{X},\boldsymbol{w},\boldsymbol{Z}), f_{\boldsymbol{\theta}_0}(\boldsymbol{y},\boldsymbol{X},\boldsymbol{w},\boldsymbol{Z})\right) \leq \frac{1}{4}. \tag{89}$$

1. To establish the lower bound for the normalized inner product, we calculate the distance $d$ as

   $$d = \left| \frac{\|\boldsymbol{\alpha}\|_2^2}{\left(\sqrt{\eta_0^2 + \|\boldsymbol{\alpha}\|_2^2}\right)^2} - 0 \right| \geq c \min\left\{ \frac{1}{\eta_0^2} \frac{k \log p}{n}, 1 \right\}. \tag{90}$$

   Applying Lemma 5, we establish (71).

2. To establish the lower bound for the inner product, we take $\eta_0 = 0$ and hence $\mathcal{H}_0, \mathcal{H}_1 \subset \Theta_3(k)$. We calculate the distance

   $$d = \left| \|\boldsymbol{\alpha}\|_2^2 - 0 \right| \geq c \frac{k \log p}{n}.$$

   Applying Lemma 5, we establish (78).

# C Supplementary Proof

In this section, we prove the extra lemmas. The proofs of Lemma 1 and 2 follow from the results and analysis in Cai and Guo (2016$a$); Ren et al. (2013); Sun and Zhang (2012); Ye and Zhang (2010).

**Proof of Lemma 1**

In the following, we establish upper bounds for $\|\widehat{\boldsymbol{\beta}} - \boldsymbol{\beta}\|_1$, $\|\widehat{\boldsymbol{\beta}} - \boldsymbol{\beta}\|_2$ and the corresponding proofs for $\|\widehat{\boldsymbol{\gamma}} - \boldsymbol{\gamma}\|_1$ and $\|\widehat{\boldsymbol{\gamma}} - \boldsymbol{\gamma}\|_2$ are similar and omitted here. The proof has been established in Sun and Zhang (2012); Ye and Zhang (2010) for fixed designs under certain assumptions for the design. In the following, we establish that the conditions are satisfied with high

probability for the Sub-gaussian random designs considered in this paper.

By normalizing the columns of $\boldsymbol{X}$ and the true sparse vector $\boldsymbol{\beta}$, the linear regression model can be expressed as

$$\boldsymbol{y} = \boldsymbol{H}\boldsymbol{d} + \epsilon, \quad \text{with } \boldsymbol{H} = \boldsymbol{X}\boldsymbol{D},\ \boldsymbol{d} = \boldsymbol{D}^{-1}\boldsymbol{\beta} \text{ and } \boldsymbol{\epsilon} \sim N(0, \sigma_1^2 \mathbf{I}), \tag{91}$$

where $\boldsymbol{D} = \operatorname{diag}\left(\sqrt{n}/\|\boldsymbol{X}_{\cdot j}\|_2\right)_{j\in[p]}$ denotes the $p \times p$ diagonal matrix with $(j, j)$ entry to be $\sqrt{n}/\|\boldsymbol{X}_{\cdot j}\|_2$. Take $\delta_0 = 1.0048$, we have $\lambda_0 = (1.01)\sqrt{2\delta_0 \log p}$. Take $\epsilon_0 = 202$, $C_1 = 2.25$, $c_0 = 1/6$ and $C_0 = 3$. Rather than use the constants directly in the following discussion, we use $\delta_0, \epsilon_0, C_1, C_0$ and $c_0$ to represent the above fixed constants in the following discussion. We also assume that $\log p/n \leq 1/25$ and $\delta_0 \log p > 2$. We review the following definitions of restricted eigenvalue introduced in Bickel et al. (2009),

$$\kappa(\boldsymbol{X}, k, \alpha_0) = \min_{\substack{J_0 \subset \{1,\cdots,p\},\\ |J_0| \leq k}} \min_{\substack{\boldsymbol{\eta} \neq 0,\\ \|\boldsymbol{\eta}_{J_0^c}\|_1 \leq \alpha_0 \|\boldsymbol{\eta}_{J_0}\|_1}} \frac{\|\boldsymbol{X}\boldsymbol{\eta}\|_2}{\sqrt{n}\|\boldsymbol{\eta}_{J_0}\|_2}. \tag{92}$$

and

$$\kappa(\boldsymbol{X}, k, s, \alpha_0) = \min_{\substack{J_0 \subset \{1,\cdots,p\},\\ |J_0| \leq k}} \min_{\substack{\boldsymbol{\eta} \neq 0,\\ \|\boldsymbol{\eta}_{J_0^c}\|_1 \leq \alpha_0 \|\boldsymbol{\eta}_{J_0}\|_1}} \frac{\|\boldsymbol{X}\boldsymbol{\eta}\|_2}{\sqrt{n}\|\boldsymbol{\eta}_{J_{01}}\|_2}, \tag{93}$$

where $J_1$ denotes the subset corresponding to the $s$ largest in absolute value coordinates of $\boldsymbol{\eta}$ outside of $J_0$ and $J_{01} = J_0 \cup J_1$. Define $\sigma_1^{\text{ora}} = \|\boldsymbol{y} - \boldsymbol{X}\boldsymbol{\beta}\|_2/\sqrt{n}$, and

$$\tau = \sqrt{1+\epsilon_0}\frac{\sqrt{k}\lambda_0}{\sqrt{n}\kappa(\boldsymbol{H}, k, 1+2\epsilon_0)}. \tag{94}$$

Let $\boldsymbol{\Omega} = \boldsymbol{\Sigma}^{-1}$ denote the precision matrix of the design matrix. To facilitate the proof, we define the following events for the random design $\boldsymbol{X}$ and the error $\boldsymbol{\epsilon}$,

$$G_1 = \left\{\frac{2}{5}\frac{1}{\sqrt{M_1}} < \frac{\|\boldsymbol{X}_{\cdot j}\|_2}{\sqrt{n}} < \frac{7}{5}\sqrt{M_1} \text{ for } 1 \leq j \leq p\right\},$$

$$G_2 = \left\{\left|\frac{(\sigma_1^{ora})^2}{\sigma_1^2} - 1\right| \leq 2\sqrt{\frac{\log p}{n}} + 2\frac{\log p}{n}\right\},$$

$$G_3 = \left\{\frac{1}{2} \leq \frac{\|\boldsymbol{X}\boldsymbol{v}\|_2}{\sqrt{n}\|\boldsymbol{\Omega}^{\frac{1}{2}}\widehat{\boldsymbol{\gamma}}\|_2} \leq \frac{3}{2}\right\}, \text{ where } \boldsymbol{v} = \boldsymbol{\Omega}\widehat{\boldsymbol{\gamma}},$$

$$G_4 = \left\{\kappa(\boldsymbol{X}, k, k, \alpha_0) \geq \frac{c}{\sqrt{\lambda_{\max}(\boldsymbol{\Omega})}}\right\},$$

$$G_5 = \left\{ \frac{\|\boldsymbol{H}^{\intercal}\boldsymbol{\epsilon}\|_{\infty}}{n} \leq \sigma_1 \sqrt{\frac{2\delta_0 \log p}{n}} \right\},$$

$$S_1 = \left\{ \frac{\|\boldsymbol{H}^{\intercal}\boldsymbol{\epsilon}\|_{\infty}}{n} \leq \sigma^{ora} \lambda_0 \frac{\epsilon_0 - 1}{\epsilon_0 + 1} (1 - \tau) \right\},$$

$$S_2 = \{(1 - \nu_0)\, \hat{\sigma} \leq \sigma \leq (1 + \nu_0)\hat{\sigma}_1\},$$

$$B_1 = \left\{ \|\widehat{\boldsymbol{\gamma}}^{\intercal}\boldsymbol{\Omega}\widehat{\boldsymbol{\Sigma}} - \widehat{\boldsymbol{\gamma}}^{\intercal}\|_{\infty} \leq \|\widehat{\boldsymbol{\gamma}}\|_2 \lambda_1 / \sqrt{n} \right\}, \quad \text{where} \quad \lambda_1 = 4C_0 M_1^2 \sqrt{\log p}.$$

where $\widehat{\boldsymbol{\Sigma}} = \boldsymbol{X}^{\intercal}\boldsymbol{X}/n$ and $\widehat{\boldsymbol{\gamma}}$ is the scaled Lasso estimator defined in (5) in the main paper. Define $G = \cap_{i=1}^{5} G_i$ and $S = \cap_{i=1}^{2} S_i$. We will check the assumptions in Corollary 1 in Sun and Zhang (2012) and apply equation (23) in Sun and Zhang (2012). By the definition of $\tau^*$ in Sun and Zhang (2012), we have $\tau^* \leq \tau$ where $\tau$ is defined in (94). Hence, on the event $S_1$, equation (23) in Sun and Zhang (2012) holds. By equation (27) and the discussion after (27) in Ye and Zhang (2010), we obtain

$$\|\widehat{\boldsymbol{\beta}} - \boldsymbol{\beta}\|_2 \leq C \frac{\sqrt{k}\lambda_0 \sigma_1}{\sqrt{n}\kappa^2(H, k, k, 1 + 2\epsilon_0)}. \tag{95}$$

Similar to the proof of Lemma 13 in Cai and Guo (2016*b*), we establish

$$\kappa^2\left(\boldsymbol{H}, k, k, 1 + 2\epsilon_0\right) \geq \frac{n}{\max \|\boldsymbol{X}_{\cdot j}\|_2^2} \kappa^2 \left( \boldsymbol{X}, k, k, (1 + 2\epsilon_0) \left( \frac{\max \|\boldsymbol{X}_{\cdot j}\|_2}{\min \|\boldsymbol{X}_{\cdot j}\|_2} \right) \right). \tag{96}$$

Hence, on the event $G \cap S$, we establish the upper bound for $\|\widehat{\boldsymbol{\beta}} - \boldsymbol{\beta}\|_2$ in (36) in the main paper. By the discussion in Sun and Zhang (2012); Ye and Zhang (2010) or Lemma 5 in Cai and Guo (2016*a*), we can establish the upper bound for $\|\widehat{\boldsymbol{\beta}} - \boldsymbol{\beta}\|_1$ in (35) in the main paper. The remaining proof of Lemma 1 relies on the following lemma, which controls the probability of events $G$, $S$ and $B_1$.

Lemma 10. *Suppose the assumption* (A1) *in the main paper holds and* $k \log p/n \to 0$ *with* $k = \max\{\|\boldsymbol{\beta}\|_0, \|\boldsymbol{\gamma}\|_0\}$*. Then*

$$\mathbb{P}(G) \geq 1 - \frac{6}{p} - 2p^{1-C_1} - \frac{1}{2\sqrt{\pi\delta_0 \log p}} p^{1-\delta_0} - c' \exp(-cn), \tag{97}$$

*and*

$$\mathbb{P}(G \cap S) \geq \mathbb{P}(G) - 2\exp(-cn) - c'' \frac{1}{\sqrt{\log p}} p^{1-\delta_0}, \tag{98}$$

*and*

$$\mathbb{P}\left(B_1\right) \geq 1 - 2p^{1-c_0 C_0^2}, \tag{99}$$

*where $c$, $c'$ and $c''$ are universal positive constants.*

**Proof of Lemma 2**

In the following, we establish the upper bound for $\left|\left(\widehat{\boldsymbol{u}}_1^{\intercal}\widehat{\boldsymbol{\Sigma}} - \widehat{\boldsymbol{\gamma}}^{\intercal}\right)\left(\boldsymbol{\beta} - \widehat{\boldsymbol{\beta}}\right)\right|$,$|\widehat{\boldsymbol{u}}_1^{\intercal}\boldsymbol{X}^{\intercal}\boldsymbol{\epsilon}/n|$ and $\left|\widehat{\boldsymbol{u}}_1^{\intercal}\boldsymbol{X}^{\intercal}\left(\boldsymbol{y} - \boldsymbol{X}\widehat{\boldsymbol{\beta}}\right)/n - \langle\widehat{\boldsymbol{\gamma}}, \boldsymbol{\beta} - \widehat{\boldsymbol{\beta}}\rangle\right|$ and the corresponding proofs for $\left|\left(\widehat{\boldsymbol{u}}_2^{\intercal}\widehat{\boldsymbol{\Gamma}} - \widehat{\boldsymbol{\beta}}^{\intercal}\right)\left(\boldsymbol{\gamma} - \widehat{\boldsymbol{\gamma}}\right)\right|$, $|\widehat{\boldsymbol{u}}_2^{\intercal}\boldsymbol{Z}^{\intercal}\boldsymbol{\delta}/n|$ and $\left|\widehat{\boldsymbol{u}}_2^{\intercal}\boldsymbol{Z}^{\intercal}\left(\boldsymbol{w} - \boldsymbol{Z}\widehat{\boldsymbol{\gamma}}\right)/n - \langle\widehat{\boldsymbol{\beta}}, \widehat{\boldsymbol{\gamma}} - \boldsymbol{\gamma}\rangle\right|$ are similar and omitted here.

Note that

$$\widehat{\boldsymbol{u}}_1^{\intercal}\frac{1}{n}\boldsymbol{X}^{\intercal}\left(\boldsymbol{y} - \boldsymbol{X}\widehat{\boldsymbol{\beta}}\right) - \langle\widehat{\boldsymbol{\gamma}}, \boldsymbol{\beta} - \widehat{\boldsymbol{\beta}}\rangle = \widehat{\boldsymbol{u}}_1^{\intercal}\frac{1}{n}\boldsymbol{X}^{\intercal}\boldsymbol{\epsilon} + \left(\widehat{\boldsymbol{u}}_1^{\intercal}\widehat{\boldsymbol{\Sigma}} - \widehat{\boldsymbol{\gamma}}^{\intercal}\right)\left(\boldsymbol{\beta} - \widehat{\boldsymbol{\beta}}\right).$$

It is sufficient to establish (38) and (37) in the main paper and then (39) in the main paper follows. On the event $S \cap G \cap B_1$, we have $\|\widehat{\boldsymbol{\beta}} - \boldsymbol{\beta}\|_1 \leq Ck\sqrt{\log p/n}$ and $\|\widehat{\boldsymbol{u}}_1^{\intercal}\widehat{\boldsymbol{\Sigma}} - \widehat{\boldsymbol{\gamma}}^{\intercal}\|_\infty \leq \|\widehat{\boldsymbol{\gamma}}\|_2\lambda_1/\sqrt{n}$. The control of $\left(\widehat{\boldsymbol{u}}_1^{\intercal}\widehat{\boldsymbol{\Sigma}} - \widehat{\boldsymbol{\gamma}}^{\intercal}\right)\left(\boldsymbol{\beta} - \widehat{\boldsymbol{\beta}}\right)$ in (37) in the main paper follows from the inequality $\left|\left(\widehat{\boldsymbol{u}}_1^{\intercal}\widehat{\boldsymbol{\Sigma}} - \widehat{\boldsymbol{\gamma}}^{\intercal}\right)\left(\boldsymbol{\beta} - \widehat{\boldsymbol{\beta}}\right)\right| \leq \|\widehat{\boldsymbol{u}}_1^{\intercal}\widehat{\boldsymbol{\Sigma}} - \widehat{\boldsymbol{\gamma}}^{\intercal}\|_\infty\|\boldsymbol{\beta} - \widehat{\boldsymbol{\beta}}\|_1$ and Lemma 1. Since $\widehat{\boldsymbol{u}}_1$ only depends on $\boldsymbol{X}$, $\boldsymbol{Z}$ and $\boldsymbol{\delta}$, which is independent of $\boldsymbol{\epsilon}$, we obtain

$$\widehat{\boldsymbol{u}}_1^{\intercal}\frac{1}{n}\boldsymbol{X}^{\intercal}\boldsymbol{\epsilon} \mid \boldsymbol{Z}, \boldsymbol{X}, \boldsymbol{\delta} \sim N\left(0, \frac{1}{n^2}\|\boldsymbol{X}\widehat{\boldsymbol{u}}_1\|_2^2\sigma_1^2\right)$$

and hence

$$\begin{aligned} &\mathbb{P}\left(\left|\widehat{\boldsymbol{u}}_1^{\intercal}\frac{1}{n}\boldsymbol{X}^{\intercal}\boldsymbol{\epsilon}\right| \geq \frac{1}{n}z_{\alpha/2}\|\boldsymbol{X}\widehat{\boldsymbol{u}}_1\|_2\sigma_1\right) \\ &= \int \mathbb{P}\left(\left|\widehat{\boldsymbol{u}}_1^{\intercal}\frac{1}{n}\boldsymbol{X}^{\intercal}\boldsymbol{\epsilon}\right| \geq \frac{1}{n}z_{\alpha/2}\|\boldsymbol{X}\widehat{\boldsymbol{u}}_1\|_2\sigma_1 \mid \boldsymbol{Z}, \boldsymbol{X}, \boldsymbol{\delta}\right) d\boldsymbol{Z}d\boldsymbol{X}d\boldsymbol{\delta} = 1 - \alpha. \end{aligned} \tag{100}$$

On the event $B_1$, $\boldsymbol{\Omega}\widehat{\boldsymbol{\gamma}}$ belongs to the feasible set of (10). By the definition of $\widehat{\boldsymbol{u}}_1$ defined in (10), we have $\|\boldsymbol{X}\widehat{\boldsymbol{u}}_1\|_2^2/n \leq \|\boldsymbol{X}\boldsymbol{\Omega}\widehat{\boldsymbol{\gamma}}\|_2^2/n$. On the event $G_3$, the term $\|\boldsymbol{X}\boldsymbol{\Omega}\widehat{\boldsymbol{\gamma}}\|_2^2/n$ can be further upper bounded by $C\|\boldsymbol{\Omega}\widehat{\boldsymbol{\gamma}}\|_2^2 \leq C\|\widehat{\boldsymbol{\gamma}}\|_2^2$. Combined with (100), we establish (38) in the main paper.

**Proof of Lemma 3**

The proof of this lemma is similar to that of Lemma 2. We only provide the sketches of the proofs in the following. The proof of (40) in the main paper is similar to (38) in the main paper and follows from the fact

$$\widehat{\boldsymbol{u}}_3^{\intercal}\frac{2}{n}\left(\boldsymbol{X}^{(2)}\right)^{\intercal}\boldsymbol{\epsilon}^{(2)} \mid \boldsymbol{X}, \boldsymbol{Z}, \boldsymbol{\delta}, \boldsymbol{\epsilon}^{(1)} \sim N\left(0, \frac{4}{n^2}\widehat{\boldsymbol{u}}_3^{\intercal}\left(\boldsymbol{X}^{(2)}\right)^{\intercal}\boldsymbol{X}^{(2)}\widehat{\boldsymbol{u}}_3\right).$$

The proof of (41) in the main paper is similar to (37) in the main paper and follows from the inequality $\left|\left(\widehat{\boldsymbol{u}}_3^{\intercal}\widehat{\boldsymbol{\Sigma}}^{(2)} - \widehat{\boldsymbol{\beta}}^{\intercal}\right)\left(\boldsymbol{\beta} - \widehat{\boldsymbol{\beta}}\right)\right| \leq \|\widehat{\boldsymbol{u}}_3^{\intercal}\widehat{\boldsymbol{\Sigma}}^{(2)} - \widehat{\boldsymbol{\beta}}^{\intercal}\|_\infty\|\boldsymbol{\beta} - \widehat{\boldsymbol{\beta}}\|_1$ and the fact that $\|\widehat{\boldsymbol{u}}_3^{\intercal}\widehat{\boldsymbol{\Sigma}}^{(2)} - \widehat{\boldsymbol{\beta}}^{\intercal}\|_\infty \leq C\|\widehat{\boldsymbol{\beta}}\|_2\sqrt{\log p/n}$.

**<u>Proof of Lemma 4</u>**

By (22) in the main paper, on the event $S \cap G \cap B_1$, we have

$$\left|\sqrt{\widehat{\mathrm{Q}}(\boldsymbol{\beta})} - \|\boldsymbol{\beta}\|_2\right| \leq \frac{C\|\boldsymbol{\beta}\|_2\left(\frac{1}{\sqrt{n}} + \frac{k\log p}{n}\right) + C\frac{k\log p}{n}}{\sqrt{\widehat{\mathrm{Q}}(\boldsymbol{\beta})} + \|\boldsymbol{\beta}\|_2} \leq C\left(\frac{1}{\sqrt{n}} + \frac{k\log p}{n}\right) + C\frac{1}{\|\boldsymbol{\beta}\|_2}\frac{k\log p}{n}. \tag{101}$$

Similarly, on the event $S \cap G \cap B_1$, we have

$$\left|\sqrt{\widehat{\mathrm{Q}}(\boldsymbol{\gamma})} - \|\boldsymbol{\gamma}\|_2\right| \leq C\left(\frac{1}{\sqrt{n}} + \frac{k\log p}{n}\right) + C\frac{1}{\|\boldsymbol{\gamma}\|_2}\frac{k\log p}{n}. \tag{102}$$

Combined with Assumption (A2) in the main paper, we establish that, on the event $S \cap G \cap B_1$,

$$\left|\frac{\sqrt{\widehat{\mathrm{Q}}(\boldsymbol{\beta})}}{\|\boldsymbol{\beta}\|_2} - 1\right| \leq \frac{1}{2} \quad \text{and} \quad \left|\frac{\sqrt{\widehat{\mathrm{Q}}(\boldsymbol{\gamma})}}{\|\boldsymbol{\gamma}\|_2} - 1\right| \leq \frac{1}{2}. \tag{103}$$

By Lemma 1 and Assumption (A2) in the main paper, we can also establish that, on the event $S \cap G \cap B_1$,

$$\left|\frac{\|\widehat{\boldsymbol{\beta}}\|_2}{\|\boldsymbol{\beta}\|_2} - 1\right| \leq \frac{1}{2} \quad \text{and} \quad \left|\frac{\|\widehat{\boldsymbol{\gamma}}\|_2}{\|\boldsymbol{\gamma}\|_2} - 1\right| \leq \frac{1}{2}. \tag{104}$$

By the triangle inequality, we have

$$\left\|\frac{\widehat{\boldsymbol{\beta}}}{\sqrt{\widehat{\mathrm{Q}}(\boldsymbol{\beta})}} - \frac{\boldsymbol{\beta}}{\|\boldsymbol{\beta}\|_2}\right\|_2 = \left\|\frac{\widehat{\boldsymbol{\beta}}}{\sqrt{\widehat{\mathrm{Q}}(\boldsymbol{\beta})}} - \frac{\widehat{\boldsymbol{\beta}}}{\|\boldsymbol{\beta}\|_2} + \frac{\widehat{\boldsymbol{\beta}}}{\|\boldsymbol{\beta}\|_2} - \frac{\boldsymbol{\beta}}{\|\boldsymbol{\beta}\|_2}\right\|_2 \leq \|\widehat{\boldsymbol{\beta}}\|_2\frac{\left|\sqrt{\widehat{\mathrm{Q}}(\boldsymbol{\beta})} - \|\boldsymbol{\beta}\|_2\right|}{\sqrt{\widehat{\mathrm{Q}}(\boldsymbol{\beta})}\|\boldsymbol{\beta}\|_2} + \frac{1}{\|\boldsymbol{\beta}\|_2}\|\widehat{\boldsymbol{\beta}} - \boldsymbol{\beta}\|_2.$$

By Assumption (A2) in the main paper, (101), (103) and (104), on the event $S \cap G \cap B_1$,

$$\left\|\frac{\widehat{\boldsymbol{\beta}}}{\sqrt{\widehat{\mathrm{Q}}(\boldsymbol{\beta})}} - \frac{\boldsymbol{\beta}}{\|\boldsymbol{\beta}\|_2}\right\|_2 \leq C\left(\frac{1}{\|\boldsymbol{\beta}\|_2}\left(\frac{1}{\sqrt{n}} + \frac{k\log p}{n}\right) + \frac{1}{\|\boldsymbol{\beta}\|_2^2}\frac{k\log p}{n} + \frac{1}{\|\boldsymbol{\beta}\|_2}\sqrt{\frac{k\log p}{n}}\right) \leq C\frac{1}{\|\boldsymbol{\beta}\|_2}\sqrt{\frac{k\log p}{n}}. \tag{105}$$

Similarly, on the event $S \cap G \cap B_1$, we have

$$\left\| \frac{\widehat{\boldsymbol{\gamma}}}{\sqrt{\widehat{\mathrm{Q}}\left(\boldsymbol{\gamma}\right)}} - \frac{\boldsymbol{\gamma}}{\|\boldsymbol{\gamma}\|_2} \right\|_2 \leq C \frac{1}{\|\boldsymbol{\gamma}\|_2} \sqrt{\frac{k \log p}{n}}. \tag{106}$$

Hence, (43) in the main paper follows from (105) and (106). Note that

$$\begin{aligned}
&\left\langle \frac{\widehat{\boldsymbol{\beta}}}{\sqrt{\widehat{\mathrm{Q}}\left(\boldsymbol{\beta}\right)}} - \frac{\boldsymbol{\beta}}{\|\boldsymbol{\beta}\|_2}, \frac{\boldsymbol{\gamma}}{\|\boldsymbol{\gamma}\|_2} \right\rangle + \frac{\widetilde{\mu}_1}{\sqrt{\widehat{\mathrm{Q}}\left(\boldsymbol{\beta}\right)}\sqrt{\widehat{\mathrm{Q}}\left(\boldsymbol{\gamma}\right)}} = \left( \frac{\widetilde{\mu}_1}{\sqrt{\widehat{\mathrm{Q}}\left(\boldsymbol{\beta}\right)}\sqrt{\widehat{\mathrm{Q}}\left(\boldsymbol{\gamma}\right)}} - \left\langle \frac{\widehat{\boldsymbol{\gamma}}}{\sqrt{\widehat{\mathrm{Q}}\left(\boldsymbol{\gamma}\right)}}, \frac{\boldsymbol{\beta} - \widehat{\boldsymbol{\beta}}}{\sqrt{\widehat{\mathrm{Q}}\left(\boldsymbol{\beta}\right)}} \right\rangle \right) \\
&- \left( \left\langle \frac{\widehat{\boldsymbol{\gamma}}}{\sqrt{\widehat{\mathrm{Q}}\left(\boldsymbol{\gamma}\right)}}, \frac{\widehat{\boldsymbol{\beta}} - \boldsymbol{\beta}}{\sqrt{\widehat{\mathrm{Q}}\left(\boldsymbol{\beta}\right)}} \right\rangle - \left\langle \frac{\boldsymbol{\gamma}}{\|\boldsymbol{\gamma}\|_2}, \frac{\widehat{\boldsymbol{\beta}}}{\sqrt{\widehat{\mathrm{Q}}\left(\boldsymbol{\beta}\right)}} - \frac{\boldsymbol{\beta}}{\|\boldsymbol{\beta}\|_2} \right\rangle \right).
\end{aligned} \tag{107}$$

By (39) in the main paper, we have

$$\left| \widetilde{\mu}_1 - \langle \widehat{\boldsymbol{\gamma}}, \boldsymbol{\beta} - \widehat{\boldsymbol{\beta}} \rangle \right| \leq C \|\widehat{\boldsymbol{\gamma}}\|_2 \left( \frac{1}{\sqrt{n}} + \frac{k \log p}{n} \right), \tag{108}$$

and hence

$$\begin{aligned}
\left| \frac{\widetilde{\mu}_1}{\sqrt{\widehat{\mathrm{Q}}\left(\boldsymbol{\beta}\right)}\sqrt{\widehat{\mathrm{Q}}\left(\boldsymbol{\gamma}\right)}} - \left\langle \frac{\widehat{\boldsymbol{\gamma}}}{\sqrt{\widehat{\mathrm{Q}}\left(\boldsymbol{\gamma}\right)}}, \frac{\boldsymbol{\beta} - \widehat{\boldsymbol{\beta}}}{\sqrt{\widehat{\mathrm{Q}}\left(\boldsymbol{\beta}\right)}} \right\rangle \right| \leq & C \frac{\|\widehat{\boldsymbol{\gamma}}\|_2}{\sqrt{\widehat{\mathrm{Q}}\left(\boldsymbol{\beta}\right)}\sqrt{\widehat{\mathrm{Q}}\left(\boldsymbol{\gamma}\right)}} \left( \frac{1}{\sqrt{n}} + \frac{k \log p}{n} \right) \\
\leq & C \frac{1}{\|\boldsymbol{\beta}\|_2} \left( \frac{1}{\sqrt{n}} + \frac{k \log p}{n} \right),
\end{aligned} \tag{109}$$

where the last inequality follows from (103) and (104). Note that

$$\begin{aligned}
&\left\langle \frac{\widehat{\boldsymbol{\gamma}}}{\sqrt{\widehat{\mathrm{Q}}\left(\boldsymbol{\gamma}\right)}}, \frac{\widehat{\boldsymbol{\beta}} - \boldsymbol{\beta}}{\sqrt{\widehat{\mathrm{Q}}\left(\boldsymbol{\beta}\right)}} \right\rangle - \left\langle \frac{\boldsymbol{\gamma}}{\|\boldsymbol{\gamma}\|_2}, \frac{\widehat{\boldsymbol{\beta}}}{\sqrt{\widehat{\mathrm{Q}}\left(\boldsymbol{\beta}\right)}} - \frac{\boldsymbol{\beta}}{\|\boldsymbol{\beta}\|_2} \right\rangle \\
=& \left\langle \frac{\widehat{\boldsymbol{\gamma}}}{\sqrt{\widehat{\mathrm{Q}}\left(\boldsymbol{\gamma}\right)}} - \frac{\boldsymbol{\gamma}}{\|\boldsymbol{\gamma}\|_2}, \frac{\widehat{\boldsymbol{\beta}} - \boldsymbol{\beta}}{\sqrt{\widehat{\mathrm{Q}}\left(\boldsymbol{\beta}\right)}} \right\rangle + \left\langle \frac{\boldsymbol{\gamma}}{\|\boldsymbol{\gamma}\|_2}, \left( \frac{\widehat{\boldsymbol{\beta}} - \boldsymbol{\beta}}{\sqrt{\widehat{\mathrm{Q}}\left(\boldsymbol{\beta}\right)}} \right) - \left( \frac{\widehat{\boldsymbol{\beta}}}{\sqrt{\widehat{\mathrm{Q}}\left(\boldsymbol{\beta}\right)}} - \frac{\boldsymbol{\beta}}{\|\boldsymbol{\beta}\|_2} \right) \right\rangle.
\end{aligned} \tag{110}$$

We further have

$$\left| \left\langle \frac{\widehat{\boldsymbol{\gamma}}}{\sqrt{\widehat{\mathrm{Q}}\left(\boldsymbol{\gamma}\right)}} - \frac{\boldsymbol{\gamma}}{\|\boldsymbol{\gamma}\|_2}, \frac{\widehat{\boldsymbol{\beta}} - \boldsymbol{\beta}}{\sqrt{\widehat{\mathrm{Q}}\left(\boldsymbol{\beta}\right)}} \right\rangle \right| \leq \left\| \frac{\widehat{\boldsymbol{\gamma}}}{\sqrt{\widehat{\mathrm{Q}}\left(\boldsymbol{\gamma}\right)}} - \frac{\boldsymbol{\gamma}}{\|\boldsymbol{\gamma}\|_2} \right\|_2 \frac{\|\widehat{\boldsymbol{\beta}} - \boldsymbol{\beta}\|_2}{\sqrt{\widehat{\mathrm{Q}}\left(\boldsymbol{\beta}\right)}} \leq C \frac{1}{\|\boldsymbol{\beta}\|_2 \|\boldsymbol{\gamma}\|_2} \frac{k \log p}{n}, \tag{111}$$

where the last inequality follows from (103) and (106). Note that

$$
\begin{aligned}
&\left|\left\langle \frac{\gamma}{\|\gamma\|_2}, \left(\frac{\widehat{\boldsymbol{\beta}}-\boldsymbol{\beta}}{\sqrt{\widehat{\mathrm{Q}}\left(\boldsymbol{\beta}\right)}}\right) - \left(\frac{\widehat{\boldsymbol{\beta}}}{\sqrt{\widehat{\mathrm{Q}}\left(\boldsymbol{\beta}\right)}} - \frac{\boldsymbol{\beta}}{\|\boldsymbol{\beta}\|_2}\right)\right\rangle\right| \leq \|\boldsymbol{\beta}\|_2 \left|\frac{1}{\sqrt{\widehat{\mathrm{Q}}\left(\boldsymbol{\beta}\right)}} - \frac{1}{\|\boldsymbol{\beta}\|_2}\right| \\
&\leq C\frac{1}{\|\boldsymbol{\beta}\|_2}\left(\frac{1}{\sqrt{n}} + \frac{k\log p}{n}\right) + C\frac{1}{\|\boldsymbol{\beta}\|_2^2}\frac{k\log p}{n},
\end{aligned} \tag{112}
$$

where the first inequality follows from Cauchy inequality and the last inequality follows from (101) and (103). Combining (109), (110), (111) and (112), we establish (44) in the main paper. Similarly, we establish (45) in the main paper.

**Proof of Lemma 7** By (7.24) and (7.25) in Cai and Guo (2016*a*), if $2\|\boldsymbol{\beta}^* - \boldsymbol{\beta}\|_2^2/\sigma_0^2 < \log 2/2$, then we have

$$
\chi^2\left(f_{\boldsymbol{\theta}_1}\left(\boldsymbol{y},\boldsymbol{X}\right), f_{\boldsymbol{\theta}_0}\left(\boldsymbol{y},\boldsymbol{X}\right)\right) = \left(1 - \frac{2\|\boldsymbol{\beta}^*-\boldsymbol{\beta}\|_2^2}{\sigma_0^2}\right)^{-\frac{n}{2}} - 1 \leq \exp\left(\frac{2n\|\boldsymbol{\beta}^*-\boldsymbol{\beta}\|_2^2}{\sigma_0^2}\right) - 1, \tag{113}
$$

where the first equality follows from the moment generating function of $\chi^2$ distribution and the second inequality follows from the inequality $1/(1-x) \leq \exp(2x)$ for $x \in [0, \log 2/2]$. Taking $\bar{\epsilon} = \sqrt{\log\left(17/16\right)/2}$, we establish (65) in the main paper.

**Proof of Lemma 9**

By the definition of $L_1$ distances and the independence between $(\boldsymbol{X},\boldsymbol{y})$ and $(\boldsymbol{Z},\boldsymbol{w})$, we have

$$
\begin{aligned}
&L_1\left(f_\pi\left(\boldsymbol{X},\boldsymbol{y},\boldsymbol{Z},\boldsymbol{w}\right), f_{\boldsymbol{\theta}_0}\left(\boldsymbol{X},\boldsymbol{y},\boldsymbol{Z},\boldsymbol{w}\right)\right) = \int \left|f_\pi\left(\boldsymbol{X},\boldsymbol{y},\boldsymbol{Z},\boldsymbol{w}\right) - f_{\boldsymbol{\theta}_0}\left(\boldsymbol{X},\boldsymbol{y},\boldsymbol{Z},\boldsymbol{w}\right)\right| d\boldsymbol{X}d\boldsymbol{y}d\boldsymbol{Z}d\boldsymbol{w} \\
&= \int \left|f_\pi\left(\boldsymbol{X},\boldsymbol{y}\right) f_\pi\left(\boldsymbol{Z},\boldsymbol{w}\right) - f_{\boldsymbol{\theta}_0}\left(\boldsymbol{X},\boldsymbol{y}\right) f_{\boldsymbol{\theta}_0}\left(\boldsymbol{Z},\boldsymbol{w}\right)\right| d\boldsymbol{X}d\boldsymbol{y}d\boldsymbol{Z}d\boldsymbol{w}.
\end{aligned}
$$

By the triangle inequality, the last term in the above inequality can be further upper bounded by

$$
\begin{aligned}
&\int \left|f_\pi\left(\boldsymbol{X},\boldsymbol{y}\right)\left(f_\pi\left(\boldsymbol{Z},\boldsymbol{w}\right) - f_{\boldsymbol{\theta}_0}\left(\boldsymbol{Z},\boldsymbol{w}\right)\right)\right| d\boldsymbol{X}d\boldsymbol{y}d\boldsymbol{Z}d\boldsymbol{w} + \int \left(\left|f_\pi\left(\boldsymbol{X},\boldsymbol{y}\right) - f_{\boldsymbol{\theta}_0}\left(\boldsymbol{X},\boldsymbol{y}\right)\right) f_{\boldsymbol{\theta}_0}\left(\boldsymbol{Z},\boldsymbol{w}\right)\right| d\boldsymbol{X}d\boldsymbol{y}d\boldsymbol{Z}d\boldsymbol{w} \\
&= L_1\left(f_\pi\left(\boldsymbol{X},\boldsymbol{y}\right), f_{\boldsymbol{\theta}_0}\left(\boldsymbol{Z},\boldsymbol{w}\right)\right) + L_1\left(f_\pi\left(\boldsymbol{Z},\boldsymbol{w}\right), f_{\boldsymbol{\theta}_0}\left(\boldsymbol{Z},\boldsymbol{w}\right)\right),
\end{aligned} \tag{114}
$$

where the equality follows from $\int \left|f_\pi\left(\boldsymbol{X},\boldsymbol{y}\right)\right| d\boldsymbol{X}d\boldsymbol{y} = 1$ and $\int \left|f_{\boldsymbol{\theta}_0}\left(\boldsymbol{Z},\boldsymbol{w}\right)\right| d\boldsymbol{Z}d\boldsymbol{w} = 1$.

**Proof of Lemma 10**

The proof of Lemma 10 is a generalization of Lemma 4 in Cai and Guo (2016*a*). In the following, we extend the Gaussian design in Cai and Guo (2016*a*) to Sub-gaussian design considered in this paper. Since the error of the regression is still assumed to be Gaussian, it is sufficient to establish the probability bound of $G_1, G_3$ and $G_4$ for the Sub-gaussian design matrix. We first control the probability of $G_1$. By Corollary 5.17 in Vershynin (2012) and the union bound, we have

$$\mathbb{P}\left(\max_{1\leq j\leq p}\left|\frac{1}{n}\left(\|\boldsymbol{X}_{\cdot j}\|_2^2-\mathbf{E}\|\boldsymbol{X}_{\cdot j}\|_2^2\right)\right|\geq\epsilon\right)\leq 2p\exp\left(-\frac{1}{6}\min\left\{\frac{\epsilon^2}{K^2},\frac{\epsilon}{K}\right\}n\right),$$

where $K=4M_1$. Taking $\epsilon=12M_1\sqrt{\log p/n}$, we have

$$\mathbb{P}\left(\max_{1\leq j\leq p}\left|\frac{1}{n}\left(\|\boldsymbol{X}_{\cdot j}\|_2^2-\mathbf{E}\|\boldsymbol{X}_{\cdot j}\|_2^2\right)\right|\geq 12M_1\sqrt{\frac{\log p}{n}}\right)\leq 2p^{-\frac{1}{2}}, \tag{115}$$

and hence $\mathbb{P}(G_1)\geq 1-2p^{-\frac{1}{2}}$.

By Theorem 1.6 in Zhou (2009), if $n\geq c'M_1^4/\theta^2\times\max\left\{9(2+\alpha_0)^2M_1k\log(5ep/k),9\log p\right\}$, then with probability at least $1-2\exp\left(-c\theta^2n/M_1^4\right)$, for all $\boldsymbol{\eta}$ such that there exists $|J_0|\leq k$ and $\|\boldsymbol{\eta}_{J_0^c}\|_1\leq\alpha_0\|\boldsymbol{\eta}_{J_0}\|_1$,

$$1-\theta\leq\frac{\|\boldsymbol{X}\boldsymbol{\Omega}\boldsymbol{\eta}\|_2}{\sqrt{n}\|\boldsymbol{\Omega}^{\frac{1}{2}}\boldsymbol{\eta}\|_2}\leq 1+\theta,\quad\text{and}\quad 1-\theta\leq\frac{\|\boldsymbol{X}\boldsymbol{\eta}\|_2}{\sqrt{n}\|\boldsymbol{\Sigma}^{\frac{1}{2}}\boldsymbol{\eta}\|_2}\leq 1+\theta. \tag{116}$$

By the discussion after equation (7) in Sun and Zhang (2012), the scaled Lasso estimator $\widehat{\boldsymbol{\gamma}}$ can be viewed as the solution of Lasso algorithm with tuning parameter $\sqrt{2.01\log p/n}\widehat{\sigma}_2$. Combined with Corollary B.2 in the paper Bickel et al. (2009), we can show that the Lasso estimator satisfies the property that $\|\widehat{\boldsymbol{\gamma}}_{J_0^c}\|_1\leq\alpha_0\|\widehat{\boldsymbol{\gamma}}_{J_0}\|_1$ where $J_0=\mathrm{supp}(\boldsymbol{\gamma})$ and hence

$$\left|\frac{\|\boldsymbol{X}\boldsymbol{\Omega}\widehat{\boldsymbol{\gamma}}\|_2}{\sqrt{n}\|\boldsymbol{\Omega}^{\frac{1}{2}}\widehat{\boldsymbol{\gamma}}\|_2}-1\right|\leq\sup_{\substack{J_0\subset\{1,\cdots,p\},\\|J_0|\leq k}}\ \sup_{\substack{\boldsymbol{\eta}\neq 0,\\ \|\boldsymbol{\eta}_{J_0^c}\|\leq\alpha_0\|\boldsymbol{\eta}_{J_0}\|_1}}\left|\frac{\|\boldsymbol{X}\boldsymbol{\Omega}\boldsymbol{\eta}\|_2}{\sqrt{n}\|\boldsymbol{\Omega}^{\frac{1}{2}}\boldsymbol{\eta}\|_2}-1\right|. \tag{117}$$

By taking $\theta=1/2$ and $\boldsymbol{\eta}=\widehat{\boldsymbol{\gamma}}$, if $n\geq Ck\log p$, then we establish $\mathbb{P}(G_3)\geq 1-c\exp(-c'n)$ by the first half of (116) and (117). By the definition of (93) and the second half of (116), we establish $\mathbb{P}(G_4)\geq 1-2\exp(-cn)$.